\documentstyle[12pt]{article}

\input epsf
\newcommand{\mysection}[2]{\section[#1]{#2}\setcounter{equation}
{0}}

\newcommand{\be}{\small\begin{equation}}
\newcommand{\ee}{\end{equation}\normalsize\vspace*{-0.1ex}}
\newcommand{\bea}{\small\begin{eqnarray}}

\newcommand{\eea}{\end{eqnarray}\normalsize\vspace*{-0.1ex}}
\newcommand{\bdm}{\small\begin{displaymath}}
\newcommand{\edm}{\end{displaymath}\normalsize\vspace*{-0.1ex}}
\newcommand{\beas}{\small\begin{eqnarray*}}

\newcommand{\eeas}{\end{eqnarray*}\normalsize\vspace*{-0.1ex}}

\newcommand{\n}{\noindent}
\newcommand{\eps}{\epsilon}
\newcommand{\intl}{\int\limits}
\newcommand{\MS}{\overline{\rm MS}}
\newcommand{\qmu}{\frac{Q^2}{\mu^2}}

\newcommand{\dd}{{\mbox d}}

\begin{document}
\begin{titlepage}
\renewcommand{\thefootnote}{\fnsymbol{footnote}}
\makebox[2cm]{}\\[-1in]
\begin{flushright}
\begin{tabular}{l}
CERN-TH/95-26\\
UM-TH-95-3\\
hep-ph/9502300\\
February 1995
\end{tabular}
\end{flushright}
\vskip1.1cm
\begin{center}
{\Large\bf
Resummation of $(\beta_0 \alpha_s)^n$ Corrections in QCD:
Techniques and Applications to the $\tau$ Hadronic Width and the
Heavy Quark Pole Mass
\\[4pt]
}

\vspace{1.4cm}

Patricia Ball$^1$,
M. Beneke$^2$ and
V.M.\ Braun$^3$\footnote{On leave of absence from St.\ Petersburg
Nuclear Physics Institute, 188350 Gatchina, Russia.}

\vspace{1.4cm}

$^1${\em CERN, Theory Division, CH--1211 Gen\`{e}ve 23, Switzerland}\\[0.5cm]
$^2${\em Randall Laboratory of Physics,
University of Michigan, Ann Arbor, Michigan 48109, USA}\\[0.5cm]
$^3${\em DESY, Notkestra\ss\/e 85, D--22603 Hamburg, Germany}

\vspace{2cm}
{\bf Abstract:\\[5pt]}
\parbox[t]{\textwidth}{
We propose to resum exactly any number of one-loop vacuum polarization
insertions into the scale of the coupling of
lowest order radiative corrections. This makes maximal
use of the information contained in one-loop perturbative corrections
combined with the one-loop running of the effective coupling and
provides a natural extension of the familiar BLM scale-fixing
prescription to all orders in the perturbation theory. It is suggested
that the remaining radiative corrections should be reduced after
resummation. In this paper we implement this resummation
by a dispersion technique and indicate a possible generalization
to incorporate two-loop evolution. We investigate in some detail
higher order perturbative corrections to the $\tau$ decay width and
the pole mass of a heavy quark. We find that these corrections tend
to reduce $\alpha_s(m_\tau)$ determined from $\tau$ decays by
approximately 10\% and increase the difference between the bottom pole
and $\MS$-renormalized mass by 30\%.}

\end{center}
\end{titlepage}

\newpage
\renewcommand{\thefootnote}{\arabic{footnote}}
\setcounter{footnote}{0}


\mysection{Introduction}{Introduction}

During the past decade QCD has turned from qualitative descriptions
to quantitative predictions for the strong interactions. The influx
of more accurate experimental data represents a constant stimulus to
improve theoretical predictions. While an understanding of
confinement remains as evasive as ever, these predictions
rely on the validity of
nonperturbative factorization in processes governed by a large momentum
scale. This allows to isolate the presently uncalculable infrared
dynamics in a set of universal parameters or functions. One is then led to
the conclusion that relations of physical quantities can be calculated
in perturbation theory. For certain observables, such as those
derived from totally inclusive processes with no hadrons in the initial
state, one obtains parameter-free predictions (except for the
strong coupling) to logarithmic accuracy
in the hard scale.

Much effort has been invested in the computation of higher order QCD
radiative corrections. For a few observables for which quark masses
are unimportant, second order corrections in the strong coupling
$\alpha_s$ are known. Third order expressions
exist for hadroproduction in $e^+ e^-$-annihilation
and $\tau$-lepton decays and for the Gross-Llewellyn-Smith and Bjorken
sum rule in deep inelastic scattering. On the other hand, observables
involving quark masses, for example decay widths of hadrons containing
a heavy quark, are known only to first order. In most cases, the extension
of present results is not simply a question of algebraic complexity and
algorithms for handling it. The existing calculations have exploited
presently known techniques to the frontier beyond which new
methods need to be designed, a
task which might not be completed soon. In this situation it is
interesting to explore possible sources of systematically
large corrections in higher orders, which, once identified, might then be
taken into account exactly to all orders. In doing so, one
may hope to reduce
the uncertainty due to ignorance of exact higher order coefficients.

\begin{figure}[t]
   \vspace{-2cm}
   \epsfysize=28cm
   \epsfxsize=20cm
   \centerline{\epsffile{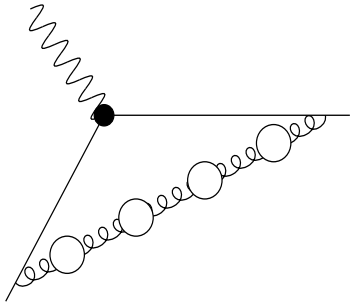}}
   \vspace*{-20cm}
\caption{\label{ex} A typical diagram with multiple fermion loop
insertions into the lowest order correction to a generic physical
quantity.}
\end{figure}

Resummations of this type are familiar and necessary for many
problems involving disparate mass scales. Large perturbative corrections
in higher orders are associated with a logarithm of the ratio
of these scales and can often be summed with renormalization
group techniques, which allow to obtain the dominant power of that
logarithm to all orders by one-loop calculations. Although in practice
it is not always clear whether the logarithms dominate the constant
pieces -- in particular, since the coefficients of logarithms grow
geometrically with order of $\alpha_s$, whereas the constants grow
factorially~--, such a resummation is controlled by an ``external''
parameter (the logarithm of two scales) that can be varied, at least
in fictitious limits. For the problem at hand, we assume that such
renormalization group improvement has already been done or consider
observables that depend only on a single scale. We are thus interested
in systematically large contributions to constant terms with no external
scale at our disposal.
\phantom{\ref{ex}}

The estimation of constant terms of yet uncalculated coefficients, which is
closely related to the problem of scale- and scheme-setting for truncated
perturbative expansions, is often thought to attempt
the impossible. While indeed for any particular observable it is
impossible to assess with rigour the quality of a certain scale-setting
prescription, some prescriptions may be supported by general physical
arguments and turn out to be closer to exact results {\em a posteriori}.
The most distinguished prescription of this kind has been formulated
by Brodsky, Lepage and Mackenzie \cite{BRO83}. Observing that in QED
all scale-dependence of the coupling results from photon vacuum
polarization, they suggested that the effect of fermion loop insertions
into a photon line in higher orders be absorbed into the scale of the
coupling of a given order and proposed to apply this criterion to
QCD as well. In practical applications, this suggestion has mainly
been realized in second order in $\alpha_s$. To this accuracy,
only a single
fermion loop insertion needs to be calculated and the relevant contribution
can be traced by its dependence on the number of light fermions $N_f$.
In general, at order $\alpha_s^{n+1}$, the effect of one-loop evolution
of the coupling can still be obtained by the highest power of $N_f$ in
the flavour-dependence of coefficients. The highest power of $N_f$
originates solely from diagrams with $n$ fermion loop insertions, such
as in Fig.~\ref{ex}, which are much easier to calculate than
flavour-independent terms. In QCD, the identification of contributions
related to one-loop
renormalization of the coupling implies that the coefficient
of the highest power in $N_f$ should in fact not multiply $N_f$
to some power,
but the combination $N_f-33/2$, as it appears in the Gell-Mann-Low
function to leading order. Computing diagrams as in Fig.~\ref{ex} and
then replacing $N_f$ by $N_f-33/2$, one takes into account
partial contributions
from other diagrams, which are much harder to evaluate exactly (the
precise identification of these contributions with particular
diagrams is not straightforward and gauge-dependent). It turns out
that in all cases where comparison with exact second order results is
possible, this replacement approximates the exact coefficient
amazingly well in the $\MS$-scheme \cite{BROT94}. In what follows we
shall refer  as ``Naive Nonabelianization''
(NNA) \cite{BG94} to the hypothesis that a substantial part of
higher order
radiative corrections can be accounted for by running of the coupling,
in the sense of substitution of $N_f\rightarrow N_f-33/2$ in the term
with the highest power of $N_f$. We should mention that
BLM scale-setting and NNA
are technically identical and the distinction is rather a matter of
interpretation: The BLM scale-setting by its physical motivation
does not necessarily claim that genuine higher order corrections
missed by the above substitution should be small, whereas NNA assumes
this stronger assertion.

This paper is concerned with an exposition of details of the recent
proposal \cite{BB94b} of two of us
to extend the BLM procedure to higher orders and, in particular,
to sum all contributions associated with one-loop running of the
coupling. We emphasize that this procedure has repeatedly been
suggested in the past \cite{BRO83,LEP93}, but has apparently not
been pursued to the point of practical implementation. A summation
similar to, but technically different from \cite{BB94b} was also
presented in \cite{Nnew}. We believe that this resummation
is useful, because the effect of evolution of the coupling is
a systematic source of potentially large perturbative coefficients.
In such a situation it is advantageous to incorporate this effect
in theoretical predictions even if it transcends a fixed-order
perturbative approximation.

To become definite, we use the following convention for the QCD
$\beta$-function\footnote{Since we discuss corrections proportional
to $(-\beta_0\alpha_s)^n$, the reader accustomed to the normalization
$11-(2/3) N_f$ may think of an expansion in terms of $\alpha_s/(4\pi)$.}:

\bea
     \mu^2\frac{d\alpha_s}{d\mu^2}&=&\beta(\alpha_s)=\beta_0\alpha_s^2
    +\beta_1\alpha_s^3 +\ldots\,;
\nonumber\\
    \beta_0 &=& -\frac{1}{4\pi}\left(11-\frac{2}{3} N_f\right)\,,
\hspace{0.5cm}
    \beta_1 = -\frac{1}{(4\pi)^2}\left(102-\frac{38}{3} N_f\right)\,.
\eea

\n For a generic physical observable $R$ (which for the sake of illustration
we assume to depend only on a single, large scale $Q$), we may eliminate
the $N_f$-dependence in radiative corrections in favour
of the $\beta$-function
coefficients $\beta_0$, $\beta_1$ etc. For now (and most parts of this
paper) we restrict ourselves to one-loop running, induced by $\beta_0$.
The (truncated) perturbative expansion of $R$ is written as

\bea
 R-R_{tree} &=& \sum_{n=0}^N[r_{n0}+r_{n1}N_f
+\ldots + r_{nn}N_f^n]\alpha_s(Q)^{n+1}
\nonumber\\&=&
 r_0 \alpha_s(Q)\sum_{n=0}^N\left[\delta_n+d_n (-\beta_0)^n\right]
\alpha_s(Q)^n\,,
\eea

\n where

\be\label{dn}
    d_n = (-6\pi)^n \frac{r_{nn}}{r_0}\,.
\ee

\n Let us stress again that $d_n$ is unambiguously determined by diagrams
with $n$ insertions of fermion bubbles as in Fig.~\ref{ex} and only
$\delta_n$ requires a true $n+1$-loop calculation. The $\delta_n$
($n\ge 2$) could
be further separated into contributions from two-loop evolution and
the effect of one-loop evolution on the genuine two-loop corrections
(see \cite{BRO94} for $\delta_2$). We introduce

\be\label{M}
M_N(-\beta_0\alpha(Q)) = 1+\sum_{n=1}^N d_n (-\beta_0\alpha(Q))^n
\ee

\n as a measure of how much the lowest order correction is modified by
summing $N$ one-loop vacuum polarization insertions.
To make an explicit connection to BLM scale-setting, we define

\be\label{blmscalesdef}
\label{scale} \alpha_s(Q^*_N) = \alpha_s(Q)\,M_N(-\beta_0 \alpha_s(Q))\,
\ee

\n absorbing the effect of vacuum polarization in the scale $Q_N^*$ of the
coupling. For $N=\infty$,
$\alpha_s(Q^*)$ ($Q^*\equiv Q^*_\infty$) is transparently interpreted as the
running coupling averaged over the lowest order corrections. Isolating the
integration over gluon virtuality in the lowest-order diagram, we may write

\be r_0\alpha_s(Q)=\alpha_s(Q)\int\mbox{d}^4 k\,F(k,Q)\,\frac{1}{k^2} .
\ee

\n Then

\be \label{average}
r_0\alpha_s(Q^*) = \frac{r_0\alpha_s(Q)}
{1-\beta_0\alpha_s(Q)\ln({Q^*}^2/Q^2)}
= \int\mbox{d}^4 k\,F(k,Q)\,\frac{\alpha_s\left(
k\exp[C/2]\right)}{k^2},
\ee

\n where  $C$ is a the scheme-dependent subtraction constant for the
fermion loop. The difference between the
leading-order BLM scale $Q_1^*$ and $Q^*$ is
precisely the difference between averaging the running coupling itself,
or the logarithm of its argument, over the lowest-order diagram.
The values of $Q^*$ and $Q^*_N$ for $N>1$
depend on the value of $\alpha_s(Q)$. Such a dependence is intrinsic to any
generalization of the BLM prescription beyond
leading-order and has previously
been noted in \cite{GRU92,BRO94}. As a result we may express any physical
quantity as

\be R-R_{tree} = r_0\left(\alpha(Q)
M_\infty(-\beta_0\alpha_s(Q))+\sum_{n=1}^\infty\delta_n\alpha_s(Q)^
{n+1}\right)\,.
\ee

Separating radiative corrections in this way and summing a partial set
of them into $M_\infty(-\beta_0\alpha_s)$
can be motivated in different ways. First of all, as already
mentioned, this procedure is supported empirically at the two-loop
level.
Provided the three-loop coefficient is sizeable, one might again
expect that a significant portion of it can already be accounted for
by $M_2(-\beta_0\alpha_s)$ and, in higher orders, by
$M_N(-\beta_0\alpha_s)$.

It must be noted that
such a statement depends on the choice of scheme and, since the $\delta_n$
depend on $N_f$ for $n>1$, also on $N_f$ for a given scheme. Although
$Q^*$ is defined such that $\alpha_s(Q^*)$ is scheme-independent if
one consistently uses the one-loop $\beta$-function, an arbitrary
$N_f$-independent finite renormalization can always ruin the
supposed dominance of $\beta_0^n\alpha_s^{n+1}$-terms by implying
a definition of the coupling that leads to anomalously large values
of the remaining terms $\delta_n$. This remark applies of course to any
partial resummation, whether it is based on a systematic (external)
parameter or not as in the case at hand\footnote{We do not consider
$\beta_0$ on the same footing as, say, $\ln Q^2/m^2$, since it is
an intrinsic parameter of QCD. However, we shall somewhat loosely refer
to the approximation of ignoring the $\delta_n$ as the ``large-$\beta_0$
limit''.}. In this respect a resummation utilizing one-loop running
is similar to resummation of $\pi^2$'s that arise from analytic
continuation from the space-like to the time-like region of, for
instance, the Sudakov form factor \cite{STE87} or the coupling as
in $e^+ e^-$-annihilation or $\tau$ decays \cite{DIB92},
and, in fact, comprises the resummation for $\tau$ decays as a
special case. Since empirically
the $\beta_0$-term dominates second order radiative corrections for
many observables in the $\MS$ scheme,
one would not expect the resummation of $\beta_0\alpha_s^{n+1}$-terms to
be numerically useful in schemes that
differ from $\MS$ by a redefinition of the coupling that changes
significantly the relative importance of flavour-dependent and
flavour-independent terms.

The dominance of
$\beta_0^n\alpha_s^{n+1}$-terms is lended further
support from the behaviour of
perturbative coefficients in large orders (renormalons). It is
precisely the effect of vacuum polarization, which leads to the
expectation that in sufficiently large orders, the coefficients should
indeed have the form

\be r_n\sim K\,(a\beta_0)^n n! n^b\,.
\ee

\n The series is thus divergent and, provided it is asymptotic at all,
cannot approximate the true result to any desired accuracy, even if the
exact coefficients could be computed to arbitrary order. The
evaluation of diagrams as in Fig.~\ref{ex} provides some insight into
this ultimate limitation of perturbative QCD and into the approach to
the asymptotic limit. In this paper, however,
we are not so much
interested in large orders and effects suppressed by powers
of the hard scale $Q$. Summation of effects associated with one-loop
running of the
coupling should prove useful in intermediate orders (say,
$n=2 - 6$), when perturbation theory is still reliable. It is amusing
to speculate whether the empirical dominance of $\beta_0$ at the
two-loop
level indicates that one is already close to the asymptotic behaviour.
The latter is derived from a saddle point
expansion and the large contribution
arises from internal momenta either much smaller or much larger than
the external scale $Q$. Proximity to the asymptotic behaviour -- if true --
would indicate that the distribution of internal momenta is
already well-approximated by a Gaussian even though the main contribution
to the Feynman integral is not from small or large momenta.

The consideration of large orders provides also useful guidance to
the limitations that arise from the restriction to one-loop evolution
effects. As a matter of fact, diagrams associated with one-loop
evolution do not provide the correct constants $b$ and $K$ in the
large-$n$ behaviour. At large $n$, the effect of two-loop evolution
on a single gluon line
and one-loop evolution on two gluon lines becomes equally important
and should eventually be taken into account\footnote{
For example, an infinite number of two-loop insertions is necessary
to obtain the correct value of $b$ \cite{ZAK92}.}.
In practical applications,
we will often find that the series has to be truncated due to its
divergence at rather low $n$. Thus, many diagrams which are required
to establish the formal large-$n$ behaviour never become relevant.
Moreover, provided $\delta_1$
is already not large, one should also expect that vacuum polarization
insertions into two gluon lines will remain comparatively small.

Let us repeat that the resummation of $\beta_0^n\alpha_s^{n+1}$-terms
can only be judged {\em a posteriori} and can fail to provide a
good estimate of higher order corrections for any particular quantity.
Even in this case, we believe that this resummation is physically
motivated and that it is appropriate to absorb this particular class
of higher order corrections into the scale of the coupling in lowest
order. Application of BLM scale-setting to leading order (absorbing
the contribution from one fermion loop) yields $Q_1^*=Q e^{-d_1/2}$.
Upon re-expansion of $\alpha_s(Q_1^*)$ to one-loop accuracy for the
$\beta$-function, this pretends a geometric
growth $d_n^{\rm BLM}=d_1^n$ to be compared with the factorial growth
of the exact coefficients. Resummation to all orders corrects for this
discrepancy by adjusting $Q^*_\infty$ such that the expansion of
$\alpha_s(Q^*_\infty)$ gives the correct values of $d_n$. Thus to
the accuracy of one-loop evolution the result of resummation is
equivalently expressed as $\alpha_s(Q) M_\infty(-\beta_0\alpha_s)$
and we often prefer this form of presentation. It is no longer
equivalent to $\alpha(Q^*_\infty)$, if one adopts two-loop accuracy
for $\alpha_s$. Since for large $n$ the true $d_n$ will always
outgrow $d_1^n$, one might conclude that the usual BLM scale-setting
underestimates higher order radiative corrections. In practice,
the effect is often just the opposite: The most
important contributions come from intermediate $n$ and it turns out that
in many cases $d_1$ is comparatively large ($Q_1^*$ is small), so
that $d_1^n > d_n$ in this region. The usual BLM scale-setting therefore
typically overestimates the size of radiative corrections associated
with one-loop running.

The remainder of the paper is organized as follows: In Sect.~2 we
develop in detail the technique to implement the resummation.
We use a dispersion technique to reduce the problem to a one-dimensional
integral over lowest order radiative corrections computed with
finite gluon mass which is suitable to numerical evaluation.
Thus compared to the complications of a complete higher
order calculation, this resummation can be performed with little
computational expense. Compared to a similar implementation of the
standard BLM scale-setting \cite{SV94}, the computation of higher
order coefficients $d_n$ comes with no additional expense at all. It is
convenient to introduce the Borel transform as a generating function
of higher order radiative corrections. The principal value definition
of the Borel integral serves as a starting point for the summation of
the series within a certain accuracy, indicated by the presence of
infrared renormalons. We shall also see that this definition requires
all kinematic invariants to be large compared to $\Lambda_{\rm QCD}$,
reflecting the inapplicability of perturbation theory as a starting
point for summation, if this requirement is not met. In this Section
we also generalize the results of \cite{BB94b} to quantities with
anomalous dimension and include quarks with finite masses
in the loops.

In Sect.~3 we apply the resummation to the
hadronic width of the $\tau$
lepton and find a 10\% decrease of $\alpha_s(m_\tau)$ due to
four- and higher loop corrections.
We discuss the possibility of $1/m_\tau^2$-corrections in the
light of our resummation. In general we prefer to be agnostic about
power corrections and stick to perturbation theory. An exception to
this rule is that we do not want to introduce power corrections in
conflict with the operator product expansion in euclidian space
without
good reasons (which we do not have). Under this assumption we show that
principal value resummation does not introduce $1/m_\tau^2$-terms to the
decay width, provided the coupling is chosen appropriately. Sect.~4
contains a detailed discussion of the resummation of
$(-\beta_0)^n\alpha_s^{n+1}$-terms for the pole mass
of a heavy quark. We combine
the resummation with the exact two-loop result to give an estimate
of the difference between the pole mass and the $\MS$-renormalized
mass. We keep finite quark masses inside loops which allows to
trace the origin of large coefficients to the relevant regions of
internal momentum.

In Sect.~5 we formulate one possible extension of
our resummation to incorporate partially the effect of two-loop
running resulting in $\beta_1 \beta_0^{n-2}$-corrections.
This extension is again guided by the flavour-dependence
of coefficients and can be considered as an extension of recent work
by Brodsky and Lu \cite{BRO94}. The size of corrections is
illustrated for the
vector correlation function relevant to $\tau$ decays and for the
pole mass of a heavy quark. Sect.~6 contains conclusions. Three Appendices
deal with technical issues: In Appendix A we derive a simple expression for
subtractions needed for logarithmically ultraviolet divergent quantities
and collect the details of results presented in Sect.~2. Appendix B
contains analytic formulae
for the lowest order radiative corrections to $R_{e^+ e^-}$ and
hadronic $\tau$ decays with finite gluon mass. In Appendix
C we list exact results for some abelian five-loop diagrams to the
vector correlation function.

In a companion paper \cite{BBB} we shall discuss the implications
of resummation for heavy quark decays and the determination of
$|V_{bc}|$.


\mysection{Techniques}{Techniques}
\setcounter{equation}{0}

The aim of this Section is to develop a systematic approach to
the calculation of diagrams with an
arbitrary number of fermion loop insertions,
such as in Fig.~1.
We assume a generic physical (short-distance)
quantity $R$ and a renormalization
scheme that does not introduce an artificial $N_f$ dependence, such
as $\MS$. It is also assumed that lowest order radiative corrections
to $R$ do not involve the gluon self-coupling.
Subtracting the tree-level contribution,
we are left with the (truncated) perturbative expansion

\be\label{rtree}
  R-R_{tree} = \sum_{n=0}^N r_n\alpha_s^{n+1}\,.
\ee

\n The coefficients $r_n$ are polynomials in $N_f$\,,

\be\label{flavourseries}
 r_n=r_{n0}+r_{n1} N_f+\ldots+r_{nn} N_f^n,
\ee

\n and we will calculate the coefficients $r_{nn}$, which
originate from $n$ fermion loop insertions into the lowest-order
diagram. We then write

\be\label{defd_n}
 r_n= r_0\left[\delta_n+(-\beta_0)^n d_n\right],
\ee

\n where $d_n=(-6\pi)^n r_{nn}/r_0$ absorbs the term with the largest power
of $N_f$. The effect of one-loop evolution of the coupling on lowest order
radiative corrections is then entirely contained in the $d_n$ and in the
remainder of the paper we will consider the $\delta_n$ as
corrections to the approximation of ``Naive Nonabelianization''. As a
measure of how much the lowest order radiative correction
is modified by
including $N$ vacuum polarization insertions, we define

\bea\label{defM}
M_N(-\beta_0\alpha_s) &\equiv& 1+\sum_{n=1}^N d_n
(-\beta_0\alpha_s)^n\,,
\nonumber\\
M_\infty(-\beta_0\alpha_s)&\equiv& M_{N\to\infty}(-\beta_0\alpha_s)\,.
\eea

\n Taken literally, the limit $N\rightarrow\infty$ does not exist, because
the series diverges. It will be interpreted in the sense of
Eq.~(\ref{borelintegral}) below.
For further use, we introduce the shorthand notation

\be
a_s(\mu) = -\beta_0 \alpha_s(\mu)\,,
\ee

\n where $\mu$ is the normalization scale, which often will
be suppressed for brevity. In this Section
we are concerned only
with technical aspects of the calculation of $M_N(a_s)$ and
$M_\infty(a_s)$.

\subsection{Borel transform vs. finite gluon mass}

A convenient way to deal with diagrams with multiple loop
insertions is to calculate the Borel transform

\be\label{borelsum}
B[R](u) \equiv \sum_n \frac{r_n}{n!}(-\beta_0)^{-n} u^n\,,
\ee

\n which can be used as a generating function for the fixed-order
coefficients \cite{BEN93}:

\be\label{genfunction}
 r_n = (-\beta_0)^n \frac{d^n}{du^n} B[R](u)_{|_{u=0}}\,.
\ee

\n Another advantage of presenting the results in form of the
Borel transform is that the result for the sum of all diagrams
can easily be recovered by the integral representation

\be\label{borelintegral}
r_0 M_\infty(a_s) = \frac{1}{a_s}
\int_0^\infty du\, e^{-u/a_s} B[R](u)\,,
\ee

\n where the integration goes over positive values of the Borel
parameter $u$. Note that we define the Borel parameter $u$
with an additional factor $(-\beta_0)$ compared to the conventional
definition. In fact, Eq.~(\ref{borelintegral})
requires some explanation: As it stands the integral
is not defined, because the Borel transform generally
has singularities on the integration path, known as infrared
renormalons. We shall adopt a definition of the integral based
on deforming the contour above or below the real axis or on
a principal value prescription. These prescriptions are not unique
and their difference, which is exponentially small in the
coupling, must be considered as an uncertainty which
can not be removed within perturbation theory. This will
be discussed in more detail. A second question
concerns the existence of the principal value
integral and the behaviour of the
Borel transform at $u=\infty$. If we consider a physical quantity
that depends only on a single scale $Q$, then, to one-loop running
accuracy, renormalization group invariance entails that the Borel
transform can be written as $(\mu^2/Q^2)^u$ times a $Q$- and
$\mu$-independent function $F(u)$. Combining the factor
$(\mu^2/Q^2)^u$ with $e^{-u/a_s(\mu)}$ in Eq.~(\ref{borelintegral}),
we deduce that the principal value integral exists, provided
that $Q^2$ is sufficiently large compared to $\Lambda_{\rm QCD}^2$
and $F(u)$ does not increase faster than any exponential as
$u$ approaches
positive infinity. Since the second property is satisfied in all
examples which we shall meet (and we may conjecture that this
is generally true for the Borel transform computed from higher
order corrections due to vacuum polarization), we shall assume
in general that all kinematic invariants on which $R$ depends
explicitly are sufficiently large compared to $\Lambda_{\rm QCD}^2$
(to be precise, the difference of two invariants is also an
invariant).
This is not an additional assumption needed to take the Borel
integral. Without it perturbative methods are not applicable
and there is no perturbation theory to start with. In particular,
we do not know whether the Borel transform, which is useful in
connection with short-distance expansions, can serve as a
{\em bona fide} starting point to summation in the strong coupling
regime\footnote{If some kinematic invariants are small,
one might still be able to define the Borel
integral as an analytic
continuation. However, in this regime all power corrections
in a short-distance expansion are of the same importance and
the analytic continuation is useless unless the summation of
the short-distance expansion is understood.}.

In simple cases the Borel transform can be calculated directly.
This is due to the fact that the evaluation of diagrams with multiple
fermion bubble insertions in Landau gauge corresponds to the evaluation
of the lowest-order diagram with the effective propagator

\be\label{propagator}
 D^{AB}(k) = i\delta^{AB}\frac{k_\mu k_\nu-k^2 g_{\mu\nu}}{k^4}
\frac{1}{1+\Pi(k^2)}\,,
\ee

\n where

\be
\Pi(k^2) = a_s \ln\left(-\frac{k^2}{\mu^2}e^C\right)
\ee

\n and $C$ is a scheme-dependent finite renormalization constant.
In the $\MS$-scheme $C=-5/3$, in the V-scheme \cite{BRO83}
$C=0$.

For a chain of fermion loops contributing to a physical amplidude with
euclidian external momenta, one can separate the integration over the
gluon momenta to write it as

\be\label{bs}
r_0 \alpha_s(\mu) M_\infty(a_s)
=\int d^4k\, F(k,Q)\frac{1}{k^2}\frac{\alpha_s(\mu)}{1+\Pi(k^2)}\,,
\ee

\n where the transverse projector that appears in the gluon propagator
in Landau gauge is assumed to be included in the function $F(k,Q)$ and
$Q$ stands for a collection of kinematic invariants.

A crucial simplification arises since for diagrams with only one
fermion bubble
chain the Borel transformation applies to the expansion in $\alpha_s$
of the propagator in Eq.~(\ref{propagator}) itself, rather than to
the set of diagrams as a whole. The effective
(Borel-transformed) gluon propagator
is \cite{BEN93}:

\be\label{gluonprop}
B[\alpha_s D_{\mu\nu}^{AB}(k)](u)=
i\delta^{AB} \left(\frac{e^C}{\mu^2}\right)^{-u} \frac{
k_\mu k_\nu - k^2 g_{\mu\nu}}{(-k^2)^{2+u}}\,.
\ee

\n Thus, the task of calculating the Borel transform of Feynman
diagrams with bubble insertions reduces to the calculation of the
leading-order diagram with the usual gluon propagator raised to
an arbitrary
power, which is familiar from analytic regularization \cite{SPE68}.
This trick suffices to derive an expression for the Borel
transform of the polarization operator with light quarks
\cite{BEN93,BRO93,BB94}, of the
heavy quark self-energy \cite{BB94} and several more complicated cases in
connection with heavy quark expansions, which can be
found in \cite{BB94,NS94,LMS94,BG94}.

However, in most phenomenologically interesting cases an analytic
expression for the Borel transform is difficult to obtain, especially
for observables involving more than one scale. Even if the
exact Borel transform is obtainable, taking $n$ derivatives
to evaluate fixed-order coefficients (see Eq.~(\ref{genfunction}))
may turn out
to be a complicated task. In this paper we shall work out a different
technique, which extracts the desired information on higher orders from the
lowest-order diagrams, calculated with a finite
gluon mass \cite{BB94b,SV94}.
Calculations of one-loop diagrams with finite gauge boson mass have
become routine for electroweak radiative corrections, which allows
to hope that this technique is generally applicable to a wide range
of observables in QCD. In this way one obtains a concise integral
representation for the Borel transform.

For a while let us restrict our discussion to euclidian quantities,
which are not sensitive to the gluon self-coupling to leading
order.\footnote{With the latter restriction, we do not have problems
with gauge invariance, which otherwise prohibits introduction of a finite
gluon mass, unless QCD is embedded, for example, in an $SU(3)$ Higgs
model.}
Call $r_0(\lambda^2)$ the leading-order radiative correction calculated
with finite gluon mass $\lambda$ and $r_0\equiv r_0(0)$.
To be precise, we define $r_0(\lambda^2)$ as the sum of all Feynman
diagrams to leading order, which in general may be ultraviolet
divergent and need to be renormalized.
In this Section we restrict our discussion to
cases where no explicit renormalization is needed, which is the case,
e.g. for the derivative of the polarization
operator in Eq.~(\ref{polaroper}),
or transition amplitudes related to heavy quark decays with on-shell
mass renormalization.
It is easy to show that this assumption is equivalent to the
requirement that
$r_0(\lambda^2)$ vanishes as $\lambda^2\to\infty$.
Equally, in the case of Borel representation for the diagrams with
fermion bubbles, we assume that fermion loops are renormalized, and
no additional explicit renormalization is necessary.
This assumption can easily be relaxed.
A detailed discussion of
renormalization is given in Appendix A, where we work out
the missing overall subtractions for the general case.

 We keep the standard gauge-fixing and work with the propagator

\be\label{massgluonprop}
-i\delta^{AB}\frac{1}{k^2-\lambda^2+i\epsilon}
\left[g_{\mu\nu} - (1-\xi) \frac{k_\mu k_\nu}{k^2-\xi\lambda^2
+i\epsilon}\right] \,.
\ee

\n The relation to the Borel transform of the massless propagator
in Eq.~(\ref{gluonprop}) (in Landau gauge, $\xi=0$, or Feynman gauge,
$\xi=1$) is established
by an (inverse) Mellin representation

\be
\frac{1}{k^2-\lambda^2} = \frac{1}{2\pi i}\,\frac{1}{k^2}
\!\!\!\int\limits_{-1/2-i\infty}^{-1/2+i\infty}
\!\!\!\!\!\mbox{d} u\,\Gamma(-u)
\Gamma(1+u)\left(-\frac{\lambda^2}{k^2}\right)^u\, .
\ee

\n Writing down the leading-order radiative correction
with non-vanishing gluon mass $\lambda$ (in Landau gauge) as

\be\label{finitemass}
r_0(\lambda^2)= \int d^4k\, F(k,Q)\frac{1}{k^2-\lambda^2}\,.
\ee

\n and comparing to Eq.~(\ref{bs}), one finds the identity \cite{BBZ94}

\be\label{relation}
r_0(\lambda)\,=\,\frac{1}{2\pi i}
\!\!\!\int\limits_{-1/2-i\infty}^{-1/2+
i\infty}\!\!\!\!\!\mbox{d} u \,\Gamma(-u)\Gamma(1+u)\left(\frac{
\lambda^2}{\mu^2} e^C\right)^u\,B[R](u)
\ee

\n  Taking the inverse, we get \cite{BB94b}

\bea\label{final1}
B[R](u) &=&
 -\frac{\sin(\pi u)}{\pi }\int_0^\infty \frac{d\lambda^2}{\lambda^2}\,
\left(\frac{\lambda^2}{\mu^2}e^C\right)^{-u}
 \left[r_0(\lambda^2)-r_0(0)\right]
\nonumber\\&=&
 -\frac{\sin(\pi u)}{\pi u }\int_0^\infty d\lambda^2\,
\left(\frac{\lambda^2}{\mu^2}e^C\right)^{-u}
 r'_0(\lambda^2)
\eea

\n where $ r'_0(\lambda^2) =(d/d \lambda^2)r_0(\lambda^2)$. If $r_0(\lambda^2)
\sim \lambda^{2 a}$ for small $\lambda^2$, the first line exists, when
$0 < u < a$. The second line provides the analytic continuation to an interval
about $u=0$, so that derivatives at zero can be taken. Thus fixed-order
coefficients $r_n$ can  be
expressed in terms of the integrals

\be
J_k\equiv\int_0^\infty d\lambda^2\,\ln^k(\lambda^2/\mu^2)\,r'_0(\lambda^2)
\qquad k\le n\,.
\ee

\n For instance (cf.~Eqs.~(\ref{defd_n}) and (\ref{genfunction})),
\bea
d_0 &=& \frac{1}{r_0}\left[-J_0\right] = \,1\nonumber\\
d_1 &=& \frac{1}{r_0}\left[J_1+C J_0\right]\nonumber\\
d_2 &=& \frac{1}{r_0}\left[-J_2-2 C J_1-\left(C^2-\frac{\pi^2}{3}
\right) J_0\right]\\
d_3 &=& \frac{1}{r_0}\left[J_3+3 C J_2+(3 C^2-\pi^2) J_1+
(C^3-C\pi^2) J_0\right]\nonumber
\eea

\n etc. and $r_{n}=r_0\,(-\beta_0)^n d_n$.
$r_{1}$ coincides with the result of Smith and Voloshin \cite{SV94}
(who use a scheme with $C=0$), integrating of their
expression by parts. \\

Next, we evaluate the Borel integral in Eq.~(\ref{borelintegral}) to
obtain a compact answer for the sum of diagrams with fermion bubble
insertions to arbitrary order, up to the ambiguities caused by renormalons
as mentioned below Eq.~(\ref{borelintegral}). For the following
derivation we assume that the Borel integral is defined with a contour
slightly above the positive real axis. We also recall that this integral
exists, if all kinematic invariants are large compared to
$\Lambda_{\rm QCD}^2$, which we assume. To find a
representation for the so-defined sum in terms of an integral over
$\lambda^2$, one needs to insert
Eq.~(\ref{final1}) in Eq.~(\ref{borelintegral}) and take the $u$-integral
explicitly. The $u$-integration is elementary, but the interchange of
orders of integration in $u$ and $\lambda^2$ can not be done
immediately, because the $\lambda^2$-integral in Eq.~(\ref{final1})
is not defined for all $u$ on the integration path.
It is convenient to use  the first  representation
in Eq.~(\ref{final1}) and write it as\footnote{The separation of the
two terms in the following equation reintroduces the pole at $u=0$
in each of the two $\lambda^2$-integrals. The pole cancels in the sum
of both terms, and both terms can be manipulated separately
regardless of this pole. This can be justified as follows:
One splits the $\lambda^2$-integral before
separation into the two terms at some sufficiently large
$\lambda_T^2$. The contribution from $\lambda^2_T$ to infinity
can be handled without difficulty. For the integral form 0
to $\lambda_T^2$ one can proceed as following Eq.~(\ref{split})
and both terms have no singularity at $u=0$.}

\bea\label{split}
   r_0  M_\infty(a_s) &=&
 -\frac{1}{2\pi i}\int_0^\infty du\, e^{-u/a_s}
\int_0^\infty \frac{d\lambda^2}{\lambda^2}\,
\left(\frac{\lambda^2e^{-i\pi}}{\mu^2}e^C\right)^{-u}
 \left[r_0(\lambda^2)-r_0(0)\right]
\nonumber\\&&\hspace{-1.5cm}+\,
 \frac{1}{2\pi i}\int_0^\infty du\, e^{-u/a_s}
\int_0^\infty \frac{d\lambda^2}{\lambda^2}\,
\left(\frac{\lambda^2e^{i\pi}}{\mu^2}e^C\right)^{-u}
 \left[r_0(\lambda^2)-r_0(0)\right]
\eea

\n The integration
over $u$ in the first of the two terms above can easily be taken,
rotating the contour to the positive imaginary axis. The quarter-circle
at infinity does not contribute, because of the behaviour of the
Borel-transform at infinity as discussed above.
The $\lambda^2$-integral exists for positive imaginary $u$ and
the order of integrations
can be interchanged. The $u$-integral in the second
term would equally easily be taken by rotating
the contour to the negative imaginary axis, but
the infrared renormalon singularities on the
real axis obstruct this deformation.
It can only be done at the price of picking
up the residues from the infinite
number of poles. A more elegant solution is to rotate first the
integration contour over $\lambda^2$ to the second
sheet: $\lambda^2\to\lambda^2 e^{-i(\pi+\eps)}$. Here we note that
for euclidian quantities, there are no singularities of
the $\lambda^2$-integrand in the lower complex plane.
Next, the $u$-integration can again be performed by rotation
to the positive imaginary axis, as above. This integration gives

\be\label{howpoleappears}
\int_0^{i\infty} du\, \exp\{-u [1/a_s+\ln(|\lambda^2|/\mu^2e^C)]
-i\epsilon\} =
\frac{a_s}{1+a_s\ln(|\lambda^2|/\mu^2e^C)-i\epsilon}
\ee

\n and introduces a pole singularity,
located at

\be\label{Landau}
\lambda_L^2 =- \mu^2\exp[-1/a_s-C]\,,
\ee

\n which is simply the
position of the Landau pole in the running coupling (in the
V-scheme). Now one can rotate the $\lambda^2$-integral back
from the second sheet to the real positive
axis. In this way one encounters the pole in Eq.~(\ref{howpoleappears}),
whose residue has to be added.
Collecting all the terms, we get after some algebra:

\be\label{firstversion}
   r_0  M_\infty(a_s) = -\!\int_0^\infty \frac{d\lambda^2}{\lambda^2}
\frac{a_s}{|1+a_s\ln(-\lambda^2/\mu^2 e^C)|^2}
\left[r_0(\lambda^2)\!-\!r_0(0)\right] +
\frac{1}{a_s}\left[r_0(\lambda_L^2\!-i\epsilon)\!-\!r_0(0)\right]
\ee

\n where the $i\epsilon$  prescription corresponds to defining the
Borel integral above the real axis: The contours are rotated
in the opposite directions if the Borel integral
is defined below the real axis, with the only modification that the
sign of the $i\eps$-prescription is reversed in the result.
Finally, integrating by parts in the first term, we obtain

\be\label{rBS}
r_0 a_s(\mu) M_\infty(a_s(\mu)) =
\int_0^\infty d\lambda^2\, \Phi(\lambda^2)\,r'_0(\lambda^2)
+[r_0(\lambda_L^2-i\eps)-r_0(0)]
\ee

\n where

\be\label{Phi}
\Phi(\lambda^2)= -\frac{1}{\pi}\arctan\left[\frac{a_s(\mu)\pi}
{1+a_s(\mu)\ln(\lambda^2/\mu^2 e^C)}\right] -
\,\Theta(-\lambda_L^2-\lambda^2)\,,
\ee

\n which coincides with the result of \cite{BB94b}, obtained by
a different method.
Note that the term with the $\Theta$-function exactly cancels the
jump of the $\arctan$ at $\lambda^2=-\lambda_L^2$.

Eq.~(\ref{rBS}) presents the desired answer for the sum of
diagrams with any number of fermion bubbles in terms of an integral
over the gluon mass. This relation is
one of the main technical tools which
we suggest in this paper, and we discuss its structure
in detail.

\begin{figure}[t]
   \vspace{-3cm}
   \epsfysize=14cm
   \epsfxsize=10cm
   \centerline{\epsffile{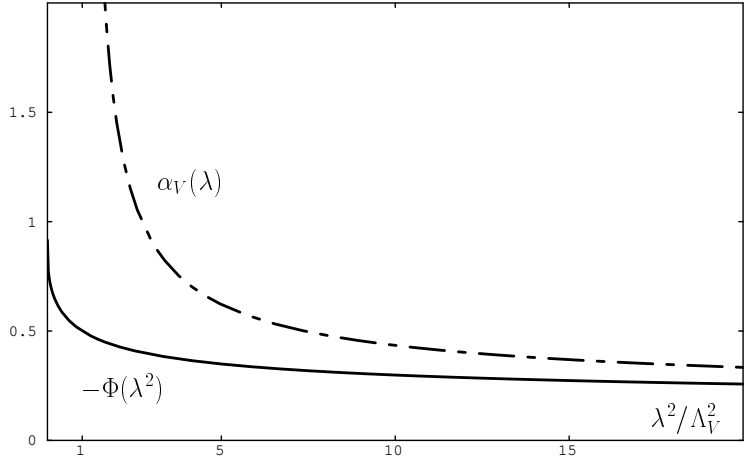}}
   \vspace*{-3cm}
\caption{\label{coupfig} One-loop running coupling in the V-scheme,
$C=0$, (broken line) and effective
coupling $-\Phi(\lambda^2)$ (solid line) as functions
of $\lambda^2/\Lambda_V^2$ .}
\end{figure}

First, notice that Eq.~(\ref{rBS}) has a very intuitive
and simple interpretation: The
quantity $r'_0(\lambda^2)$ (with certain reservations) can be considered
as the contribution
to the integral from gluons of virtuality of order $\lambda^2$, and
the function $-\Phi(\lambda^2)$ can be understood as an effective charge.
At large scales,  $-\Phi(\lambda^2)$ essentially coincides with
 $\alpha_V(\lambda)$, the QCD coupling in the V-scheme
\cite{BRO83}, but in distinction to
it remains finite at small $\lambda^2$, see Fig.~\ref{coupfig}.
 The absence of a Landau pole
in this effective coupling implies that
 the integral is a well-defined number and the fact
that we have started with the expression in
Eq.~(\ref{borelintegral}) that is ill-defined
due to infrared renormalon singularities
 (equivalently, attempted to sum a non-Borel
summable series) is isolated in the Landau pole contribution
$r_0(\lambda_L^2)$. Whenever infrared
renormalons are present, $r_0(\lambda^2)$
develops a cut at negative $\lambda^2$. The
real part of the above expression for the bubble sum coincides
with the principal value of the Borel integral and,
in particular, coincides with the Borel sum, when it exists. The
imaginary part provided by $r_0(\lambda_L^2)$ coincides with the
the imaginary part of the Borel integral, when the contour is deformed
above (or below) the positive real axis.
We therefore adopt the
real part of Eq.~(\ref{rBS}) as a definition of $M_\infty(a_s)$, and
consider the imaginary part (divided by $\pi$)
as an estimate of the intrinsic
uncertainty from summing a (non-Borel summable) divergent series
without any additional nonperturbative input. Note that this
imaginary part is proportional to a power of $\lambda^2_L/Q^2\sim
\exp(-1/a_s(Q))$ and therefore is suppressed
by powers of $\Lambda_{\rm QCD}/Q$.

Second, notice that the result in Eq.~(\ref{rBS}) is manifestly
scale- and scheme-invariant, provided the running of the coupling
is consistently implemented with the one-loop $\beta$-function:
$a_s(\mu_1)=a_s(\mu_2)/(1+a_s(\mu_2)\ln(\mu_1^2/\mu_2^2))$.
In particular, $\alpha_s(Q^*)$
is formally independent of the finite renormalization $C$ for the
fermion loop.
 To be precise, in schemes
that do not introduce artificial flavour dependence, the coefficients of
the expansion that relates the couplings in two different schemes
have itself an expansion of the type of Eq.~(\ref{flavourseries}).
Within the restriction to fermion loops with subsequent restoration
of $\beta_0$ one must again keep only the highest power of $N_f$
in these coefficients. If couplings are related in this way, the numerical
value for $\alpha_s(Q^*)=\alpha_s(Q) M_\infty(a_s(Q))$
is scheme-independent.

In practice, working with the full QCD $\beta$-function one introduces
a scheme-dependence in higher orders in $1/N_f$ similarly as
usually with higher orders in $\alpha_s(Q)$.
The quality of the NNA as compared to exact coefficients
depends on the numerical magnitude of
terms that are formally suppressed by powers of $N_f$ (or $\beta_0$)
and therefore
depends on the scheme. Since empirically the success of NNA in low orders
is observed in the $\MS$ scheme, we cannot
expect it to hold in schemes that differ from $\overline{\rm MS}$ by
terms that are formally of higher order but numerically large. Thus,
although the resummation accounts exactly
for the contribution of fermion loops
in any scheme, we believe that its phenomenological relevance
is tied to the $\overline{\rm MS}$ scheme, or others that
are ``reasonable'' in the above sense.

Third, we remark that Eq.~(\ref{rBS}) applies without
modification to quantities like inclusive decay rates, which can be
obtained starting from a suitable amplitude in euclidian space and
taking the total imaginary part upon analytic continuation
to Minkowski space. The structure of the $\lambda^2$-integral remains
unaffected, and it is only the quantity $r'_0(\lambda^2)$ that should be
substituted by the corresponding decay rate calculated with finite
gluon mass (For heavy quark decays,
no explicit renormalization is needed, when
the decay rates are expressed in terms of pole masses.).
Similarly, Eq.~(\ref{rBS}) is applicable to various semi-inclusive
quantities, with the only formal restriction that a weight-function
chosen to specify the final state does not resolve quark-antiquark
pairs in fermion bubbles, that is, their phase space must be
completely integrated. It is not applicable to quantities
like hadronic event shapes.

The derivation of Eq.~(\ref{rBS}) is only slightly modified, when
$r_0(\lambda^2)$ represents a physical cross section. In this case,
$r_0(\lambda^2)$ is written as a sum of virtual and real gluon
emission, where real gluon emission occurs only when $\lambda^2$ is
below a certain threshold $\lambda_T(Q)^2$,

\be \label{cross}
r_0(\lambda^2) = r_{virt}(\lambda^2) + r_{real}(\lambda^2)\,
\Theta(\lambda_T(Q)^2-\lambda^2)
\,.\ee

\n When all kinematic invariants, represented by $Q$, are large
compared to $\Lambda_{\rm QCD}^2$, the same is true for $\lambda_T(Q)^2$.
It is simple to show that Eq.~(\ref{final1}) holds unmodified
for the Borel transform of a cross section with Eq.~(\ref{cross}).
When deriving Eq.~(\ref{rBS}) one splits the $\lambda^2$-integral
in Eq.~(\ref{final1})
at $\lambda_T(Q)^2$. The contribution from 0 to $\lambda_T(Q)^2$
can be treated in the same way as before (adding and
subtracting a contribution from a semicircle with radius
$\lambda_T(Q)^2$). For the contribution
from $\lambda_T(Q)^2$ to infinity, one may interchange the
order of $\lambda^2$- and $u$-integrals directly, since the first
exists for all $u$. Then, the $u$-integral can be taken, since
$\lambda_T(Q)^2$ is large compared to
$\Lambda_{\rm QCD}^2$ (and therefore larger than $-\lambda_L^2$).
After the $u$-integral is taken, we combine both pieces. The final
result is identical to
Eq.~(\ref{rBS}).

We should also mention that the Borel transform of physical cross
sections generically contains a factor $\sin\pi u$, which seems
to obstruct the rotation of the $u$-integral to the imaginary
axis. Closer inspection of Eq.~(\ref{split}) shows that the two
separate $\lambda^2$-integrals there correspond in fact to the
Borel transform
of

\be
e^{\pm i \pi u}\,\frac{B[R](u)}{\sin\pi u}\,,
\ee

\n which is sufficient to guarantee convergence at the
imaginary $u$ relevant to the two terms in Eq.~(\ref{split}).

\subsection{Quantities involving renormalization}

So far the derivation was restricted to quantities, where the lowest
order radiative correction $r_0$ is ultraviolet (and infrared) finite.
Then, in diagrams involving insertion of fermion loops into the gluon
line, counterterms are needed only for the fermion loop subdiagrams, but
no overall subtraction for the diagram itself
(after summation of all diagrams that contribute to $r_n$) is
required. We consider now the more general case, that the quantity
$R(Q)$ has an anomalous dimension. One can still use Eq.~(\ref{propagator}),
but the resulting Borel transform is singular at $u=0$. This singularity
is compensated by adding a function $S_R(u)$, which accounts for the
missing counterterms. The function $S_R(u)$ in
the $\MS$ scheme is given in Eq.~(\ref{olres}).

When expressed in terms of the lowest order radiative correction with
non-zero gluon mass, the necessity of additional subtractions is reflected
in a divergence of the $\lambda^2$-integral at large $\lambda^2$ for $u=0$
in the second line of Eq.~(\ref{final1}), since $r_0(\lambda^2)$ grows
logarithmically at large $\lambda^2$ (We assume a logarithmic ultraviolet
divergence.). The additional counterterms not associated
with fermion loop insertions amount to the subtraction
of the leading term in
the large-$\lambda^2$ expansion of $r_0(\lambda^2)$
and to the addition of some scheme specific
contributions. In the $\MS$ scheme, Eqs.~(\ref{final1}) and (\ref{rBS})
are replaced by ($C=-5/3)$

\bea\label{bofin}
B[R](u) &=& -\frac{\sin\pi u}{\pi u} \intl_0^\infty
d \lambda^2 \left(\frac{\lambda^2}{\mu^2} e^C
\right)^{-u} \left[r_0^\prime(\lambda^2)-\frac{r_\infty}{\lambda^2}
\,\Theta(\lambda^2-\mu^2 e^{-C})\right]\nonumber\\
&&\,+ \frac{1}{u}\left(\tilde{G}_0(u)-r_\infty\,\frac{\sin\pi u}
{\pi u}\right)\,,
\eea

\n and

\bea\label{rBSren}
r_0 a_s M_\infty(a_s) &=&
\int_0^\infty d\lambda^2\, \Phi(\lambda^2)\,\left(r'_0(\lambda^2) -
\frac{r_\infty}{\lambda^2}\,\Theta(\lambda^2-\mu^2 e^{-C})\right)
+[r_0(\lambda_L^2)-r_0(0)]\nonumber\\
&&\hspace*{-2cm}
\,+ \intl_0^{\,a_s}\frac{d u}{u}\left(G_0(u)-r_\infty\right) + r_\infty
\left[\frac{\arctan(\pi a_s)}{\pi a_s}+\frac{1}{2} \ln\left(
1+\pi^2 a_s^2\right) - 1\right]\,.
\eea

\n Here, as always, $a_s=a_s(\mu)$. The derivation of this result
together with the definition of all new quantities that appear in the
above equation is given in Appendix A. In particular, $G_0(u)$
is essentially the anomalous dimension of $R$
(to leading order in $1/N_f$).

\subsection{Quarks with finite masses}

In practical applications it can be important to trace the number of
active fermion flavours, which may effectively decrease in high orders
because important integration regions shift towards decreasing momenta,
if the series is dominated by infrared renormalons in large or
intermediate orders. Thus, it is worthwhile to generalize the above
technique to include
massive quarks in fermion loops. For definiteness, we shall consider
here the case of one massive, and $N_f-1$ exactly massless flavours.
The generalization to several massive flavours is then obvious.
$\beta_0$ is
to be taken with $N_f$ flavours, including the massive one, since we
have in mind a situation, where the quark mass is non-zero, but smaller
than the external momenta $Q$. We
study the decoupling
of quarks with finite masses in higher orders on a particular
example in Sect.~4.

The vacuum polarization $\Pi(k^2)$ in Eq.~(\ref{propagator}) includes
summation over flavours. With one massive quark of mass $m$,
it is modified
to

\bea
\Pi(k^2) &=& a_s\Big[ \ln(-k^2/\mu^2)+C -\Delta(k^2,m^2)\Big]\,,
\nonumber\\
\Delta(k^2,m^2)&=& \frac{1}{6\pi\beta_0}\int_0^\infty
\frac{ds}{s-k^2}\big[\rho(s,m^2)-1\big]\,,
\eea

\n where

\be
   \rho(s,m^2) = \left(1+\frac{2m^2}{s}\right)\sqrt{1-\frac{4m^2}{s}}
\,\Theta(s-4m^2)\,.
\ee

\n Performing the integral gives

\bea
\Delta(k^2,m^2)&=& \frac{1}{6\pi\beta_0}\left\{
\ln\left(\frac{-k^2}{m^2}\right)
+\frac{4m^2}{k^2}
-\sqrt{1-\frac{4m^2}{k^2}}
\left(1+\frac{2m^2}{k^2}\right)
\right.\nonumber\\&&{}\left.\times
\left[\ln\left(\frac{-k^2}{m^2}\right)
+\ln\left[\frac{1}{2}\left(1-\frac{2m^2}{k^2}+\sqrt{1-\frac{4m^2}{k^2}}
\right)\right]\right]\right\}\,.
\eea

\n To derive the expression for the sum of fermion loop insertions,
we follow the method of \cite{BB94b,SV94} and substitute the
effective propagator $(1+\Pi(k^2))^{-1}$
in Eq.~(\ref{bs}) by the dispersion relation

\be\label{disprel}
\frac{1}{1+\Pi(k^2)} = \frac{1}{\pi}\int_0^\infty d\lambda^2\,
\frac{1}{k^2-\lambda^2} \frac{{\rm Im}\,
\Pi(\lambda^2)}{|1+\Pi(\lambda^2)|^2}
+\frac{1}{k^2-\lambda_L^2}
\frac{1}{\Pi'(\lambda_L^2)}
\ee

\n where $\Pi'(\lambda^2) \equiv d/d\lambda^2\Pi(\lambda^2)$ and
$\lambda_L^2<0$ is the solution of

\be
         1+\Pi(\lambda^2_L)=0
\ee

\n provided it exists. If no solution exists, the second term in
Eq.~(\ref{disprel}) is absent. We now use

\be
      \frac{1}{\pi}\mbox{\rm Im}\, \Pi(\lambda^2) =
-a_s\left[
1+\frac{1}{6\pi\beta_0}(\rho(\lambda^2,m^2)-1)\right]
\ee

\n  and write

\be
\frac{1}{k^2-\lambda^2} =\frac{1}{\lambda^2}\left(
\frac{k^2}{k^2-\lambda^2} -1\right)\,.
\ee

\n Interchanging the order of integrations in $k$ and $\lambda^2$,
we arrive at

\be\label{rBSmass}
r_0 M_\infty(a_s) = -a_s\int_{-\infty}^\infty\frac{d\lambda^2}{\lambda^2}
\frac{r_0(\lambda^2)-r_0(0)}{|1+\Pi(\lambda^2)|^2}
\left[1+\frac{1}{6\pi\beta_0}(\rho(\lambda^2,m^2)-1)\right]
+\frac{r_0(\lambda^2_L)-r_0(0)}{\lambda_L^2\Pi'(\lambda_L^2)}
\ee

\n which coincides with Eq.~(\ref{firstversion}) in the limit $m\to 0$.\\

A representation of the Borel transform which allows the calculation of
fixed-order perturbative coefficients is obtained in a similar way.
Starting from Eq.~(\ref{bs}), one only needs

\be
   \mbox{Im}\, B\left[\frac{\alpha_s}{1+\Pi(\lambda^2)}\right](u) =
   \mbox{Im}\,e^{-u \Pi(\lambda^2)/a_s} = e^{-u \mbox{\scriptsize\rm Re}
                                   \Pi(\lambda^2)/a_s}
    \sin\left[-u \mbox{Im}\Pi(\lambda^2)/a_s\right] \,.
\ee

\n Then we proceed as in the massless case and get

\bea
B[R](u) &=&
 -\frac{1}{\pi }\int_0^\infty \frac{d\lambda^2}{\lambda^2}\,
 \left[r_0(\lambda^2)-r_0(0)\right]
 \sin\Big\{ u\pi\Big[1+ \frac{1}{6\pi\beta_0}(\rho(\lambda^2,m^2)-1)\Big]\Big\}
\nonumber\\&&{}\times
\left(\frac{\lambda^2}{\mu^2}e^C\right)^{-u}
\exp\left\{u \mbox{Re}\Delta(\lambda^2,m^2)\right\}\,.
\eea

\n For $m^2=0$ we
recover the old result in Eq.~(\ref{final1}). To avoid
possible bad behaviour at $\lambda^2\to\infty$
and to separate the mass dependence, we add and subtract the
expression for $m^2=0$. Defining

\be
   T(u,\lambda^2,m^2) =
\sin\Big\{ u\pi\Big[1+ \frac{1}{6\pi\beta_0}
(\rho(\lambda^2,m^2)-1)\Big]\Big\}
\left(\frac{\lambda^2}{\mu^2}e^C\right)^{-u}
\exp\left\{u \mbox{Re}\Delta(\lambda^2,m^2)\right\}\,,
\ee

\n we obtain, finally

\bea\label{final2}
B[R](u) &=&
 -\frac{\sin(\pi u)}{\pi u }\int_0^\infty d\lambda^2\,
\left(\frac{\lambda^2}{\mu^2}e^C\right)^{-u}
 r'_0(\lambda^2)
\nonumber\\&&{}\hspace*{-1.5cm}
 -\frac{1}{\pi }\int_0^\infty \frac{d\lambda^2}{\lambda^2}\,
 \left[r_0(\lambda^2)-r_0(0)\right]
\Big[T(u,\lambda^2,m^2) -T(u,\lambda^2,0) \Big]\,.
\eea

\n The second line gives the correction due to finite quark masses
inside loops. We derived this result for quantities that do not
need renormalization.
However, in the $\MS$ scheme subtractions are mass-independent.
Therefore, only the first term in Eq.~(\ref{final2}) is affected by
subtractions, and in precisely the same way as for $m^2=0$
(cf. Sect.~2.2).

We should point out that Eq.~(\ref{rBSmass}) for the sum of fermion
loops has been derived through a dispersion relation. For massless
quarks, comparison of the derivation of Eq.~(\ref{rBS}) with the one
in \cite{BB94b} shows that the result obtained by a dispersion relation
coincides with the principal value
prescription of the Borel integral. We do not offer such an equivalence for
massive quarks and there is reason to doubt that it is true in minimal
subtraction schemes for the fermion loops. To illustrate this, suppose
there were only a single massive particle inside loops, which produces
a negative contribution to the $\beta$-function. For any finite value
of the mass of this particle, the vacuum polarization behaves as $k^2/m^2$
at very small virtuality $k$. Therefore the factorial growth of coefficients
from the infrared region of integration is eliminated at very large
order. There are no infrared renormalon poles and therefore no ambiguity
in the Borel transform (though it may be sharply peaked at those values,
where singularities appear as $m$ approaches zero). On the other
hand $1/(1+\Pi(k^2))$ does have a pole in the euclidian, when $m$ is
smaller than a critical value and $\Pi(k^2)$ is defined by minimal
subtraction. This leads to an ambiguity in the second
term in Eq.~(\ref{rBSmass}). The critical value is of order $\Lambda_{\rm
QCD}$ and the discrepancy will be noticeable only for such small
quark masses. However, for quark masses of order $\Lambda_{\rm QCD}$,
the quark mass must be considered as an infrared regulator and
one can no longer straightforwardly identify an infrared parameter,
such as the gluon condensate, with non-analytic terms in a finite
gluon mass. As is well-known, the gluon condensate must be redefined
to absorb the non-analyticities in light quark masses as well.
This is indicated by the highly singular behaviour of the Borel
transform as the quark mass goes to zero.
In our practical application the mentioned
discrepancy is numerically irrelevant,
and we
do not pursue this point further. Notice that such a difficulty does
not appear in finite order perturbative coefficients, derived from
Eq.~(\ref{final2}), and affects only the Landau pole term, in which
we are mainly interested as a measure of intrinsic uncertainty.

\subsection{Renormalon ambiguities and extended Bloch-Nordsieck
cancellations}

We want to emphasize the close relation of renormalon singularities
to non-analytic terms in the expansion of lowest-order radiative
corrections in powers of the  gluon mass \cite{BBZ94}.
A formal relation between singularities in the Borel plane and
non-analytic terms in $\lambda^2$ is established by Eq.~(\ref{relation}).
Each non-analytic term proportional to $(\sqrt{\lambda^2})^{2n+1}$
in the expansion
of $r_0(\lambda^2)$ at small $\lambda^2$ is in one-to-one correspondence
to a single-pole singularity of $B[R](u)$ at  positive half-integers
$u=n+1/2$. Each non-analytic term proportional to
$\lambda^{2n} \ln\lambda^2$
corresponds to a single pole at positive integer $u=n$.
Likewise, non-analytic terms in the expansion at large $\lambda^2$ of
type
$\sqrt{\lambda^2}^{(-2n-1)}$ or $\lambda^{-2n} \ln\lambda^2$
correspond to a single-pole singularities
of the Borel transform at {\em negative}
$u=-n-1/2$ and $u=-n$, respectively. For double or higher poles, higher
powers of $\ln\lambda^2$ appear.

We recall that through the presence of singularities for real positive
values  of the Borel parameter the perturbation series signals its
deficiency: Explicit non-perturbative corrections must be added to
make the full answer unambiguous. In fact, the infrared
renormalon problem is
just one manifestation of a generally accepted wisdom: Perturbative
calculations are only reliable if the
essential integration regions include
momenta much larger than $\Lambda_{\rm QCD}$. The above relation between
infrared renormalons and non-analytic terms in
the expansion in powers of the
infrared regulator like the gluon mass is thus natural and expected.
A comment is necessary, however, to explain why only {\em non-analytic}
terms in the expansion at small $\lambda^2$ are related to the
infrared behaviour, and simple power-like terms, $\lambda^{2n}$, are not.

A small gluon mass $\lambda^2\sim \Lambda_{\rm QCD}^2$ not only eliminates
contributions of small momenta $k^2 \sim \Lambda_{\rm QCD}^2$, but also
modifies the gluon propagator at virtualities of order $k^2\sim Q^2$.
In this region of momenta $1/(k^2-\lambda^2)$ can be expanded and produces
(infrared insensitive) corrections of the form $\lambda^2/k^2\sim
\lambda^2/Q^2$. In perturbative calculations these terms are
unimportant, since the corrections
are suppressed by powers of $Q^2$.
Hence the common practice to use the finite gluon
mass as an IR regulator in calculations of various QCD observables.
The famous Bloch-Nordsieck cancellations guarantee that
terms proportional to $\ln \lambda^2$,
which appear at intermediate steps of the calculation, cancel
in final answers for inclusive quantities.
Calculations aiming at {\em power-like} accuracy must trace
accurately the fate of power-like terms in the gluon mass.
Since analytic terms $\lambda^{2n}$ come entirely
from the expansion of the gluon propagator at large
virtualities of order
$Q^2$, they disappear when the regulator is removed. Only non-analytic terms
are related to the infrared behaviour, and indicate the failure
of perturbation theory to account for this region properly.
Thus, tracing non-analytic terms in the expansion of leading-order
radiative correction at small gluon masses, one can trace the
sensitivity of particular quantities to the IR region, and, in particular,
judge upon existence of non-perturbative
corrections suppressed by particular powers of $Q^2$.

Stated otherwise, the absence of particular power-suppressed corrections
in physical quantities can be understood as an extension of the
Bloch-Nordsieck cancellations. For example, the fact that in
Wilson's operator product expansion for the correlation function
$\Pi(Q^2)$ the $1/Q^2$-corrections are
absent, corresponds in this language to the cancellation of
corrections of order $\lambda^2\ln \lambda^2$ in the correlation
function.
Note once more that analytic terms
proportional to $\lambda^2$ do not necessarily
cancel, and in fact they are present in a quantity, closely related
to $\Pi(Q^2)$, the $\tau$-lepton hadronic width, see below.
In turn, the existence of a gluon condensate contribution,
$\langle G^2\rangle/Q^4$, implies that contributions of order
$\lambda^4 \ln \lambda^2$ do not cancel.


\mysection{Hadronic $\tau$ decays}{Hadronic $\tau$ decays}

In this Section we apply the summation of one-loop running coupling
effects developed above to observables
related by analyticity to the correlation function of vector
(axial-vector) currents

\be\label{polaroper}
\Pi_{\mu\nu}(q) = (q_\mu q_\nu-q^2 g_{\mu\nu})\,\Pi(Q^2) =
i\int\mbox{d}^4 x\,e^{i q x}\,\langle 0|T\{j_\mu^\dagger(x) j_\nu(0)\}| 0
\rangle\,, \qquad Q^2=-q^2\,.
\ee

\n Quantities of prime physical interest related to
$\Pi(Q^2)$ are the cross section of
$e^+e^-$ annihilation

\be
 R_{e^+e^-}(s)\equiv \frac{\Gamma(e^+e^-\to {\rm hadrons})}
                      {\Gamma(e^+e^-\to \mu^+\mu^-)} =12\pi
                      \mbox{Im}\,\Pi_V(s)
\ee

\n (neglecting $Z$-boson exchange) and the $\tau$-lepton total
hadronic width

\bea\label{rep1}
 R_{\tau} &\equiv& \frac{\Gamma(\tau^-\to \nu_\tau+{\rm hadrons})}
                      {\Gamma(\tau^-\to \nu_\tau e^-\bar\nu_e)}
\\
&=&\nonumber 12\pi\intl_0^{m_\tau^2} \frac{d s}{m_\tau^2} \left(
1-\frac{s}{m_\tau^2}\right)^2\left[\left(1+2\frac{s}{m_\tau^2}
\right) \mbox{Im}\,\Pi^{(1)}(s) +  \mbox{Im}\,\Pi^{(0)}(s)
\right]\,,
\eea

\n where $s=q^2$ and we introduced the decomposition

\be
\Pi^{\mu\nu}_{V/A}(q) = \left(q_\mu q_\nu-g_{\mu\nu} q^2\right)
\Pi^{(1)}_{V/A}(q^2) +q_\mu q_\nu \Pi^{(0)}_{V/A}(q^2)
\ee

\n for the vector and axial-vector correlation function
and defined
$\Pi^{(i)}(s)=\Pi^{(i)}_V(s)+\Pi^{(i)}_A(s)$. For the purpose of
our discussion we neglect the strange quark mass and omit
the overall CKM factor $|V_{ud}|^2+|V_{us}|^2\approx 1$.

 The {\em exact} (nonperturbative) correlation functions should be
analytic in the complex $s$-plane cut along the positive axis.
Exploiting this property, we may transform Eq.~(\ref{rep1}) into
\cite{BRA92}

\be\label{rep2}
R_\tau= 6\pi i \intl_{|s|=m_\tau^2} \frac{d s}{m_\tau^2} \left(
1-\frac{s}{m_\tau^2}\right)^2\left[\left(1+2\frac{s}{m_\tau^2}
\right) \Pi^{(1)}(s) + \Pi^{(0)}(s)
\right]\,.
\ee

\n The same analytic property holds to any {\em finite} order in
perturbation theory in $\alpha_s(\mu)$
(although the discontinuity is arbitrarily
wrong at small $s$). Eqs.~(\ref{rep1}) and (\ref{rep2}) are
equivalent if the correlation functions are substituted either
by the exact values or finite order perturbative expansions. In
addition, in perturbation theory, the vector and axial-vector
contributions coincide. The equivalence of
Eqs.~(\ref{rep1}) and (\ref{rep2}) does not hold in
renormalization group improved perturbation theory or after fermion loop
summation. This will be discussed extensively below.

 In what follows we concentrate on $\tau$ decays, which despite
the low energy scale involved, are considered to provide
one of the most reliable determinations of the QCD coupling
\cite{Revs}.
The state-of-the-art perturbative calculations yield \cite{GOR91}
in the $\MS$ scheme:

\begin{eqnarray}\label{tauexact}
  R_\tau &=&
 3 \Bigg[1+
\left(\frac{\alpha_s(m_\tau^2)}{\pi}\right)
+\left(\frac{\alpha_s(m_\tau^2)}{\pi}\right)^2(6.3399-0.3792 N_f)
\nonumber\\&&{}
+\left(\frac{\alpha_s(m_\tau^2)}{\pi}\right)^3
(48.5832-7.8795N_f+0.1579N_f^2)\Bigg]
+O(\alpha_s^4)\,.
\end{eqnarray}

\n The leading non-perturbative corrections due to contributions of local
operators of dimension 6  have been estimated
\cite{BRA92,Revs} and turn out to be small, below 1\%
compared to the tree-level unity in brackets above. Thus, the principal
uncertainty of the determination of $\alpha_s$ from
the $\tau$ hadronic width
comes from unknown higher-order terms in the perturbative expansion.

The purpose of this Section is twofold:
We analyze summation of $(-\beta_0\alpha_s)^n$
perturbative corrections and conclude -- with certain {\em caveats} --
that the cumulative effect of higher order corrections beyond
order $\alpha_s^3$ is somewhat
larger than the exact order $\alpha_s^3$-correction. Resummation
of vacuum polarization reduces the value of $\alpha_s(m_\tau)$
extracted from $\tau$ decays by about 10\%.
Second, we endeavour to
clarify and speculate on a conceptual point:
Several authors \cite{ALT92,DOM94,Revs} considered
(with different conclusions) the possibility that
power corrections proportional to
$1/m_\tau^2$ may creep into $R_\tau$ from
various sources (summation of large orders in perturbation
theory -- infrared and ultraviolet renormalons~--, freezing of a
physical
coupling in the infrared, violations of duality), whose absence in
the approach reviewed above is crucial to ascertain its power
for the determination of $\alpha_s$. We elaborate on two aspects,
which are sometimes omitted from the discussion: The necessity to
define a coupling parameter to power-like (in $1/m_\tau$) accuracy
and the distinction between hypothetical $1/m_\tau^2$-corrections
to the OPE of correlation functions in the euclidian and to the
$\tau$ decay width itself.

\subsection{Higher orders in $(-\beta_0)\,\alpha_s$}

To start with, let us demonstrate that Naive
Nonabelianization provides indeed
an excellent approximation in low orders. The NNA
approximation is obtained from Eq.~(\ref{tauexact})
keeping the term with the
leading power of $N_f$ only, and restoring the full $\beta_0$ by the
substitution $N_f\to N_f-33/2$. The result is

\bea\label{tauNNA}
  R_\tau^{NNA} &=&
 3 \Bigg[1+
\left(\frac{\alpha_s(m_\tau^2)}{\pi}\right)
+\left(\frac{\alpha_s(m_\tau^2)}{\pi}\right)^2(6.2568-0.3792 N_f)
\nonumber\\&&{}
+\left(\frac{\alpha_s(m_\tau^2)}{\pi}\right)^3
(42.9883-5.2107N_f+0.1579N_f^2)\Bigg]
+O(\alpha_s^4)\,.
\eea

\n The accuracy is impressive: For the practical case $N_f=3$ NNA
predicts a coefficient $28.7773$ for the $(\alpha_s/\pi)^3$
correction, to be compared to the exact coefficient $26.3658$. We
should mention, however, that this coincidence is also slightly
misleading, since a substantial part of the exact coefficient is
from a combination of $\beta$-function coefficients and lower order
coefficients of the correlation function $\Pi(s)$ and results from
contour integration in Eq.~(\ref{rep2}).

To estimate higher orders, we recall that for the
polarization operator, Eq.~(\ref{polaroper}), an analytic expression is
available for the Borel transform of the
sum of diagrams with fermion loop insertions
\cite{BEN93,BRO93,BENTH,BB94}. To get rid of
an (irrelevant) overall subtraction, it is convenient to take one
derivative with respect to $Q^2$ and to consider

\be\label{defD}
     D(Q^2)\equiv Q^2 \frac{d\Pi(Q^2)}{d Q^2}\,.
\ee

\n Then (the simple representation quoted here is adapted
from \cite{BRO93}, see \cite{BENTH,BB94})

\be\label{borelpolaroper}
B\big[D(Q^2)\big](u)\,=\,\frac{8}{3\pi^3}
\left(\qmu e^C\right)^{-u} \frac{u}{1-(1-u)^2} \sum_{k=2}^
\infty \frac{(-1)^k k}{(k^2-(1-u)^2)^2}
\ee

\n and \cite{BENTH}

\be\label{taubtexact}
B\big[R_\tau\big](u)\,=\, 12 \pi^2 B\big[D(Q^2)\big](u)\sin(\pi u)
\left[\frac{1}{\pi u}+\frac{2}{\pi(1-u)}-\frac{2}{\pi(3-u)}+
\frac{1}{\pi(4-u)}\right]
\ee

\begin{table}[t]
\addtolength{\arraycolsep}{0.2cm}
$$
\begin{array}{|l||c|c|c||c|c|}
\hline
n & d_n^{\tau,\MS} &d_n^{\tau,V} & M_n^{\tau,\MS} & d_n^{D,\MS} &
M_n^{\tau,LDP} \\ \hline\hline
0 & 1      &       1   & 1   & 1 & 1.329   \\
1 & 2.2751 & 0.6084    &1.521 & 0.6918 & 1.578 \\
2 & 5.6848 & 0.8788    &1.819 & 3.1035 & 1.855 \\
3 & 13.754 & -0.3395   &1.984  & 2.1800 & 1.898 \\
4 & 35.147 & 3.7796    &2.081 & 30.740 & 2.027  \\
5 & 84.407 & -14.680   &2.134 & -34.534 & 2.000 \\
6 & 248.83 & 99.483    &2.170 & 759.74 & 2.094 \\
7 & 525.38 & -664.00   &2.187 & -3691.4 & 2.041 \\
8 & 3036.0 & 5400.06   &2.210 & 42251 & 2.042 \\ \hline
\end{array}
$$
\caption{
Coefficients for $n$ fermion loop insertions into one-loop
radiative corrections for the $\tau$-lepton hadronic width
and partial sums of the perturbation theory for
$\alpha_s(m_\tau)=0.32$ [$a_s(m_\tau)=0.229$]. See text.
}
\label{tab1}
\end{table}

\n Taking derivatives of $B\big[R_\tau\big](u)$ (see
Eq.~(\ref{genfunction}))
it is easy to evaluate
fixed-order perturbative coefficients. In Table~\ref{tab1},
second to fourth column,
we give values for the
coefficients $d_n$ defined in Eq.~(\ref{defd_n}) for $n\le8$
in the $\overline{\rm MS}$- and V-scheme\footnote{
Note that both series are far from the expected asymptotic behaviour.
There are two reasons for this: First, for the current
correlation function, the formal large-$n$ behaviour is dominated
by ultraviolet renormalons, whose suppression in the $\MS$ scheme
with $C=-5/3$ is significant in low orders. Therefore, one does not
expect to see sign-alternating behaviour in low orders. Second,
the contour integration rearranges coefficients and
postpones the onset of the asymptotic regime, see
below.} ($C=-5/3$ and $C=0$),
and partial sums
of the perturbative series $M_n(a_s(m_\tau))$ in the
$\overline{\rm MS}$-scheme, defined in Eq.~(\ref{defM}), for
$\alpha_s^{\overline{\rm MS}}(m_\tau) = 0.32$ \cite{Revs} and
taking $N_f=3$ massless flavours.

\begin{figure}[t]
   \vspace{-4cm}
   \epsfysize=16.8cm
   \epsfxsize=12cm
   \centerline{\epsffile{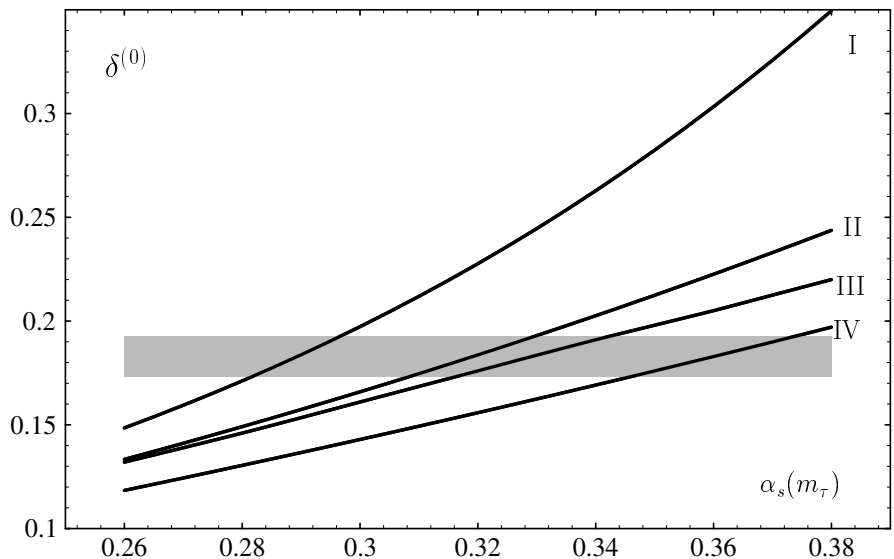}}
   \vspace*{-4cm}
\caption{Perturbative corrections to the $\tau$ hadronic width. I:
After resummation of one-loop running effects; II: Exact order
$\alpha_s^3$-approximation; III: Exact order
$\alpha_s^3$-approximation including the resummation of running
coupling effects along a circle in the complex plane (taken from
Pich); IV: Exact order $\alpha_s^2$-approximation for
comparison. The
shaded bar gives the experimental value with experimental errors
only.}
\label{taufig}
\end{figure}

\phantom{\ref{taufig}}
Truncating the expansion of $R_\tau$ at its minimal term,
one deduces from the fourth column of
Table~\ref{tab1} that the effect of summation of
$(-\beta_0\alpha_s(m_\tau))^n$-corrections increases the leading order
correction by factor $M_{7}^{\tau,\MS}(0.229)\simeq 2.19$, which is to be
compared with 1.803, obtained from the exact coefficients up to
order $\alpha_s^3$. The value of $M_{7}^{\tau,\MS}(a_s)$ from the truncated
series is in good
agreement with the result of resummation\footnote{We have obtained
this value in three different ways: (1) Starting from Eq.~(\ref{rBS})
with the leading order corrections to hadronic $\tau$ decays with finite
gluon mass collected in Appendix B; (2) Taking the principal value
Borel integral of Eq.~(\ref{taubtexact}) and (3) Computing the
principal value integral of Eq.~(\ref{borelpolaroper}) along a
circle of radius $m_\tau^2$ in the complex $s$-plane and then
taking the contour integral along the circle according to
Eq.~(\ref{rep2}). The intrinsic uncertainty due to IR renormalons
is small, $\pm 0.02$, since the leading pole at $u=2$ disappears
in $R_\tau$ in the large-$\beta_0$ limit.},

\be\label{sumtau} M_\infty^{\tau,\MS}(0.229) = 2.233\,.
\ee

\n Taken at face value, higher order vacuum polarization effects lead
to a significant increase of radiative corrections beyond the
$\alpha_s^3$-approximation. The cumulative effect amounts to somewhat
more than the $\alpha_s^3$-correction itself.

Before we discuss its impact on the determination of $\alpha_s$, we
note that it has become customary not to use a fixed order approximation
for $R_\tau$, but to employ the approach of Le~Diberder and Pich (LDP)
\cite{DIB92}, who resum exactly the effect of the running coupling
along the circle in the complex plane, but not on $Q^2 d\Pi/dQ^2$
itself. This resummation is motivated by the observation that evolution
along the circle generates a series of large higher order corrections,
which is convergent, but only barely so at the actual value of
$\alpha_s(m_\tau)$. The resummation of Le~Diberder and Pich (restricted
to one-loop running of the coupling)
is automatically included in our resumation, since the
running coupling is not expanded in $\alpha_s(m_\tau)$ in the
derivation of Eq.~(\ref{rBS}). It is yet instructive
to apply the procedure of \cite{DIB92} to fixed order
approximations for $Q^2 d\Pi/dQ^2$. The successive approximations for
$R_\tau$ are then given by

\bea\label{mlp}
M_N^{\tau,LDP}(a_s(m_\tau))&\equiv& \sum_{n=0}^N d_n^D\,
A_n(a_s(m_\tau))\,a_s(m_\tau)^n \\
\nonumber A_n(a_s(m_\tau)) &=&
\intl_{|s|=m_\tau^2}\frac{d s}{s}\left(1-2\frac{s}{m_\tau^2}+
2\frac{s^3}{m_\tau^6}-\frac{s^4}{m_\tau^8}\right)\frac{1}{[1+a_s(m_\tau)
\ln(-s/m_\tau^2)]^{n+1}}\,,
\eea

\n where $d_n^D$ are the expansion coefficients of $Q^2 d\Pi/dQ^2$,
defined as in Eq.~(\ref{defd_n}). $M_N^{\tau,LDP}(a_s(m_\tau))$ and
$d_n^{D,\MS}$ are given in the last two columns of Table~\ref{tab1}. From
this we would conclude from truncation of the series that
$M_\infty^{\tau,\MS}$ is about 2.05 or even less since the
coefficient $d_2^{D,\MS}=3.10$ is almost a factor three larger
than the exact $\alpha_s^3$-coefficient 1.26 for $Q^2 d\Pi/dQ^2$,
which indicates that keeping vacuum polarization alone is not a
good approximation\footnote{The discrepancy is partially removed,
when two-loop running is incorporated, see Sect.~5.2.}
for $Q^2 d\Pi/dQ^2$ beyond order $\alpha_s^2$.

The difference between $M_\infty^{\tau,\MS}(0.229) = 2.233$ and
2.05 can be explained by the different weight of infrared and
ultraviolet regions of internal momenta for $R_\tau$ as compared
to $Q^2 d\Pi/dQ^2$. Since the $A_n$ in Eq.~(\ref{mlp}) are positive
numbers of order one, the behaviour of $M_N^{\tau,LDP}(a_s(m_\tau))$
as function of $N$
is controlled by the series for $Q^2 d\Pi/dQ^2$. The point is
that the series for this quantity becomes dominated by ultraviolet
regions of integration much earlier than $R_\tau$ itself, for
which the contour integral suppresses the ultraviolet renormalon
singularities as well as their residues. For instance, the ratio
of the contribution of the leading infrared renormalon pole and the
leading ultraviolet renormalon pole to the coefficients of the
expansion for $R_\tau$ at order $n+1$ is given by

\be
e^{20/3}\,\frac{20}{9}\left(\frac{1}{3}\right)^n n
\ee

\n and the crossover to the asymptotically expected ultraviolet
renormalon dominance takes place at $n\approx 8-9$. For
$Q^2 d\Pi/dQ^2$ we obtain instead

\be
e^{5}\,\frac{2}{3}\left(\frac{1}{2}\right)^n \frac{1}{n}
\ee

\n with crossover at $n\approx 4-5$, which is indeed confirmed by the
coefficients for $d_n^{D,\MS}$ given in Table~\ref{tab1}. The problem
with the onset of ultraviolet renormalon dominance is that then
one is forced to construct the analytic continuation of the Borel
transform in order to overcome the formal $1/m_\tau^2$-uncertainty,
associated with the truncation of the series
due to ultraviolet renormalons. Since for $\alpha_s=0.32$
this onset of divergence is expected at $n\sim 1/a_s\sim 4-5$, we
see that, in the $\MS$ scheme,
the resummation of Le~Diberder and Pich can not be continued
beyond $n\approx 5$ without running into this $1/m_\tau^2$-uncertainty.
It is the suppression of ultraviolet regions of integration that
allows a fixed-order approximation of $R_\tau$ itself to higher
$n$ than five, which explains the proximity of $M_7^{\tau,\MS}(0.229)$
from truncation to 2.233 in this case. Recall that after resummation,
there is no uncertainty left due to ultraviolet renormalons,
because the sign-alternating behaviour is summable.

Let us turn to the determination of $\alpha_s$. We may write the
(normalized) hadronic $\tau$ decay width as

\be
R_\tau = 3\,(|V_{ud}|^2+|V_{us}|^2)\, S_{EW} \left\{
1+\delta^{(0)}+\delta_{EW}+\delta_{p}\right\}\,,
\ee

\n where $S_{EW}$ and $\delta_{EW}$ are electroweak corrections,
$\delta^{(0)}$ is the perturbative QCD correction (with quark masses
set to zero) and $\delta_p$ denotes power suppressed corrections
in $1/m_\tau^2$, including quark mass and condensate terms. We have
borrowed the values for $S_{EW}$, $\delta_{EW}$ and $\delta_p$
from \cite{BRA92,Revs}. To a very good approximation we can neglect
the variation of $\delta_p$ with $\alpha_s$ and we have evaluated
it at $\alpha_s(m_\tau)=0.32$. Then the experimental value
$R_\tau=3.56\pm 0.03$ (quoted from \cite{Revs}),
obtained from an average of branching ratio and
lifetime measurements translates into a constant experimental
value\footnote{The latest experimental data, which we have not
taken into account, seem to indicate a
larger value for $R_\tau$. In this case, too,
resummation reduces $\alpha_s(m_\tau)$ obtained
without resummation by $10\%-15\%\,$.}

\be
\delta^{(0)}_{exp}=0.183\pm 0.010
\ee

\n for the perturbative QCD corrections. In Fig.~\ref{taufig} we
have plotted the prediction for $\delta^{(0)}$ as a function
of $\alpha_s(m_\tau)$ after resummation of one-loop running
effects (recall for $R_\tau$ the exact $\alpha_s^2$- and
$\alpha_s^3$-coefficients are very well approximated by one-loop
running) as compared to the $\alpha_s^2$- and
$\alpha_s^3$-approximation as well as the $\alpha_s^3$-approximation,
including the resummation of \cite{DIB92}. We conclude that
resummation of one-loop running reduces the
central value for $\alpha_s(m_\tau)$ by
approximately 10\% to

\be \label{estimate}
\alpha_s(m_\tau) = 0.29\,.
\ee

There is some difficulty in assigning an error to this value,
which is related to the extent to which one trusts
the restriction to one-loop running effects as a good estimate
of higher order coefficients, not known exactly at present.
In view of the fact that one-loop
running overestimates the coefficient $d_2$ of $Q^2 d\Pi/dQ^2$,
one might consider $M_\infty^{\tau,\MS}(a_s)$ as an upper
estimate for the effect of higher order perturbative
corrections and the quoted value for
$\alpha_s(m_\tau)$ as a lower estimate for the central value.
On the other hand, we shall see in Sect.~5.2, that a partial
inclusion of two-loop running points towards even a lower value
of $\alpha_s(m_\tau)$. It seems safe to conclude that higher
order corrections add up constructively and a cumulative effect
is most likely to shift $\alpha_s(m_\tau)$ towards the above value.
Optimistically, one could even hope for a reduction of the theoretical
uncertainty from the unknown residual perturbative corrections. A
conservative evaluation would not
exclude the entire region from 0.27 to 0.34 for $\alpha_s$ at the scale
$m_\tau$. The lower bound reflects experimental and other than
perturbative theoretical uncertainties and the upper bound follows
from an analysis that takes into account only the completely
known perturbative corrections up to cubic order.

The implications of the expected ultraviolet renormalon dominance
for $R_\tau$ in large orders, which we have briefly touched upon
above, have recently been investigated in Ref.~\cite{ALT95}, whose
authors use conformal mapping techniques to construct the analytic
continuation of the Borel transform beyond its radius of convergence
set by the first ultraviolet renormalon. As noted in \cite{ALT95},
such a
technique is not particularly successful, when the series is not already
close to the asymptotic ultraviolet
renormalon behaviour (which is not the
case for $R_\tau$ and $Q^2 d\Pi/dQ^2$ up to
cubic order), because then the intricate
cancellations necessary to push the ultraviolet
renormalon further away from
the origin of the (conformally mapped) Borel plane than the first
IR renormalon do not take place. The variation of results obtained
from different mapping functions can then be considered analogous
to the variations induced by different choices of
renormalization schemes and as
in the latter case it is difficult to decide to what extent such a
(in principle arbitrary) variation should be considered as a
theoretical error. Based on the evidence presented above, we believe
that ultraviolet renormalons are sufficiently suppressed in fixed-order
perturbative approximations to $R_\tau$ (but potentially not to
$Q^2 d\Pi/dQ^2$) to be safely ignored at present.

\subsection{$(\Lambda_{\rm QCD}/m_\tau)^2$-corrections}

In this Subsection we investigate whether resummation of perturbative
corrections can introduce power corrections, which elude the
operator product expansion. We will exclude from the discussion
the effect of renormalons, which have received main attention
in this context \cite{BRO92,ZAK92,B92,BEN93,DUN94,ALT95,Revs}.
As far as evidence
from explicit calculations is available, infrared renormalons are
in correspondence with condensates in the OPE as expected. There
is no indication for explicit power corrections of dimension two
from this source, which could turn out to be numerically
significant. The effect
of the dominant ultraviolet renormalon divergence is taken into
account automatically by the error estimate due to unknown higher
order perturbative corrections. After resummation it disappears
completely in principle\footnote{This is seen explicitly in the
representation Eq.~(\ref{rBS}) for the Borel integral, which avoids
construction of the
analytic continuation of the Borel transform beyond its radius of
convergence set by the first ultraviolet renormalon.}, although
in practice this is not simple to implement \cite{ALT95} without
approximations (like large $\beta_0$).
It is also conceptually important
to distinguish the statement that $1/Q^2$-corrections are absent in
the OPE of correlation functions at euclidian momenta
from that that $1/m_\tau^2$-corrections are absent in $R_\tau$.
Validity of the first needs
not imply the second, though the second will hardly be true without
the first.

\subsubsection{Definition of the coupling}

Our first point of concern is the definition of the coupling parameter
inherent to perturbative expansions and therefore to their resummations.
To emphasize that this question is not connected with the minkowskian
nature of $R_\tau$, we work with the derivative of the correlation
function $D(Q^2)$ rather than with $R_\tau$ itself.

Let us first take a closer look at the Borel sum (to be definite,
let us assume a principal value prescription throughout this
subsection, which corresponds to
taking the real part of $r_0(\lambda_L^2)$)
in the representation as an integral over finite gluon mass in
the large-$\beta_0$ (NNA) approximation. Let us denote the two terms
on the right hand side of Eq.~(\ref{rBS}) by $I(\alpha_s)$ for the
integral and $L(\alpha_s)$ for the contribution from the Landau
pole in the dispersion relation for the running coupling (using the
technique of Sect.~2.3 or \cite{BB94b}). The leading (two-loop)
radiative correction with finite $\lambda$ can
be obtained from the Borel transform by a Mellin transformation, see
Eq.~(\ref{relation}). Taking
the integral analytically can be difficult, but for our present purpose
we are interested only in the coefficients in the small-$\lambda^2$
expansion. This expansion can be obtained easily, since particular
terms proportional to $(\lambda^2/m_\tau^2)^n$ (modulo logarithms)
correspond to contributions from
singularities of the integrand in the right $u$-plane at $u=n$.
For example,
to pick up the term of order $\lambda^2/m_\tau^2$ one can evaluate the
integrand in Eq.~(\ref{relation}) at $u=1$, except for $\Gamma(-u)$.
For the $D$-function defined in Eq.~(\ref{defD}) we obtain

\begin{eqnarray}\label{Dexpand}
   4 \pi^2 D(Q^2,\lambda^2) &=&1+\frac{\alpha_s}{\pi}\Bigg\{
1-\left[\frac{32}{3}-8\zeta(3)\right]\frac{\lambda^2}{Q^2}-
\left[2 \ln (Q^2/\lambda^2)+\frac{20}{3}-8\zeta(3)\right]
\frac{\lambda^4}{Q^4}
\nonumber\\&&{}+O\left(\lambda^6\ln^2 \lambda^2/Q^2\right)\Bigg\}\,.
\end{eqnarray}

\n The presence of quadratic terms,
$\lambda^2/Q^2$, is not in conflict with the operator product expansion,
since such terms come from the region of large momenta: As emphasized
in Sect.~2.4, only non-analytic terms in the small $\lambda^2$-expansion
can unambiguously be identified with infared contributions.  The leading
non-analytic term is proportional to $\lambda^4\ln\lambda^2$
and produces a correction proportional to $1/Q^4$, which can
be related to the
contribution of the gluon condensate \cite{SVZ} in the OPE.
This contribution agrees with the calculation
of the gluon condensate with finite gluon mass in \cite{CHE88}, after the
corresponding Wilson coefficient is extracted\footnote{The full contribution
of order $\lambda^4/Q^4$ which is multiplied by $\ln(Q^2/\lambda^2)+
const$ should be schematically decomposed as $\ln(\mu^2/\lambda^2)$,
contributing to the gluon condensate, and $\ln(Q^2/\mu^2)+const$, contributing
to the coefficient function to $\alpha_s$ accuracy. Here $\mu$ is the
scale separating small and large distances. If one considers $\lambda$ as
as an infrared cutoff, which is natural, then  the full
contribution should be ascribed
to the coefficient function. Hence the rationale of ascribing the constant
term to contributions of large momenta (small distances). In general,
the separation of matrix elements and coefficient functions is of course
factorization scheme-dependent and constant terms
can be reshuffled. We see once more
that only non-analytic terms (logarithmic in this example) can be
used to trace infrared contributions.}.
(The difference in the constant $20/3$ arises, because we consider the
derivative of $\Pi(Q^2)$.).

The presence of a $\lambda^2/Q^2$-correction in Eq.~(\ref{Dexpand})
implies that the Landau-pole contribution in Eq.~(\ref{rBS}) is

\be
L(\alpha_s(Q)) \propto \exp\left(\frac{1}{\beta_0\alpha_s(Q)}
\right) \propto \frac{\Lambda_{\rm QCD}^2}{Q^2} \,.
\ee

\n Numerically, this term is quite substantial. Taking $\alpha_s(Q)=
0.32$ (which corresponds to $Q\simeq m_\tau$), separation of the
two terms in Eq.~(\ref{rBS}) for $D(Q^2)$ amounts to

\be
M_\infty^D(0.229) = 1.48 = \mbox{integral over gluon mass} + 0.31\,.
\ee

\n Note that $L(\alpha_s)$ has identically vanishing perturbative
expansion. Thus, without any additional information, keeping
$I(\alpha_s)$ alone in Eq.~(\ref{rBS}) provides an equally
legitimate summation of the original series, which differs from
Borel summation by
terms of order $1/Q^2$, which are not related to renormalons or
any particular regime of small or large momenta. We conclude,
at this stage, that statements about power corrections are
meaningful only with respect to particular summation prescriptions.

Physically, dropping the contribution $L(\alpha_s)$ is equivalent
to dropping the Landau pole contribution to the dispersion relation
for the running
coupling which can be interpreted
as a redefinition of the coupling, such that the new coupling
has no Landau pole. Since Borel summation
in our limit of large $\beta_0$ coincides
(in the V-scheme, $C$=0) with averaging
the running coupling $\alpha_s(k) = \alpha_s(Q)/(1-\beta_0
\alpha_s(Q) \ln(k^2/Q^2))$, it is readily seen from
Eq.~(\ref{disprel}) that neglecting $L(\alpha_s)$ corresponds to
averaging with a coupling $\alpha^{\rm eff}_s(Q)$, related
to $\alpha_s(Q)$ by

\be \label{effcoup}
\alpha^{\rm eff}_s(Q) = \alpha_s(Q)-\frac{\lambda_L^2}{\beta_0
(Q^2+\lambda_L^2)} = \alpha_s(Q)-\frac{1}{\beta_0}\,e^{1/(\beta_0
\alpha_s(Q))} + O\left(e^{2/(\beta_0
\alpha_s(Q))}\right)\,.
\ee

\n This coupling has no Landau pole and freezes to a finite value as
$Q^2$ approaches zero. However, it has $1/Q^2$-corrections to its
evolution and correspondingly to the $\beta$-function:

\be
\beta^{\rm eff}(\alpha_s^{\rm eff}) = \beta_0(\alpha_s^{\rm eff})^2
+ \left(\frac{1}{\beta_0}+2\alpha_s^{\rm eff}\right) e^{1/(\beta_0
\alpha_s^{\rm eff})} + O\left(e^{2/(\beta_0
\alpha_s^{\rm eff})}\right)
\ee

\n Note that the absence of a Landau pole in $\alpha_s^{\rm eff}$
is seen only after summation of an {\em infinite} number of power
corrections in $1/Q^2$ (exponentially small terms in $\alpha_s(Q)$)
in Eq.~(\ref{effcoup}). Further, the coefficients of perturbative
expansions of any quantity are the same whether one uses $\alpha_s(Q)$
or $\alpha_s^{\rm eff}(Q)$ and diverge although the average

\be \label{aveff}
\int d^4 k \,F(k,Q) \frac{\alpha^{\rm eff}_s(k)}{k^2} \sim
I(\alpha_s(Q))
\ee

\n has no Landau pole ambiguities. We emphasize
once more that averaging
one-loop radiative corrections with
this freezing coupling differs
from Borel summation of the series in $\alpha_s(Q)$
(which gives $I(\alpha_s(Q))+L(\alpha_s(Q))$
by $1/Q^2$-terms. So does Borel summation of
the identical series in $\alpha^{\rm eff}_s(Q)$,
because the couplings differ by such terms. We shall argue that
couplings like
$\alpha_s^{\rm eff}(Q)$ obscure the relation to the operator product
expansion in the sense that explicit $1/Q^2$-terms must be added
to the resummed result (such as $L(\alpha_s(Q)$), had one used
$\alpha_s^{\rm eff}(Q)$) in order to cancel spurious $1/Q^2$-effects
in the large $Q^2$-expansion of the resummed result.
This remark applies identically to a freezing coupling of
type $1/(-\beta_0 \ln(c+Q^2/\Lambda^2))$ and thus to the procedure
of \cite{Nnew}, where such a coupling has been used to
estimate the size of infrared contributions for quantities related
to heavy quark expansions.

We will now try to make precise the statement that $1/Q^2$-terms
should be absent in the OPE. In order to talk about power corrections
we have to attach some meaning to the divergent perturbative
expansion at leading order in $1/Q^2$. A natural (though one might
still ask whether it is justified to prefer Borel summation-type
schemes to any other) definition apparently is

\be \label{interpretation1}
\left| D(Q^2)_{\rm exact} - BS[D(\alpha_s(Q))_{\rm pert}]\right|
\sim O\left(\frac{1}{Q^4}\right)\,,
\ee

\n where $BS[D(\alpha_s(Q))_{\rm pert}]$ denotes the Borel
integral (with principal value prescription) of the perturbative
expansion of $D$ in some $\alpha_s(Q)$. However, this statement
is still ambiguous, because it implies knowledge of the
$Q^2$-dependence in the coupling, which is arbitrary to a large
extent.
Moreover, it is not sufficient to appeal to the usual ambiguities
in the choice of perturbative renormalization schemes, since the
coupling and its evolution must be specified to power-like
accuracy. We can bypass this point, noting that QCD with massless
fermions has only a single free parameter. Since we are discussing
asymptotic expansions in a dimensionful parameter $1/Q^2$, it seems
most natural to choose a physical mass scale as this parameter,
say $m_\rho^2$. This is especially natural in the context of lattice
definitions of QCD, where one would trade the bare coupling for
$m_\rho^2$. Then, if the operator product expansion exists
nonperturbatively, this suggests the existence of a double expansion
in $1/\ln(Q^2/m_\rho^2)$ and $m_\rho^2/Q^2$ at
large $Q^2$ and the interpretation
of the statement that there are no $1/Q^2$ terms could be\footnote{
This is still a simplification. In fact, $\ln\ln (Q^2/m_\rho^2)$ is
also expected to appear. We will restrict the discussion to the
large-$\beta_0$ limit, which might be considered as an analytic
continuation to a large negative number of massless fermion flavours.
In this limit, only $\ln(Q^2/m_\rho^2)$ appears and Borel transformation
with respect to  $\ln(c Q^2/m_\rho^2)$ has a unique meaning.}

\be \label{interpretation2}
\left| D(Q^2)_{\rm exact} - BS[D(Q^2)_{\rm pert}]\right|
\le K(\delta,R) \left(\frac{m_\rho^2}{c Q^2}\right)^2\,.
\ee

\n Here $BS[D(Q^2)_{\rm pert}]$ denotes the Borel sum (with
principal value prescription) of the leading term in the
expansion of $D(Q^2)$ at large $Q^2$, which is an infinite series
in $1/\ln(c Q^2/m_\rho^2)$, where
$c$ is a constant to be specified
later. The constant $K(\delta,R)$ depends on
the opening angle $\delta$ and radius $R$ of the sector in the
complex $1/Q^2$-plane, where
Eq.~(\ref{interpretation2})
is supposed to hold. In general, $K(\delta,R)$ will neither be
continuous nor bounded as a function of these two parameters.
In particular, one can not expect a uniform bound in the
entire cut plane.
In the limit $\delta\rightarrow \pi$,
this is related to violations of duality. We stress that as far
as mathematical rigour is concerned, the validity of
Eq.~(\ref{interpretation2}) must be regarded as purely hypothetical.
We wish to present it as a mathematical formulation of the
{\em assumption} that the operator product expansion holds
at {\em euclidian} momenta (i.e. $\delta > 0$)
and no $1/Q^2$-terms are present in the
asymptotic expansion. We note that the
condition Eq.~(\ref{interpretation2})
is stronger than the condition that long-distance contributions
can be factorized into condensates, which does not exclude the
presence of power-like corrections, in particular $1/Q^2$, to
coefficient functions, in particular of the unit operator, from
short distances.

After we have chosen $m_\rho$ as the fundamental parameter of
QCD, we may define the coupling by its beta-function and an
overall scale. We define (again, to leading-$\beta_0$ accuracy)

\be \label{defcoup2}
\alpha_s(Q)=\frac{1}{-\beta_0 \ln(c Q^2/m_\rho^2)} \equiv
\frac{1}{-\beta_0 \ln(Q^2/\Lambda_V^2)}\,,
\ee

\n where we have fixed $c$ by matching the large $Q^2$ behaviour
with the $V$-scheme\footnote{This is a matter of
convenience, since it
eliminates writing $C$ in the large-$\beta_0$ approximation. We
could also have matched to the perturbative $\MS$ coupling. We
also note that the change of variables from $m_\rho^2$ to
$\alpha_s(Q)$ is singular due to the Landau pole at $\Lambda_V^2$.
However, this is not a restriction, since Eq.~(\ref{interpretation2})
is limited to finite $R$ anyway. The position of the pole
may be varied by the choice of $c$, which implies a reorganization
of powers in inverse logarithms, but the pole occurs always at
a finite value of $Q^2$.}.
It is easy to see that Eq.~(\ref{interpretation2}) implies
Eq.~(\ref{interpretation1}) with this coupling, which by
definition has no power-like evolution. This implication is
not valid for couplings, that incorporate $1/Q^2$-dependence
in their running. It is in this sense that we believe
the use of freezing couplings
is hazardous. {\em If} Eq.~(\ref{interpretation2}) is correct,
then the use of Borel summation for perturbative series expressed
in terms of such a coupling, or averaging lowest order
radiative corrections with such a running coupling, necessitates
the addition of explicit $1/Q^2$-corrections simply to
cancel such corrections hidden in the
definition of the coupling.

In practice, in one way or another, one relates physical quantities
and an unphysical coupling like Eq.~(\ref{defcoup2}) might be
considered as an intermediate concept only. However, the importance
of being definite with the evolution of the coupling to power-like
accuracy is not diminished by the use of physical couplings. To give
a somewhat constructed example: If one expressed $R_{e^+ e^-}$ as
an expansion in the effective coupling, defined by QCD corrections
to the Gross-Llewellyn-Smith
sum rule, one would expect $1/Q^2$-corrections to this
perturbative relation, which are imported from the definition of
the coupling.

\subsubsection{Analyticity and the Landau pole}

After renormalization group improvement, perturbative expansions are
plagued by the Landau pole. This unphysical singularity is endemic
not only to perturbative expansions, but to the operator product
expansion, truncated to any finite order. It requires some care
in defining resummations for quantities like $R_\tau$, which are
related to euclidian quantities by analyticity. For the purpose of
illustration, we shall restrict ourselves to the approximation, where
$\beta(\alpha_s)=\beta_0\alpha_s^2$ and adopt the V-scheme, $C=0$,
in all explicit formulas. Then, ignoring an irrelevant $Q^2$-independent
subtraction, we can write the Borel sum (with principal value
prescription, as usual) as

\bea
BS[\Pi](Q^2) &\equiv& \frac{1}{-\beta_0}\intl_0^\infty d u \,e^{-u/(
-\beta_0\alpha_s(\mu))} B[\Pi](u,Q^2/\mu^2)
=  \frac{1}{-\beta_0}\intl_0^\infty d u \left(\frac{\Lambda_V^2}{Q^2}
\right)^u F(u)\nonumber\\
&=& BS[\Pi]_<(Q^2,u_0) + BS[\Pi]_>(Q^2,u_0)\,,
\eea

\n where we use $\alpha_s(Q)=1/(-\beta_0 \ln(Q^2/\Lambda^2_V))$ and
$F(u)$ is $Q^2$-independent. In the second line, we have defined two
new functions by splitting the Borel integral into two regions from
0 to $u_0$ and $u_0$ to $\infty$. From the explicit form in
Eq.~(\ref{borelpolaroper}), we deduce that, for {\em any} (positive)
$u_0$, $BS[\Pi]_<(Q^2,u_0)$ is analytic in the $Q^2$-plane cut
along the negative axis. On the other hand the $u$-integral in
$BS[\Pi](Q^2)$ diverges, when $Q^2 < \Lambda_V^2$ and $Q^2=\Lambda^2_V$
is a singular point. We say that $BS[\Pi](Q^2)$ has a Landau pole,
though, in general, $Q^2=\Lambda^2_V$ will rather be a branch point.

Notice, since in practice one truncates the operator product expansion
at operators of some dimension $d$, it is perfectly consistent to replace
$BS[\Pi](Q^2)$ by $BS[\Pi]_<(Q^2,u_0)$, provided we choose $u_0 >
d/2+1$, since the difference is bounded by

\be\label{bound2}
\left| BS[\Pi](Q^2) - BS[\Pi]_<(Q^2,u_0)\right| < \tilde{K}(R)
\left(\frac{\Lambda_V^2}{Q^2}
\right)^{u_0}
\ee

\n and vanishes faster than other corrections neglected in the truncation
of the OPE. In this case, the bound can be established in a cut
circle of radius $R$ in the $1/Q^2$-plane. Again, it will not be
possible to establish uniform bounds, since typically $\tilde{K}(R)
\sim 1/\ln(R \Lambda_V^2)$. Thus, although for fixed $Q^2>1/R$,
the difference can be made vanish faster than any desired power
of $\Lambda_V^2/Q^2$ by increasing $u_0$, at fixed $u_0$ the bound
may become arbitrarily weak as $Q^2$ approaches $\Lambda_V^2$.
Still, we conclude that the presence or absence of a Landau pole
in resummed results is related to the behaviour of the Borel integral
at infinity
and is thus an effect that formally vanishes faster than any power
of $1/Q^2$.

Consider now the two different representations for
the tau decay width, Eqs.~(\ref{rep1}) and
(\ref{rep2}), in this light. We use equality of
vector and axial-vector correlators in perturbation theory and
abbreviate the weight function by $w(s/m_\tau^2)$. Then

\be\label{sum1}
R_\tau^{
\mbox{{\tiny eq.(\ref{rep1})}}}
= 24\pi\intl_0^{m_\tau^2} \frac{d s}{m_\tau^2}
\,w\!\left(\frac{s}{m_\tau^2}\right)\,\frac{1}{2 i}\mbox{disc} \,
BS[\Pi](s)\,,
\ee
\be\label{sum2}
R_\tau^{
\mbox{{\tiny eq.(\ref{rep2})}}} = 12\pi i\intl_{|s|=m_\tau^2} \frac{d
s}{m_\tau^2}
\,w\!\left(\frac{s}{m_\tau^2}\right)\,BS[\Pi](s)\,.
\ee

\n Note that in both cases summation is carried out inside the
$s$-integral. The following considerations can
easily be extended to the situation, where summation is
taken after $s$-integration (and do not lead to any of
the differences observed below). We define $R_{\tau,<}^{
\mbox{{\tiny eq.(\ref{rep1})}}}(u_0)$ and $R_{\tau,<}^{
\mbox{{\tiny eq.(\ref{rep2})}}}(u_0)$ by the replacement of
$BS[\Pi](s)$ by $BS[\Pi]_<(s,u_0)$. Using the analyticity properties
discussed above as well as $m_\tau^2 > \Lambda_V^2$, it is
straightforward to find

\be
R_\tau^{\mbox{{\tiny eq.(\ref{rep2})}}} -
R_{\tau,<}^{\mbox{{\tiny eq.(\ref{rep2})}}}(u_0)
\sim \left(\frac{\Lambda_V^2}{m_\tau^2}\right)^{u_0}
\qquad R_{\tau,<}^{\mbox{{\tiny eq.(\ref{rep1})}}}(u_0) =
R_{\tau,<}^{\mbox{{\tiny eq.(\ref{rep2})}}}(u_0)\,,
\ee

\n but

\be
R_\tau^{\mbox{{\tiny eq.(\ref{rep1})}}} -
R_\tau^{\mbox{{\tiny eq.(\ref{rep2})}}} =
-12 \pi i \intl_C \frac{d s}{m_\tau^2}
\,w\!\left(\frac{s}{m_\tau^2}\right)\,BS[\Pi]_>(s,u_0)\,,
\ee

\n where the contour $C$ runs along a circle of radius
$\Lambda_V^2$ around $s=-\Lambda_V^2$. Against appearances, the
right hand side is independent of $u_0$. Of course, one can not take
$u_0$ to infinity inside the integral and conclude that it is
zero. $BS[\Pi]_>(s,u_0)$ has a pole or branch point at
$s=-\Lambda_V^2$ and we find

\be\label{differ}
R_\tau^{\mbox{{\tiny eq.(\ref{rep1})}}} -
R_\tau^{\mbox{{\tiny eq.(\ref{rep2})}}} \sim \frac{\Lambda_V^2}{m_\tau^2}
\ee

\n and similarly

\be
R_\tau^{\mbox{{\tiny eq.(\ref{rep1})}}} -
R_{\tau,<}^{\mbox{{\tiny eq.(\ref{rep1})}}}(u_0)
\sim \frac{\Lambda_V^2}
{m_\tau^2}\,.
\ee

\n Since Eq.~(\ref{bound2}) is not valid for $s<\Lambda_V^2$,
this result should not surprise. The difference in Eq.~(\ref{differ})
arises, because the resummation introduces (or preserves) the
Landau pole singularity, which is in conflict with the analytic
properties of the exact correlation function, that have been
assumed to derive Eq.~(\ref{rep2}) from Eq.~(\ref{rep1}).

Should one conclude then that resummations of perturbative corrections
are ambiguous by terms of order $1/m_\tau^2$, even if there are
no $1/Q^2$ terms in the OPE at euclidian momenta in the strong sense of
Eq.~(\ref{interpretation2})? Do we have evidence for power corrections
not captured by the OPE, since they originate from $u=\infty$ in
the Borel integral? The answer is no. A positive answer would be
warranted, if there were no reason to prefer the prescription
Eq.~(\ref{sum1}) to Eq.~(\ref{sum2}) or vice versa, while only
the second can be used. The reason is that the region $|s| <
\Lambda_V^2$ can not be penetrated to any finite order in the
short distance expansion ($1/Q^2$-expansion). Since all summation
prescriptions of perturbative expansions are formulated within the
context of the short-distance expansion\footnote{
See the discussion below Eq.~(\ref{borelintegral}).},
they can not be applied
to $|s| < \Lambda_V^2$, unless the summation of this expansion itself
is understood. This discards Eq.~(\ref{sum1}) as a legitimate summation.
One must first use the analyticity properties of the exact correlation
functions to deform the contour outside the region $|s| < \Lambda_V^2$,
before an attempt at summing perturbative expansions can be made,
which privileges Eq.~(\ref{rep2}) as the starting point. A different
way to express this fact is to observe that the principal value Borel
integral is defined for $|s| < \Lambda_V^2$ only in the sense of
analytic continuation, but cannot be used as a numerical approximation
since all power corrections in the operator product expansion are
of the same order of magnitude in this region. It is only when
all these are taken into account that the Landau pole vanishes in
physical observables.

To conclude this Section, let us mention that a relation between
analyticity in $s$ and behaviour of the Borel integral at infinity
has been noted in a very different and much more physical
context in \cite{tHO77}. The presence of
resonances and multiple thresholds on the physical axis was observed
to be in conflict with Borel summation\footnote{
Historically, it is interesting to note that the
horn-shaped analyticity region in the coupling, which leads to
this conclusion, was
discovered for the photon propagator in \cite{AZI70}.}.
In this case, the restoration
of the correct analytic properties should also be understood
in connection
with summing the OPE itself \cite{BEN93b}. A much more elaborate
argument has been presented in \cite{SHI94}, where the presence
of resonances was connected with the divergence of OPE, which also offers
a way to understand the concept of duality and its limitations. If
the OPE is by itself only an asymptotic expansion, it is quite possible
that its application is limited to a finite phase range in the complex
$s$-plane around the negative $s$-axis. We
believe that this possibility
should be taken seriously, since it might imply the presence of
$1/s_0$-corrections in finite averages
up to $s_0$ of the discontinuity along the
physical axis (for $R_\tau$, $s_0=m_\tau^2$). Unfortunately, we
do not know how to substantiate or disprove such
a statement theoretically.



\mysection{The pole mass of a heavy quark}
{The pole mass of a heavy quark}

Although on-shell quarks do not exist and there is presumably no
natural nonperturbative definition of a pole
mass of a quark, the pole
mass has proven useful as an auxiliary concept in applications
of perturbative QCD to heavy quark physics, where physically
the quark is expected to be close to its would-be mass-shell.
Still, the relation between the pole mass and an off-shell
renormalized mass is known to have large perturbative coefficients
from small loop momenta, at least in high orders of
perturbation theory \cite{BB94,BIG94}. Since neither quark
mass definition is physical, this might appear as an irrelevant
problem. However, the behaviour of perturbative expansions of
quantities involving quark masses change accordingly with the
quark mass definition used and in
general one expects coefficients to be
significantly reduced, when the pole mass is abandoned in favour
of an off-shell renormalized mass \cite{BIG94,BBZ94}, such as $\MS$.
In the latter case, it is quite well-known that the exact two-loop
coefficient \cite{GRA90} in the relation to the pole mass
is substantial and one might wonder
whether this is coincidental.

In this Section we apply resummation and NNA to the difference
between the pole and $\MS$ mass.
Apart from its practical interest,
we also use this quantity to illustrate certain features of higher
order perturbative corrections, when the masses of fermions in
loops are finite. The reader interested only in
results may jump directly to Sect.~4.3, where our best estimate
is presented. In this Section $\alpha_s(\mu)$ always refers
to the $\MS$ coupling.

\subsection{Preliminaries}

To begin with, let us quote the exact two-loop
result from \cite{GRA90}:

\be\label{m1}
\frac{\delta m}{m}
\equiv\frac{m_{\rm pole}-m_{\MS}(m_{\MS})}{m_{\MS}(m_{\MS})}
= \frac{4}{3}\frac{\alpha_s(m_{\MS})}{\pi} \left[1 +
\left(4.68\,(-\beta_0^{N_f}) - \left\{\begin{array}{c} 0.89\\0.91\end{array}
\right\}\right) \alpha_s(m_{\MS})\right]
\ee

\n The upper entry in curly brackets refers to the calculation with
$N_f$ massless quarks and a quark of mass $m$ in loops, the
lower entry to the situation, where the quark flavour of mass $m$ is
excluded from loops\footnote{To obtain the lower value,
it is necessary to insert a missing $C_F$ in front of the mass
correction in Eq.~(17) of \cite{GRA90}, see also \cite{BG94}.}.
The superscript on $\beta_0$ indicates the value
of $N_f$ to be taken for the $\beta$-function. Since $(-\beta_0^{N_f})
\sim 2/3$ for cases of interest, $N_f=3,4$, we note that keeping
the term proportional to $\beta_0^{N_f}$ alone provides indeed a
reasonable approximation within $30-40\%$ of the exact coefficient.

Note that above we have normalized $m_{\MS}$ at the scale $m_{\MS}$,
since we prefer to eliminate $m_{\rm pole}$ in all places. When
Eq.~(\ref{m1}) is expressed in terms of the $\MS$ mass normalized at
$m_{\rm pole}$,

\be\label{m2}
\frac{m_{\rm pole}-m_{\MS}(m_{\rm pole})}{m_{\MS}(m_{\rm
pole})}
= \frac{4}{3}\frac{\alpha_s(m_{\rm pole})}{\pi} \left[1 +
\left(4.68\,(-\beta_0^{N_f}) - \left\{\begin{array}{c} 0.25\\0.28\end{array}
\right\}\right) \alpha_s(m_{\rm pole})\right]\,
\ee

\n the approximation of keeping only the term proportional to
$\beta_0^{N_f}$ is significantly improved. This ambiguity is in fact a
source of trouble within the BLM and our extended prescription. For
$\delta m$ additional renormalization scheme and scale dependence
is present from the definition of the quark mass parameter, whereas
the BLM method by construction deals only with the scale-dependence of
the coupling. Although the calculation of fermion loop insertions
and subsequent restoration of $\beta_0$ also provides the anomalous
dimension of the quark mass within the same approximation, the scheme
ambiguity in the size of neglected genuine two-loop corrections is
amplified by this additional source of scheme-dependence. If
$\mu_1^2-\mu_2^2\sim O(\alpha_s)$, then the difference
$m_{\MS}(\mu_1)-m_{\MS}(\mu_2)$ must be neglected in the
approximation of large $\beta_0$, although it can be sizeable,
if the coefficient of $\alpha_s$ in $\mu_1^2-\mu_2^2$ is so. In the
following, we will work with Eq.~(\ref{m1}), whenever relevant.

Let us first consider the (excellent) approximation, that the quark
flavour of mass $m$ is neglected in loops. All other $N_f$
quarks are massless. In this case the relevant $\beta$-function coefficient
is $\beta_0^{N_f}$ and an exact expression for the Borel transform
of the mass shift exists \cite{BB94}:

\be \label{polebt}
B\left[\delta m/m\right](u) = \frac{1}{3\pi}\left[e^{5 u/3}\,6 (1-u)
\frac{\Gamma(u)\Gamma(1-2 u)}{\Gamma(3-u)} + \frac{\tilde{G}_0(u)}{u}
\right]
\ee

\n $\tilde{G}_0(u)$ is defined in the following way (see Appendix~A): If
$g_n$ are the expansion coefficients of $G_0(u)$ in $u$, then
$\tilde{G}_0(u)$ has expansion coefficients $g_n/n!$. $G_0(u)$ can
be calculated with the methods of Appendix~A and is found to be

\be\label{subfunction}
G_0(u) = -\frac{1}{3} (3+2 u)\frac{\Gamma(4+2 u)}{\Gamma(1-u)
\Gamma^2(2+u) \Gamma(3+u)}\,.
\ee

\n Higher order perturbative corrections
with fermion loop insertions
into the one-loop diagram can then be computed according to
Eq.~(\ref{genfunction}). Equivalently, one can start from Eq.~(\ref{bofin})
with the input of the one-loop mass shift with finite gluon mass,
given by ($x\equiv \lambda^2/m^2$)

\bea\label{polefg}
r_0(\lambda^2) &=&\nonumber \frac{1}{3\pi}\Bigg[4+x-\frac{x^2}{2}\ln x-
\frac{\sqrt{x}(8+2x-x^2)}{\sqrt{4-x}}
\Bigg\{\arctan\left[\frac{2-x}{\sqrt{x(4-x)}}\right]+\\
&&\,\arctan\left[\frac{\sqrt{x}}{\sqrt{4-x}}\right]\Bigg\}\Bigg]
= \frac{4}{3\pi}\left[1+\frac{\pi}{2} \sqrt{x} + \frac{3}{4} x + O\left(
x^{3/2}\right)\right]\,.
\eea

\n We define coefficients $d_n$ as in Sect.~2 by

\be
\frac{\delta m}{m}
= \frac{4}{3}\frac{\alpha_s(m)}{\pi} \left[1+\sum_{n=1}^\infty
d_n \,(-\beta_0^{N_f} \alpha_s(m))^n\right]\,.
\ee

\n  They are listed in Table~\ref{tabpole}. Higher order perturbative
corrections grow very rapidly, as expected from the dominant asymptotic
behaviour from the pole at $u=1/2$ in Eq.~(\ref{polebt}),

\be\label{aspole}
d_n \stackrel{n\gg 1}{=} e^{5/6}\, 2^n n!\,.
\ee

\begin{table}[t]
$$
\begin{array}{|c||c|c|c|}
\hline
n & d_n & M_n\,[\mbox{c-quark}] & M_n\,[\mbox{b-quark}] \\ \hline\hline
0 & 1 & 1 & 1 \\
1 & 4.6861511 & 2.176 & 1.623  \\
2 & 17.622650 & 3.286 & 1.935  \\
3 & 109.85885 & 5.024 & 2.193  \\
4 & 873.92393 & 8.492 & 2.467  \\
5 & 8839.6860 & 17.30 & 2.835  \\
6 & 105814.28 & 43.76 & 3.420  \\
7 & 1484968.4 & 137.0 & 4.514  \\
8 & 237407365 & 511.0 & 6.838  \\ \hline
\infty & - & 1.712\pm 0.608 & 2.041\pm 0.201 \\ \hline
m^*_1 & - & 0.096 \,m_c & 0.096 \,m_b  \\
m^*_\infty & - & 0.437 \,m_c & 0.147 \,m_b \\
\hline
\end{array}
$$
\caption{Higher order perturbative corrections to the difference between
the pole and $\MS$ mass from $n$ fermion loop insertions. The second
and third columns give partial sums and summations for the charm
quark ($a_s=-\beta_0^{(3)}\alpha_s(m_c)=0.251$) and bottom quark
($a_s=-\beta_0^{(4)}\alpha_s(m_b)=0.133$). In the latter case, the c-quark
is treated as massless inside loops.}
\label{tabpole}
\end{table}

\n This asymptotic formula is in fact a very good approximation
(within 5\%) to the exact coefficient $d_n$ for all $n\ge 1$, which
seems to imply that the saddle point approximation of the loop
momentum distribution inherent to deriving the asymptotic
behaviour is a very good substitute for the exact
distribution, even if the width of the Gaussian is not small.
Given that genuine two-loop corrections are rather small compared
to $d_1$, it can be asserted that the exact two-loop coefficient
is already dominated by the first infrared renormalon at $u=1/2$.
Evidently, this observation is scheme-dependent and we do not
know why the $\MS$ scheme is preferred. Let us note, however, that
a similar coincidence can not be expected and indeed does not
happen (see Sect.~3) for quantities, whose asymptotic behaviour
is dominated by ultraviolet renormalons, like $R_{e^+ e^-}$ or
$R_\tau$, since with $C=-5/3$ in the $\MS$ scheme, the
leading ultraviolet singularity at $u=-1$ is suppressed compared
to the leading infrared singularity at $u=2$ by a factor
$e^{-5}\approx 7\cdot 10^{-3}$. Therefore the latter is expected
to dominate in intermediate orders \cite{BEN93b} (if there is
any regularity at all) and it might not be an accident, that
exact low order coefficients are indeed of same sign for these
quantities.

Table~\ref{tabpole} also shows how the one-loop radiative correction
to the pole mass is modified by inclusion of a finite number
of fermion loops and by summation according to Eq.~(\ref{rBSren}).
We have taken $\alpha_s(m_b)=0.2$ and $\alpha_s(m_c)=0.35$.
Recall that the sum corresponds to a principal value prescription for
the Borel integral. The errors quoted correspond to the imaginary
part of the Borel integral, when it is defined by deforming the
contour into the complex plane. The
imaginary part  can be taken as an estimate
of inherent uncertainty of perturbative relations. We have actually
divided this imaginary part by $\pi$, which upon inspection we
find closer to, but somewhat smaller than
the naive estimate of uncertainty by the minimal
term of the series. A more conservative estimate would be to enlarge
these errors by a factor of two. The BLM scales in Table~\ref{tabpole}
are defined by (cf.~Eq.(\ref{blmscalesdef}))

\be\label{compare}
   m^*_1 = m\,\exp\left[-\frac{1}{2a_s}(M_1(a_s)-1)\right]\,, \qquad
   m^*_\infty = m\,\exp\left[-\frac{1}{2a_s}
   \left(1-\frac{1}{M(a_s)}\right)\right]\,.
\ee

\n The uncertainty in $M_\infty(a_s)$ translates into an
uncertainty of the BLM scale $m^*_\infty$ by the previous
equation.
We want to point out that the usual BLM scale $m^*_1$, which
uses only $d_1$ as input is smaller than $m^*_\infty$ although
all higher orders $d_n$ add up positively. The reason is that
$\alpha_s(m^*_1)$ upon re-expanding in terms of $\alpha_s(m)$
implies $d_n^{\rm BLM}=d_1^n$. For the charm and bottom quark mass
the most important effect comes from $n=1,2$. But since $d_1$
is rather large, $d^{\rm BLM}_2=d_1^2\approx 21.9 > 17.6 = d_2$, and
the usual BLM prescription overestimates the size of
radiative corrections. This behaviour is quite generic to quantities
dominated by scales below a few GeV and with a leading infrared
renormalon at
$u=1/2$.

\subsection{Effect of internal quark masses}

Before proceeding to realistic charm and bottom quarks we want
to illustrate the effect of finite quark masses inside loops
on the coefficients of diagrams with quark loop insertions in
the toy example of the mass shift due to a single quark flavour
with mass $m_i$ in loops. The ``heavy'' quark of mass $m$
is again excluded from loops.

\begin{figure}[t]
   \vspace{-4.5cm}
   \epsfysize=16.8cm
   \epsfxsize=12cm
   \centerline{\epsffile{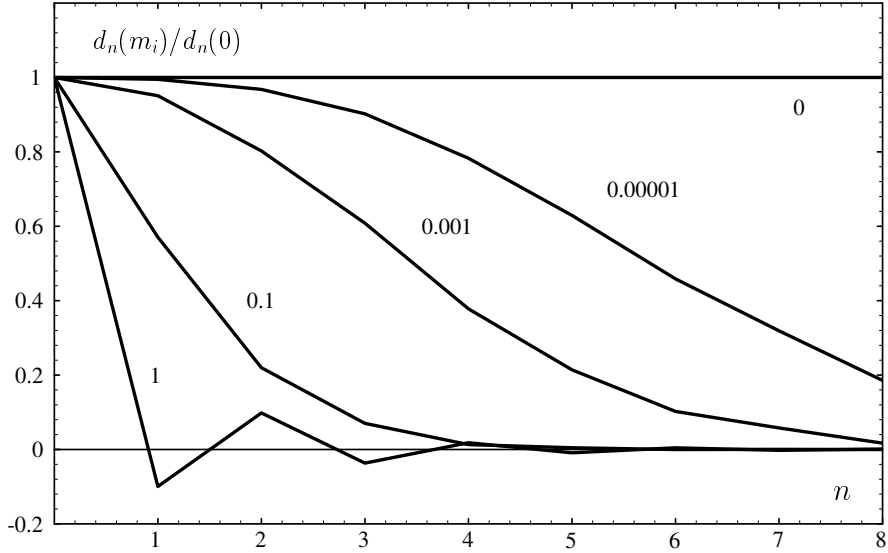}}
   \vspace*{-3.5cm}
\caption{\label{masses1} Ratio $d_n(m_i)/d_n(0)$ for different values
of internal quark masses $m_i$ as a function of the number of fermion
loop insertions. The value of $m_i^2/m^2$ is indicated to the right
of each
curve.}
\end{figure}

 The factorially large contribution from small and large loop
momenta can be traced to the logarithmic behaviour of the
vacuum polarization for a massless particle at very small and
very large virtuality. In the case of small virtuality, the
large coefficient arises from momenta $k\sim m e^{-n}\ll m$.
For a massive quark the logarithmic behaviour is cut off, because
at very small momenta its vacuum polarization is proportional
to $k^2/m_i^2$. Thus, no matter how small the quark mass, the
factorially large contribution should be eliminated when
$n$ is such that $m e^{-n} < m_i$. There are no infrared renormalon
singularities in the Borel transform in the absence of massless
particles. This expectation is verified in Fig.~\ref{masses1},
where we have plotted the ratio of the coefficient $d_n(m_i)$,
computed with internal mass $m_i$, and $d_n(0)$ for the massless
case as in Table~\ref{tabpole} as a function of the number of loops
$n$ for various ratios of $m_i/m$. Since $m_i$ serves
effectively as an infrared cutoff, Fig.~\ref{masses1} provides
some information on what proportion of the coefficient $d_n(0)$
originates from low momentum regions. Note that low momentum
means low momentum compared to $m$ and not $\Lambda_{\rm QCD}$.
Eventually, as $n$ becomes
large, coefficients will
be dominated by large momentum for any $m_i$ and the asymptotic
behaviour of Eq.~(\ref{aspole}) is replaced by a sign-alternating
behaviour due the leading
ultraviolet renormalon

\be\label{aspolem}
d_n(m_i) \stackrel{n\gg 1}{=} \left[-1+\frac{9}{2}
\frac{m_i^2}{m^2}\right] e^{-5/3}\, (-1)^n n!\,.
\ee

\n We find that for $m_i^2/m^2 \ge 0.1$, this asymptotic behaviour
practically coincides with the exact coefficient for $n>6$.

The presence of mass dependence in the coefficient of ultraviolet
renormalons requires some explanation. The singularity of the
Borel transform at $u=-1$, which is responsible for Eq.~(\ref{aspolem}),
is due to the presence of a $1/k^6$-term in the expansion of the
Feynman integrand for the one-loop mass-shift at large $k$:

\be \delta m \sim \alpha_s \,m \int d^4 k \left(\frac{a}{k^4}+
b\frac{m^2}{k^6} + \ldots\right)
\ee

\n The $1/k^4$-term creates a logarithmic ultraviolet divergence,
which is subtracted by minimal subtraction. The effect of the Borel
transformed gluon propagator (with fermion loop insertions) amounts
to insertion of (see Sect.~2.3)

\be \label{massbtprop}
\exp(-u \Pi(k^2)/a_s) = \left(-\frac{m^2}{k^2} e^{-C}\right)^{\!u}
\left[
1+6 u \frac{m_i^2}{k^2} +O\left(\frac{m_i^4}{k^4}\right)\right]
\ee

\n into the integrand. The pole at $u=-1$ arises, because close
to $u=-1$ the $1/k^6$-term is converted into a logarithmically
divergent term. But when $m_i$ is not zero a second contribution
of order $1/k^6$ arises, when $a/k^4$ combines with the first
mass correction in Eq.~(\ref{massbtprop}). This term accounts for
the mass-dependence of ultraviolet renormalons
in minimal subtraction schemes. Had one chosen a
renormalization prescription, which subtracts the $a/k^4$-term
inside the integrand, that mass-dependence would be absent for
the first ultraviolet renormalon at $u=-1$, but still present
in coefficients of singularities at $u=-2,-3\ldots$.

\subsection{Charm and bottom pole masses}

For realistic charm and bottom quarks, the numerical results of
Sect.~4.1 need to be amended in two respects: The quark, whose
mass shift is considered, should also be taken into account in
loops. This effect is tiny, but we include it for completeness.
More interesting is the effect of finite charm mass on the bottom
pole mass, because the charm mass might be larger than
the typical loop momentum already in low orders. If so, it would seem
more appropriate to use $\beta_0^{(3)}$ rather than $\beta_0^{(4)}$,
when restoring the QCD $\beta$-function coefficient from the
calculation of fermion loops. Strictly speaking, changing the
mass of one flavour is a
negligible effect in the formal large-$\beta_0$ limit (i.e.
large-$N_f$), and
should be discarded in a consistent approximation.
{\em In realiter}
the effect is numerically noticeable and we prefer to include it
as a calculable correction, keeping in mind its size as an
uncertainty in our approach.

To gain some understanding of the numerical results to follow, we
trace the decoupling of internal charm loops for the bottom pole mass
in more detail. For this purpose, we ignore again the bottom quark in
loops. When the order of perturbation theory is sufficiently large,
internal integrations are dominated by momenta much smaller
than $m_c$. Let us denote coefficients including charm loops by
$d_n^{[3+1]}(m_c)\,(-\beta_0^{(4)})^n$ and
those without charm loops
by $d_n^{[3]}\,(-\beta_0^{(3)})^n$. Then one might expect

\be \label{decoup1}
d_n^{[3+1]}(m_c)\,\left(-\beta_0^{(4)}\right)^n \alpha_s(m_b)^{n+1}
\approx
d_n^{[3]}\,\left(-\beta_0^{(3)}\right)^n \alpha_s(m_b)^{n+1}
\ee

\n if $n$ is sufficiently large up to corrections suppressed by
a power of the typical internal momentum divided by $m_c$. This is
not quite correct, because there is no manifest decoupling in the $\MS$
scheme. In the limit that we consider only the contribution
from small $\lambda^2$, appropriate for sufficiently large $n$, the
Borel transform including the massive charm in Eq.~(\ref{rBSmass})
reduces to

\be\label{decoupborel}
B\left[\frac{\delta m}{m}\right](u)
= \exp\!\left(\frac{u}{6\pi\beta_0^{(4)}}\ln
\frac{\mu^2}{m_c^2}\right)\!\left(-\frac{1}{\pi}\right)
\sin\!\left(\frac{u\beta_0^{(3)}}{\beta_0^{(4)}}\right)\!\!\intl_0^\infty
\frac{d\lambda^2}{\lambda^2}\,(r_0(\lambda^2)-r_0(0)) \!\left(
\frac{\lambda^2}{\mu^2 e^{5/3}}\right)^{\!\!u \beta_0^{(3)}/
\beta_0^{(4)}}
\ee

\n up to power corrections. If the first factor
were absent, this Borel transform would
coincide with the one for purely massless flavours up to a rescaling
of $u$ which is just what is necessary to obtain Eq.~(\ref{decoup1})
from Eq.~(\ref{genfunction}). The first factor is present, because
the vacuum polarization for charm has been renormalized in the
$\MS$ scheme
and not by zero momentum subtraction, which would lead to manifest
decoupling. The difference can be accomodated by a change of the
coupling
constant. This is most easily seen by combining the extra exponential
factor with the exponential in the Borel integral, Eq.~(\ref{borelintegral}).
We deduce that (up to power corrections in the typical loop
momentum divided by $m_c$)

\be \label{decoup2}
d_n^{[3+1]}(m_c) \,\left(-\beta_0^{(4)}\right)^n
\alpha_s^{(4)}(m_b)^{n+1} \approx
d_n^{[3]} \,\left(-\beta_0^{(3)}\right)^n \alpha_s^{(3)}(m_b)^{n+1}\,,
\ee

\n where the two couplings are related by

\be
\frac{1}{\alpha_s^{(3)}(\mu)} = \frac{1}{\alpha_s^{(4)}(\mu)} +
\frac{1}{6\pi} \,\ln\frac{\mu^2}{m_c^2}\,.
\ee

\n Alternatively, we can simply replace $u$ in the
extra exponential factor in Eq.~(\ref{decoupborel})
by the location of the closest infrared
pole of the Borel transform, $u=1/2$, and obtain
\phantom{\ref{tabmc},\ref{tabmb}}

\be \label{decoup3}
d_n^{[3+1]}(m_c) \,\left(-\beta_0^{(4)}\right)^n
\alpha_s(m_b)^{n+1} \approx
\exp\left(\frac{1}{12\pi\beta_0^{(4)}}\ln
\frac{m_b^2}{m_c^2}\right)
d_n^{[3]} \,\left(-\beta_0^{(3)}\right)^n \alpha_s(m_b)^{n+1}\,,
\ee

\n which is equivalent to Eq.~(\ref{decoup2}) for large $n$.

\begin{table}[t]
$$
\begin{array}{|c||c|c|c|c|}
\hline
n & d_n^{[3]} & d_n^{[3+1]} & d_n^{[3]}\,
(\beta_0^{(3)}/\beta_0^{(4)})^n & M_n^{[3+1]}\\ \hline\hline
0 & 1 & 1 & 1 & 1\\
1 & 4.6862 & 5.0984 & 5.0613 & 2.183  \\
2 & 17.623 & 20.514 & 20.555 & 3.291  \\
3 & 109.86 & 138.58 & 138.38 & 5.021  \\
4 & 873.92 & 1188.3 & 1189.0 & 8.470  \\ \hline
\infty & - & - & - & 1.720 \pm 0.608 \\ \hline
m^*_1 & - & - & - & 0.078 \, m_c  \\
m^*_\infty & - & - & - & 0.406 \, m_c \\
\hline
\end{array}
$$
\caption{Perturbative corrections to the pole charm quark mass with
(column 3) and without (column 2) charm inside loops. For comparison,
column 4 gives coefficients in the limit that all internal
masses are large compared to the typical loop momentum. The last
column updates the modification of one-loop corrections due to
coefficients $d_n^{[3+1]}$.}
\label{tabmc}
\end{table}

\begin{table}[t]
$$
\begin{array}{|c||c|c|c|c|}
\hline
n & d_n^{[3]} & d_n^{[3+2]} & K\,d_n^{[3]}\,
(\beta_0^{(3)}/\beta_0^{(5)})^n & M_n^{[3+2]}\\ \hline\hline
0 & 1 & 1 & 1 & 1\\
1 & 4.6862 & 5.3093 & 4.9774 & 1.648  \\
2 & 17.623 & 22.720 & 21.975 & 1.986  \\
3 & 109.86 & 159.69 & 160.82 & 2.276  \\
4 & 873.92 & 1.5\cdot 10^3 & 1502.0 & 2.601  \\ \hline
\infty & - & - & - & 2.099 \pm 0.224 \\ \hline
m^*_1 & - & - & - & 0.064\, m_b  \\
m^*_\infty & - & - & - & 0.117\, m_b \\
\hline
\end{array}
$$
\caption{Coefficients as
in the previous table, where $d_n^{[3+2]}$ now includes charm
and bottom masses with $m_c^2/m_b^2=0.1$. $K=\exp(1/(12\pi\beta_0^{(5)})
\ln(m_b^2/m_c^2))\approx 0.9$.}
\label{tabmb}
\end{table}

Numerical values for corrections to the charm pole mass and the bottom
pole mass, including the charm and bottom masses in loops, are
given in Tables~\ref{tabmc} and \ref{tabmb}. For the bottom pole mass,
the result depends only on the ratio $m_c/m_b$ and we used
$m_c^2/m_b^2=0.1$. For comparison coefficients
according to Eq.~(\ref{decoup3}) are given. The last column in both
tables displays the modification of the one-loop mass shift by
running coupling effects. Note that at the charm scale, the perturbative
series has to be truncated already at second order. The lower rows
give the result of summation according to Eq.~(\ref{rBSmass}) and the
corresponding BLM scales, Eq.~(\ref{compare}). Two technical observations
are in order: First, as mentioned in Section 2, the sum as defined by
Eq.~(\ref{rBSmass}) does not exactly coincide with the principal
value of the Borel integral. The difference is tiny for the value of
masses considered here and does not affect our conclusions which are
based on the behaviour of perturbation theory in low and intermediate
orders. Second, for the bottom quark the
BLM scale is defined without taking into account the charm threshold
in the running coupling.

 The difference between the pole mass and the $\MS$-renormalized
mass (normalized at $m_{\MS}$) for bottom quarks can now be
written as

\be
\frac{\delta m_b}{m_b} = \frac{4}{3}\frac{\alpha_s(m_b)}{\pi} \left[
M_\infty^{[3+2]}(-\beta_0^{(5)} \alpha_s(m_b))-0.91 \alpha_s(m_b)
+\mbox{higher orders}\right]\,,
\ee

\n where the second term in square brackets accounts for the genuine
gluonic two-loop corrections, cf.~Eq.~(\ref{m1}).
Unknown higher order corrections
include genuine three- and higher loop corrections, vacuum polarization
insertions into two-loop corrections as well as effects of two-loop
running on lowest order radiative corrections.
Numerically, for $\alpha_s(m_b)=0.2$,
we find

\be
\frac{\delta m_b}{m_b} = (16.3\pm 2.9\pm 1.5)\%\,.
\ee

\n Compared to the two-loop expression, the estimate obtained from
NNA increases the mass shift from about $12\%$ to $16\%$.
The second error reflects our estimate
of unknown higher order corrections, which we allow to be as large
as the genuine two-loop corrections. The first error, which
dominates the total uncertainty, represents an estimate of the
ultimate accuracy of $\delta m_b/m_b$ due to the divergence of the
perturbative series. It can not be reduced by calculating higher
orders (but of course refined -- the numerical value we quote
has been obtained by increasing
the uncertainty of $M_\infty^{[3+2]}(a_s)$
in Table~\ref{tabmb} by 50\%, upon which it is close to the
minimal term of the series at $n\approx 2 - 3$). In absolute values,
the intrinsic uncertainty of $\delta m_b$ and therefore of
the pole mass of
the bottom quark ranges between $100$ and $150\,$MeV. The above
errors do not include an uncertainty in $\alpha_s(m_b)$, which
has been fixed to 0.2.

All results were given in terms of $m_{\MS}(m_{\MS})$. To run
to a different renormalization point $\mu$, one
should use the anomalous dimension to the same large-$\beta_0$
approximation. In the $\MS$-scheme it is given by

\be
\gamma_m(\alpha_s)\equiv -\frac{\mu^2}{m}\frac{d m(\mu)}{d\mu^2}
= -\frac{\alpha_s}{3\pi}\,G_0(-\beta_0\alpha_s)
\ee

\n with $G_0$ as in Eq.~(\ref{subfunction}) in agreement with
\cite{PAL84}. Thus, in this approximation

\be
m(\mu)=m(m)\,\exp\left(\,\,\intl_{\alpha_s(m)}^{\alpha_s(\mu)}
\frac{d\alpha_s}{3\pi\beta_0\alpha_s}\, G_0(-\beta_0\alpha_s)
\right)\,.
\ee

\n To the approximation considered, one may replace the
exponential by the first two terms of its expansion. In practice,
it might be better to consider the approximation as an approximation
to the exponent and to keep the exponential.



\renewcommand{\topfraction}{1}
\renewcommand{\textfraction}{0}

\mysection{Scale-setting at next-to-leading order}
{Scale-setting at next-to-leading order}

\begin{figure}[t]
   \vspace{-1cm}
   \epsfysize=28cm
   \epsfxsize=20cm
   \centerline{\epsffile{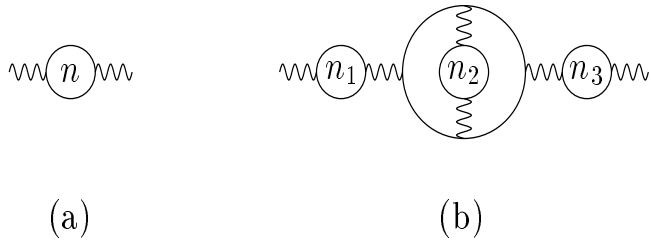}}
   \vspace*{-21.5cm}
\caption{\label{nlodiag} QED-like diagrams incorporating evolution of
the coupling to leading (a) and next-to-leading order (b). A circle with
letter $m$ denotes a chain of $m$ fermion loops. At order $\alpha^{k+1}$,
the relevant diagrams are specified by $n=k$ and $n_1+n_2+n_3=k-2$. At NLO,
the diagrams, where the $n_2$-chain forms a self-energy-type
insertion, are
not depicted.}
\end{figure}

So far we have been discussing a scale-setting procedure that utilizes only
one-loop evolution of the strong coupling. At the same number of loops,
where this procedure extends the familiar scheme of Brodsky, Lepage
and Mackenzie, one also encounters diagrams associated with two-loop
evolution, which are not suppressed compared to insertions of single
fermion loops by any small parameter. This Section is devoted to the
possibility of and difficulty in extending the scale-setting to
next-to-leading order (NLO). The precise meaning of NLO in this context
is illustrated in Fig.~\ref{nlodiag}: For any process, we will calculate
the class of higher order corrections, generated by substituting the
gluon (photon) propagator in lowest order by the
chains of Fig.~\ref{nlodiag}a
(this was done in previous Sections) and Fig.~\ref{nlodiag}b.
In the abelian theory,
the results are exact, while in the nonabelian theory, one must again
define how to restore the coefficients $\beta_0$ and $\beta_1$ of
the nonabelian $\beta$-function. Let us note that as in the case of
one-loop running this extension is heuristically motivated by the
behaviour of perturbation theory in large orders. Two-loop running
is known \cite{MUE85,ZAK92} to modify the
strength of renormalon singularities from poles to branch points
involving the ratio $\beta_1/\beta_0^2$. The class of diagrams in
Fig.~\ref{nlodiag}b is expected to dominate at large $n$ over one-loop
running by a factor $\beta_1/\beta_0^2 \ln n$ and is the first in
a series of multiple two-loop insertions that exponentiate to produce
an enhancement factor $n^{-\beta_1/\beta_0^2}$. We are thus again led
by the expectation that a class of systematically large corrections
can be identified and calculated.
Consider a physical quantity,
with perturbative expansion written as

\be R-R_{tree} = \sum_{n=0} r_n \alpha_s(Q)^{n+1}\,,\qquad
r_n=r_{n0}+r_{n1} N_f+\ldots r_{nn} N_f^n\,,
\ee

\n assuming for simplicity that $R$ depends only on a single scale
$Q$, which is equal to the renormalization scale, such that the
$r_n$ are numerical coefficients.

Since the scale-setting prescription is derived from QED (that is,
its fictitious version with $N_f$ massless fermions), it is useful
to consider first the abelian analogue of $R$. Due to the Ward
identity, the evolution of the coupling in the abelian theory is
generated by radiative corrections to the photon propagator. It is
then natural to define the scale of the coupling at each order in
perturbation theory by replacing each photon propagator $1/k^2$ by
the full propagator $1/(k^2 (1+\Pi(k^2))$ and absorbing the effect
of integrating loops with full propagators into the normalization
of the coupling constant to the order where the corresponding skeleton
diagram appears. In this way, one is led to a modified expansion
\cite{BRO83}

\be R-R_{tree} = r_0\left[\alpha_s(Q^*) + \delta_1 \alpha_s(Q^{**})^2
\ldots\right]\,.
\ee

\n Specifically, the scale of the coupling at order $\alpha_s$ is
determined by the replacement

\be \label{mom}
\alpha_R(Q) \rightarrow \alpha_V(k)\equiv
\frac{\alpha_R(Q)}{1+\Pi_R(k^2/Q^2,\alpha_R(Q))}
\ee

\n inside the loop-integration over $k$, where $k$ is the
momentum of the virtual photon. We have attached a subscript $R$ to
quantities that depend on the choice of the
renormalization scheme $R$. On
the right hand side we notice that $\alpha_V(Q)$ is the effective
coupling, defined by the potential between two static sources in momentum
space,

\be
V(Q)=-\frac{4\pi\alpha_V(Q)}{Q^2}\,,
\ee

\n which defines the V-scheme \cite{BRO83}. Thus, in {\em any} scheme
$R$, the scale $Q_R^*$ is given by averaging the lowest order radiative
corrections to the quantity $R$ with the running coupling in the
V-scheme, $\alpha_V(k)$, and not\footnote{At the level of
one-loop running, the relevant transformation of schemes is
simply a shift of scale, since, to this approximation,

\bdm
\alpha_V(k)=\frac{\alpha_R(k)}{1-\beta_0 C_R\alpha_R(k)} = \alpha_R
\left(k e^{C_R/2}\right)\,.
\edm} with $\alpha_R(k)$. By construction
$\alpha_R(Q_R^*)$ is scheme-independent.
The choice of $\alpha_V(k)$ to average the loop momentum distribution
is physically appropriate, because the photon exchanged between
static sources at distance $r$ has momentum $k\sim 1/r$. Therefore
it is in this scheme that $\alpha(k)$ can be interpreted as the
effective coupling of a virtual photon of momentum $k$.

With this remark one is prepared to adopt the same definition in
the nonabelian theory. $Q^*$ is defined as \cite{LEP93}

\be\label{defofq}
r_0\alpha_R(Q^*) \equiv \int d^4 k\,F(k,Q)\,\frac{\alpha_V(k)}{k^2}
\,,
\ee

\n where $F(k,Q)$ is the integrand of the lowest order radiative correction
and $C_F 4\pi \alpha_V(k)/k^2$ the potential between two static
colour sources in momentum space, given by a Wilson loop. The Wilson
loop is nonperturbatively well-defined (up to a distance-independent
contribution) and the above integral could be evaluated without any
ambiguity. In the following, we will use it only perturbatively, in the
sense of an expansion of $\alpha_V(k)$ in $\alpha_V(Q)$. This restriction
is not only self-imposed. Power corrections to the static potential might
be different and in particular larger than those to the quantity $R$,
in which case one would not like to use Eq.~(\ref{defofq}) with power-like
accuracy. Let us also note that one-loop gauge boson exchange exponentiates
exactly in the abelian Wilson loop but not in the nonabelian case.
Therefore re-expressing $\alpha_V(Q)$ in terms of some other $\alpha_R(Q)$
inevitably involves contributions to the Wilson loop with more than one
gluon exchanged.

The exact (perturbative) evaluation of Eq.~(\ref{defofq}) is impossible
even in the abelian theory. One has to resort to some truncation of
(a) the $\beta$-function in the $V$-scheme, needed to relate $\alpha_V(k)$
to $\alpha_V(Q)$ and (b) the relation between the coupling in the V-scheme
and some other scheme $R$, if one wishes to express the result in
terms of $\alpha_R$. To leading order, the $\beta$-function has been
replaced by $\beta_0 \alpha^2$. After expansion of $\alpha(k)$ in
$\alpha(Q)$, this corresponds to the diagrams of Fig.~\ref{nlodiag}a
in QED. In this section we discuss a truncation, which is
guided by the subleading $N_f$-dependence of coefficients and incorporates
the set of diagrams depicted in Fig.~\ref{nlodiag}b in addition to those
in Fig.~\ref{nlodiag}a.
Let us first consider $n_1=n_3=0$ in Fig.~\ref{nlodiag}b.
The diagram with $n_2=0$ gives $\beta_1\alpha^3$ to the abelian
$\beta$-function. For $n_2>0$, these diagrams give contributions
to higher order coefficients to the $\beta$-function. Again in QED,
this contribution can be written as
$b_{n_2+1} (-\beta_1) (-\beta_0)^{n_2}$
with a $N_f$-independent number $b_{n_2}$. When expanding
$\alpha_V(k)$ as

\be \label{expexp}
\alpha_V(k)= \exp\left[\ln\frac{k^2}{Q^2}\,\beta_V(\alpha^\prime)
\frac{d}{d\alpha^\prime}\right]\alpha^\prime_{\big|\alpha^\prime=
\alpha_V(Q)}\,,
\ee

\n the diagrams of Fig.~\ref{nlodiag} are recovered by
using

\be\label{betav}
\beta_V(\alpha)=\beta_0\alpha^2+\beta_1 \sum_{n=1}^\infty b_n
(-\beta_0)^{n-1} \alpha^{n+2}\,,
\ee

\n but keeping only terms with at most one power of $\beta_1$. A
different truncation could be imagined, where the $\beta$-function
is kept only up to $\beta_1\alpha^3$ but all powers of $\beta_1$ are
taken into account in the expansion of Eq.~(\ref{expexp}).
The difference to the previous
truncation appears first at order $\alpha^5$.

 The transition to the nonabelian theory is arguably uniquely
achieved at leading order by replacing the abelian $\beta_0$ by
its nonabelian value. At NLO one has to make a choice which is
less compulsory: Calculate the insertions of the
diagrams of Fig.~\ref{nlodiag}b
and express them as $c_n (-\beta_1)(-\beta_0)^{n-2}\alpha(Q)^{n+1}$
in QED. Then replace $\beta_0$ and $\beta_1$ by their nonabelian
values. In effect this implies substitution of the true $\beta$-function
in the V-scheme by Eq.~(\ref{betav}). We can now write the perturbative
coefficients of the quantity $R$ as

\be\label{conlo}
r_n = r_{n0}+\ldots r_{nn-1}N_f^{n-1}+r_{nn} N_f^n
\equiv r_0\left[\delta_n -\beta_1 (-\beta_0)^{n-2}
c_n+(-\beta_0)^n d_n
\right]\qquad n\ge2\,,\ee

\n where, according to the prescription, the last two
terms are absorbed into
$Q^*$. Both sets of coefficients $c_n$ and $d_n$ are computed
from abelian diagrams. Note that $d_n$ is constructed as to eliminate
the highest power of $N_f$. On the other hand, to NLO, the
remainder $\delta_n$ still contains subleading flavour dependence,
$N_f^{n-1}$, which arises from insertion of fermion loops into
diagrams with two gluon lines as well as the effective three-gluon
coupling generated by attaching three gluons to a fermion loop.

Before turning to the calculation of $c_n$, we want to illustrate the
prescription at order $\alpha_s^3$. Up to this order, the expansion
of $R$ in a certain scheme with coupling $\alpha_s$
can be written as

\be\label{rr}
R-R_{tree} = r_0\left[\alpha_s(Q)+\left\{\delta_1+(-\beta_0) d_1\right\}
\alpha_s(Q)^2+\left\{\delta_2 -\beta_1 c_1+(-\beta_0)^2 d_2\right\}
\alpha_s(Q)^3\right]\,.
\ee

\n From Eq.~(\ref{defofq}) we obtain

\be
\alpha_s(Q^*)=\alpha_V(Q)+(-\beta_0) d_1 \alpha_V(Q)^2 + \left\{
(-\beta_1) d_1+(-\beta_0)^2 d_2\right\}
\alpha_V(Q)^3
\ee

\n where the (scheme-dependent) $d_n$ are computed as in Section~2
from diagrams with fermion loop insertions.
By comparison we find that in the V-scheme ($\alpha_s(Q)=\alpha_V(Q)$)
$c_1=d_1$. The remaining
linear $N_f$-dependence in $\delta_2$ can then be absorbed into
a scale $Q^{**}$ of the $\alpha_s^2$-correction \cite{BRO94}.
Notice, however, that
while $Q^*$ is defined without knowledge of the
exact $\alpha_s^3$-coefficient,
the definition of $Q^{**}$ in the nonabelian theory requires
the exact $\alpha_s^3$-result.

To obtain the scale $Q^*$ as an
expansion in the coupling
different from the V-scheme, one has to relate the couplings to the same
accuracy and expand $\alpha_V(Q)$ in, say, $\alpha_{\MS}(Q)$ in the
same form as Eq.~(\ref{rr}):

\bea\label{rel}
\alpha_V(Q) &=& \alpha_{\MS}(Q)+\left\{\gamma_1+(-\beta_0)
\left(-\frac{5}{3}
\right)\right\}\alpha_{\MS}(Q)^2\nonumber\\
&&\,+\left\{\gamma_2 -\beta_1
\left(-\frac{55}{12}+4\zeta(3)\right)+(-\beta_0)^2 \frac{25}{9}
\right\}\alpha_{\MS}(Q)^3
\eea

\n One can then determine $c_1$ in the $\MS$ scheme. Once $d_1$,
$d_2$ and $c_1$ are determined, $Q^*_2$ to $\alpha_s^3$-accuracy
is given by

\be Q^*_2=Q\,e^{-d_1/2}\left(1+\left\{(-\beta_0) b_1+\tilde{b}_1
\frac{\beta_1}{\beta_0}\right\}
\alpha_s(Q)\right)\,.
\ee

\n $b_1$ and $\tilde{b}_1$ are uniquely determined by the
condition

\bea \alpha_s(Q^*_2) &=& \alpha_s(Q)+(-\beta_0) d_1 \alpha(Q)^2
+\left\{(-\beta_0)^2 (d_1^2-2 b_1)-(d_1-2 \tilde{b}_1)\beta_1
\right\} \alpha_s(Q)^3+\ldots\nonumber\\
&\equiv& \alpha_s(Q)+(-\beta_0) d_1 \alpha(Q)^2
+\left\{(-\beta_0)^2 d_2-\beta_1 c_1\right\}\alpha_s(Q)^3\,,
\eea

\n which specifies that the diagrams of Fig.~\ref{nlodiag}
relevant at order $\alpha_s^3$ are absorbed into $Q^*_2$ in
any scheme. In the V-scheme $\tilde{b}_1=0$, but in general
$\tilde{b}_1$ is non-zero.

In the V-scheme the prescription outlined here
coincides exactly with the one by Brodsky and Lu \cite{BRO94}.
When $Q^*_2$ is defined in other schemes, we suspect that
$Q^*$ defined in \cite{BRO94} does not absorb the QED diagrams of
Fig.~\ref{nlodiag}b and part of the flavour-dependence
from these diagrams is hidden in $Q^{**}$, since $\tilde{b}_1$
is always zero in \cite{BRO94}.

When we compute $c_n$ in the following Section, rather than using the
V-scheme in intermediate steps, we will perform the average of
Eq.~(\ref{defofq}) directly in an arbitrary scheme by replacing
$\alpha_V(k)$ with the appropriately truncated form of Eq.~(\ref{mom}).

\subsection{Abelian diagrams at NLO}

We now evaluate the perturbative coefficients generated by inserting
the chains of Fig.~\ref{nlodiag} into the gluon (photon) line which
appears in lowest order radiative corrections to an observable $R$.
In this subsection $\beta_0$ and $\beta_1$ refer to their abelian values:
$\beta_0=N_f/(3 \pi)$ and $\beta_1=N_f/(4\pi^2)$. These chains have
previously been considered for particular quantities: The muon anomalous
magnetic moment \cite{BRO93} and the self-energy of a static quark
\cite{BEN94}. Because vacuum polarization insertions are universal, the
results can easily be adapted to the general case. The vacuum polarization
to the present approximation is given by the diagrams with $n=1$ and
$n_1,n_3=0$ in Fig.~\ref{nlodiag} ($Q^2$ is euclidian):

\be \Pi\left(\frac{Q^2}{\mu^2},\alpha(\mu)\right) = -\beta_0\alpha(\mu)
\left[\ln\frac{Q^2}{\mu^2}+C\right] + \beta_1\Pi_1\left(
\frac{Q^2}{\mu^2},\alpha(\mu)\right)
\ee

\n When $\alpha_V(k)$ is expanded inside the
integrand of Eq.~(\ref{defofq}),
the result is expressed in terms of coefficients of the vacuum
polarization and averages of $\ln^n(k^2/Q^2)$ in the lowest
order radiative correction $F(k,Q)$, which have already been
evaluated through the insertion of the diagrams of
Fig.~\ref{nlodiag}a. Therefore, since
$\Pi_1$ is known, the extension to incorporate Fig.~\ref{nlodiag}b is
merely a combinatorical problem. To organize the combinatorics, it is
convenient to introduce the Borel transform. First, we define the
leading order and next-to-leading order Borel transform as the
result of insertion of diagrams
with a single chain (Fig.~\ref{nlodiag}a) and
the sum of both diagrams in the Fig.~\ref{nlodiag}.
{}From Eq.~(\ref{borelsum})
and Eq.~(\ref{conlo}):

\bea
B_{\rm LO}[R](u) &\equiv& r_0 \sum_{n=0}^\infty\frac{d_n}{n!} u^n
\nonumber\\
B_{\rm NLO}[R](u) &\equiv& r_0 \sum_{n=0}^\infty\frac{d_n^{\rm NLO}}
{n!} u^n\qquad d_n^{\rm NLO}=d_n-\frac{\beta_1}{\beta_0^2} c_n \quad
(n\ge 2)
\eea

\n and $d_n^{\rm NLO}=d_n$ for $n=0,1$. After Borel transformation of
Eq.~(\ref{defofq}), the truncation of $\alpha_V(k)$ discussed above
corresponds to insertion of

\be\label{vacnlo}
B\left[\frac{\alpha(Q)}{1+\Pi(k^2/Q^2,\alpha(Q))}\right](u)
= \left(\frac{k^2}{Q^2} e^C\right)^{-u} - \frac{\beta_1}{\beta_0^2}
\intl_0^u d v\, v\left(\frac{k^2}{Q^2} e^C\right)^{-v}
B\left[\frac{\Pi_1}{\alpha}\right](u-v)
\ee

\n into Eq.~(\ref{defofq}) instead of the complete $\alpha_V(k)$.
On the right hand side, we have neglected
multiple insertions of $\Pi_1$. The Borel transform of $\Pi_1/\alpha$
can be represented as

\be
B\left[\frac{\Pi_1}{\alpha}\right](u) =
\left(\frac{k^2}{Q^2}\right)^{-u}
F(u) - G(u)\,,
\ee

\n where $F(u)$ is a scheme-independent function that can be obtained
from Eq.~(\ref{borelpolaroper}) by integration with respect to $Q^2$:

\be F(u) = \frac{32}{3} \frac{1}{1-(1-u)^2} \sum_{k=2}^\infty
\frac{(-1)^k k}{(k^2-(1-u)^2)^2} \equiv \sum_{n=-1}^\infty f_n u^n
\ee

\n $G(u)$ is a scheme-dependent integration constant. One can use the
renormalization group equation obeyed by the photon
vacuum polarization to
relate the expansion coefficients of $G(u)$ to those of the $\beta$-function
in the chosen scheme. The precise relation is as follows: Let us
write the highest power of $\beta_n$ as

\be
\beta_{n\big|N_f^n} \equiv b_n \beta_1 (-\beta_0)^{n-1}\,.
\ee

\n Then

\be G(u) = \sum_{n=0}^\infty \frac{b_{n+1}}{n!} u^{n-1}\,,\qquad
g_n = \frac{b_{n+2}}{(n+1)!}\,.
\ee

\n With these preliminaries, we insert Eq.~(\ref{vacnlo}) into
Eq.~(\ref{defofq}) and obtain

\bea\label{rnlo}
B_{\rm NLO}[R](u) &=& B_{\rm LO}[R](u) - \frac{\beta_1}{\beta_0^2}
\intl_0^u\frac{d v\, v}{u-v} \left(B_{\rm LO}[R](u)-B_{\rm LO}[R](v)
\right)\nonumber\\
&&\hspace*{-2cm}
\, - \frac{\beta_1}{\beta_0^2} \intl_0^u d v\, v\left(B_{\rm LO}[R](u)
F_{\rm reg}(u-v) - B_{\rm LO}[R](v) G_{\rm reg}(u-v)
\right)\,,
\eea

\n where $F_{\rm reg}(u)$ and $G_{\rm reg}(u)$ are defined by
removing the pole
at $u=0$: $F_{\rm reg}(u)\equiv F(u)-1/u$, $G_{\rm reg}(u)\equiv G(u)-1/u$.

The coefficients $d_n^{\rm NLO}$ are given by

\be d_n^{\rm NLO}=\frac{1}{r_0}\frac{d^n}{du^n}
B_{\rm NLO}[R](u)_{|_{u=0}}\,.
\ee

\n Taking the derivatives of Eq.~(\ref{rnlo}) we obtain (for $n\ge 2$)

\bea\label{dnnlo}
d^{\rm NLO}_n &=& d_n-\frac{\beta_1}{\beta_0^2}
n\,(\psi(n+1)-\psi(2))\,d_{n-1}
\nonumber\\
&&\,-\frac{\beta_1}{\beta_0^2} \sum_{k=0}^{n-2}
\left[\left(\begin{array}{c}
n\\k\end{array}\right)
f_{n-2-k}-(k+1) g_{n-2-k}\right] (n-2-k)!\, d_k
\,,\eea

\n where $\psi(x)$ is the logarithmic derivative of the $\Gamma$-function.
Note that for large $n$ the second term on the right side indeed dominates
the first one by a factor $\beta_1/\beta_0^2\ln n$.
With correspondingly changed conventions, this equation
agrees with the corresponding one in \cite{BRO93}. The expansion
coefficients of $F(u)$ are given by \cite{BRO93}

\be f_n = -\frac{2}{3} (n+2) \left[-2 n-2-\frac{n+7}{2^{n+3}}+\frac{16}{n+2}
\sum_{k=1}^{\left[\frac{n+3}{2}\right]} k (1-2^{-2 k}) (1-2^{2 k-n-3})
\zeta(2 k+1)\right]\,,
\ee

\n where $\zeta(k)=\sum_{n=1}^\infty n^{-k}$ and $[..]$ denotes the integer
part of the number in brackets. The coefficients $g_n$ depend on the
scheme employed for the definition of $\alpha$. In the $\MS$ scheme,
we can use the $\beta$-function to the approximation required here from
\cite{PAL84,BRO93}. We then find

\be
g_n^{\MS} = \frac{1}{(n+1)! (n+2)!} \frac{d^{n+1}}{du^{n+1}} \left[
\frac{(1-u) (1+2 u) (3+2 u) \Gamma(4+2 u)}{9\Gamma(2+u)^2 \Gamma(3+u)
\Gamma(1-u)}\right]_{\big|u=0}\,.
\ee

\n In the $V$-scheme, we have

\be g_n^V = f_n\,.
\ee

\phantom{\ref{tt1},\ref{tt2},\ref{tt3}}
\n The simplest way to see this is that in this scheme by definition
$\Pi(1,\alpha_V(Q))=0$. Then the expansion of the
integrand in Eq.~(\ref{defofq}) has only non-zero powers of $\ln(k^2/Q^2)$
and $d_0$ can not appear in Eq.~(\ref{dnnlo}).

\subsection{Numerical analysis}

The transition to QCD according to the prescription formulated above
is performed by replacing $\beta_1/\beta_0^2$ in Eq.~(\ref{dnnlo})
by its QCD value. Note that contrary to $d_n$ the NLO coefficient
depends explicitly on the number of flavours. In this
Subsection, rather then presenting
values for the BLM scale $Q^*$ to NLO accuracy, we give the perturbative
coefficients in low orders generated by expansion of of $\alpha_s(Q^*)$
in the $\MS$-coupling, that is the coefficients $d_n^{\rm NLO}$ and
$M^{\rm NLO}_n(-\beta_0\alpha_s)$, defined in complete analogy with
$M_n(-\beta_0\alpha_s)$. To compare the coefficients obtained from
keeping only vacuum polarization with the exact ones, if available,
we shall also define $d_n^{\rm exact}$ as the exact coefficient,
divided by $(-\beta_0)^n$. Numerical values for the difference between
pole and $\MS$ mass are shown in Table~\ref{tt1}, and for the
derivative of the hadronic vacuum polarization $Q^2 d\Pi/dQ^2$ and
the $\tau$ decay width in Tables~\ref{tt2} and \ref{tt3}.

\begin{table}[th]
$$
\begin{array}{|c||c|c|c|c|}
\hline
n & d_n & d_n^{\rm NLO} & M_n\,[\mbox{b-quark}] &
M_n^{\rm NLO}\,[\mbox{b-quark}]\\ \hline\hline
0 & 1 & 1 & 1 & 1 \\
1 & 4.68615 & 4.68615 & 1.623 & 1.623 \\
2 & 17.6227 & 19.6884 & 1.935 & 1.972 \\
3 & 109.859 & 127.529 & 2.193 & 2.272 \\
4 & 873.924 & 1138.28 & 2.467 & 2.628 \\
5 & 8839.69 & 12085.6 & 2.835 & 3.131 \\
\hline
\end{array}
$$
\caption{\label{tt1}
Comparison of leading order and next-to-leading order
coefficients for the difference of pole and $\MS$
mass, where quark masses inside loops
are neglected and loops containing the heavy quark are excluded
(cf. Sect.~4). The comparison of the modification of lowest order
radiative corrections is for $\alpha_s(m_b)=0.2$, relevant to bottom
quarks.}
\vspace*{0.4cm}
$$
\begin{array}{|c||c|c|c||c|c|c|}
\hline
n & d_n & d^{\rm NLO}_n & d_n^{\rm exact} & M_n & M^{\rm NLO}_n
& M^{\rm exact}_n\\ \hline\hline
0 & 1 & 1 & 1 & 1 & 1 & 1 \\
1 & 0.691772 & 0.691772 & 0.728809  & 1.158 & 1.158 & 1.167\\
2 & 3.10345 & 2.31417 & 1.25847 & 1.321 & 1.271 & 1.233\\
3 & 2.18004 & 7.40378 & - & 1.347 & 1.360 & - \\
4 & 30.7398 & 18.4580 & - & 1.432 & 1.411 & - \\
5 & -34.5336 & 146.293 & - & 1.410 & 1.503 & - \\
\hline
\end{array}
$$
\caption{\label{tt2}
 Comparison of leading order, next-to-leading order and
exact coefficients of $Q^2 d\Pi(Q)/dQ^2$. The NLO and
exact values are given for
$N_f=3$ and the partial sums $M_n$ for $\alpha_s=0.32$. Exact values,
when available, include genuine higher order corrections.}
\vspace*{0.6cm}
$$
\begin{array}{|c||c|c|c||c|c|c|}
\hline
n & d_n & d^{\rm NLO}_n & d_n^{\rm exact} & M_n & M^{\rm NLO}_n
& M^{\rm exact}_n\\ \hline\hline
0 & 1 & 1 & 1 & 1 & 1 & 1 \\
1 & 2.27511 & 2.27511 & 2.31213  & 1.521 & 1.521 & 1.529\\
2 & 5.68475 & 5.98780 & 5.20806 & 1.819 & 1.835 & 1.803\\
3 & 13.7536 & 18.1248 & - & 1.984 & 2.053 & - \\
4 & 35.1470 & 54.7939 & - & 2.081 & 2.203 & - \\
5 & 84.4066 & 178.897 & - & 2.134 & 2.316 & - \\
\hline
\end{array}
$$
\caption{\label{tt3}
Same as in the
previous table for the perturbative expansion of the $\tau$
hadronic width ($\alpha_s(m_\tau)=0.32$).}
\end{table}

The main conclusion for the radiative mass shift is that the effect
of two-loop running remains small up to the order, at which the series
has to be truncated. It leads to a less then 5\% additional increase
of the one-loop radiative correction, which can be neglected in view
of the uncertainties that have been discussed in Sect.~4. This is
reassuring, because it supports the suggestion that the mass shift is
to a large extent given by one-loop running effects, although it must
be borne in mind that by incorporating two-loop running one does
not gain control over genuine two- and higher loop corrections.

 On the other hand, the effect of two-loop running is very large
for the derivative of the vacuum polarization and the efficiency of
extending the BLM prescription beyond leading order can be doubted
in this case. Although the inclusion of two-loop running improves
the estimate of $d_2$ as compared to the exact coefficient $d_2$
(Table~\ref{tt2}), for $n>3$ the effect of two-loop running is
exceedingly large and this improvement might as well be accidental.

It is interesting to observe that for the $\tau$ hadronic width
itself (Table~\ref{tt3}),
such irregularities do not occur, which is in qualitative agreement
with the conclusion of Sect.~3.1 that the perturbative
expansion of $R_\tau$ has a smoother behaviour in intermediate
orders of perturbation theory than $Q^2 d\Pi/dQ^2$. The relative
importance of two-loop running increases with $n$ as expected from the
asymptotic $\ln n$-enhancement. Though smooth, the effect of
two-loop running is significant for $R_\tau$ and points towards
an even further increase of the cumulative effect of higher order
perturbative corrections and consequently a further decrease
of $\alpha_s(m_\tau)$ from $R_\tau$.

In Appendix C, we give exact results for the five-loop diagrams that
enter the calculation of $d_3^{\rm NLO}$ for $Q^2 d\Pi/dQ^2$.



\mysection{Conclusions}{Conclusions}

In this paper we have shown  how to deal with
higher order vacuum polarization
insertions into radiative corrections to observables
in QCD. This enterprise
has been motivated by the fact that these higher order
corrections often
lead to disturbingly large coefficients in the perturbative
expansion. This
trend is systematic, but since it can be identified, we suggest
to calculate
these corrections and separate them from the remaining ones.
The behaviour
of the remaining series should then be improved. Even
if the remaining corrections
turn out not to be small, the summation of vacuum polarization
insertions
can be motivated as a physical way of scale-setting.
In this respect it
can be considered as an extension of the scale-setting proposed
by Brodsky, Lepage and Mackenzie \cite{BRO83} and Brodsky
and Lu \cite{BRO94},
as far as two-loop running is concerned. It is worth
noting that the familiar
BLM scale typically overestimates the size of higher
order vacuum polarization
insertions. The smallness of this scale needs not
necessarily indicate
a failure of perturbation theory, unless it is
indicative of its divergence
already at two-loop order.

We would like to stress the computational ease with
which the summation of
effects of one-loop running can be performed compared
to a genuine higher
order calculation. We thus have devoted considerable space
to technical aspects,
which should allow routine implementation of this
summation. The two existing
techniques can be
summarized as follows:
The first one \cite{BEN93,BRO93,BB94,BG94} amounts to
direct evaluation of the relevant
higher order Feynman
diagrams. A convenient tool to organize such a calculation
is the Borel
transform. The result can be obtained as a certain
analytically regularized
Feynman integral plus  -- if necessary -- a subtraction
function that is
easily computed directly from a large-mass expansion.
This technique
has its limits, when applied to processes with several
scales or to physical
cross sections that are more directly calculated as a
sum of real and
virtual corrections. In these cases, we have applied a
dispersion technique \cite{BBZ94,SV94,BB94b}
(or, if one starts from the Borel transform, a Mellin
transformation) to
reduce the problem to calculation of lowest order
corrections with finite
gluon mass. It is in this representation, that summation
(up to an
accuracy set by renormalons) is most easily performed.
This summation is
usually complicated by the need to analytically
continue
the Borel transform beyond its radius of convergence
and to take
the (principal value) integral
over the Borel parameter. Our Eqs.~(\ref{rBS}) and
(\ref{rBSren}) are
immediately suited to numerical evaluation. Given this
simplicity, the
computational expense seems worth the investment even
if one could only
hope to absorb a higher order radiative correction that
has the correct sign compared to the exact one.

In the present paper, we have investigated two
quantities in more detail.
Higher order $\beta_0^n\alpha_s^{n+1}$-corrections
are indeed sizeable
for the hadronic decay width of the $\tau$ lepton,
when expressed as
a series in $\alpha_s$. We compared fixed order perturbative
approximations to $R_\tau$ with those from a partial resummation
of running coupling effects due to contour integration \cite{DIB92}
and found that in intermediate orders the first one has a smoother
behaviour, since potentially large corrections from contour
integration conspire with the divergent coefficients of
$Q^2d\Pi/dQ^2$ to produce an effective suppression of ultraviolet
renormalons (Table~\ref{tab1}). Resummation of all one-loop
running coupling effects leads to an approximate 10\% decrease
of $\alpha_s(m_\tau)$ determined from hadronic $\tau$ decays to
a central value of\footnote{As an input we have taken
$R_\tau=3.56\pm 0.03$ from \cite{Revs}, see Sect.~3.1.}

\be
\alpha_s(m_\tau)\simeq 0.29\,.
\ee

\n This shift of about
one error margin of previous analyses is not caused by a single
large next-order coefficient, but by the constructive addition
of several higher order coefficients beyond the exactly known
coefficient at order $\alpha_s^3$.

The accurate extraction of $\alpha_s$ from $\tau$-decays
relies on the absence of
$\Lambda_{\rm QCD}^2/m_\tau^2$-corrections.
Although from a purely theoretical point of view,
the situation with
respect to such terms is not conclusive, given
that the limitations
of duality are only poorly understood, we conclude
that effects
associated with summation of the divergent perturbative
expansion
should be excluded as a source of such terms,
provided one accepts
their absence in the relevant current correlation
functions in euclidian
space.

For the difference between the pole and $\MS$-renormalized
mass of a
heavy quark, we conclude that the large two-loop
correction found in
\cite{GRA90} is probably not accidental but the first one
in a rapidly
divergent series of higher order corrections dominated
indeed by one-loop
running of the coupling. For the charm quark, the
divergence is such that
the perturbative series can certainly not be improved
beyond two-loop
order. For the bottom pole mass, our estimate incorporating
one-loop
running reads

\be
\frac{\delta M_b}{M_b} = (16.3\pm 2.9\pm 1.5)\%\,,
\ee

\n which is about 30\% larger than the two-loop result
without resummation.
We expect this difference to be significant in phenomenological
applications. The first error quoted is associated with the divergence
of the series that relates the two mass definitons and is irreducible.
Numerically this error amounts to an uncertainty somewhat
larger than 100 MeV
for the mass difference, a value in between the estimates initially
reported in \cite{BIG94,BB94}. Short-distance observables containing
quark masses, when expressed in terms of pole masses will generally
exhibit large coefficients implanted by the use of this parameter. Thus,
although both mass definitions are ultimately unphysical, we expect that
the $\MS$ mass can be determined to better accuracy through perturbative
relations\footnote{This argument does not apply to pole mass differences.
The large coefficients associated with the pole mass can be understood
as a universal (flavour-independent) additive mass renormalization,
which cancels in the difference.}.

Finally, let us mention that the calculation of vacuum polarization
can never replace an exact calculation. Until this is completed, we
find it worthwhile to incorporate the systematic effects exhibited
by one-loop running. Surely, the authors are among those who await
further exact results with suspense.\\

{\bf Note added.} While this paper was written,
we were informed by M.~Neubert on work of his, which partially
overlaps with the present one. We acknowledge exchange of manuscripts
and are very grateful for the ensuing discussions which
helped to clarify
our presentation. We disagree with the conclusion of \cite{NEU95}
that the difference of various summation prescriptions for cross
sections should be interpreted as a sign and quantification of
the failure of the operator product expansion in the physical
region. As emphasized in Sect.~3.2, once one abandones Borel-summation
type prescriptions, which are distinguished by their relation to
the OPE, one can introduce $1/Q^2$-differences in euclidian
space just as well and there is no discrimination between the
euclidian and minkowskian region from the point of view of
summation of perturbation theory, given the present state of
knowledge.\\

{\bf Acknowledgements.} M.~B. would like to thank D.~Broadhurst and
A.~Pich for stimulating discussions. M.~B. is supported by the
Alexander von Humboldt-foundation.

\newpage
\begin{appendix}
\n {\Large\bf Appendices}

\mysection{Subtractions}
{Subtractions}

In general, we will also be interested in ultraviolet divergent
quantities. Examples include the difference between the pole mass
and the $\MS$-renormalized quark mass and the correlation function
of heavy-light currents in heavy quark effective theory. Denoting
a generic quantity by $R(\alpha)$, the renormalized Borel transform
has the form

\be
B[R](u) = B[R]_0(u) + S_R(u)\,,
\ee

\n where $B[R]_0(u)$ is the bare Borel transform, calculated from
diagrams without overall subtraction,
using the Borel parameter itself as a
regulator. This corresponds to analytic regularization \cite{SPE68}.
Consequently, the bare Borel transform is
singular at the origin (Recall
that the perturbative series is generated by derivatives of the
Borel transform at the origin.).
This singularity is subtracted by the function $S_R(u)$, which
depends on the renormalization scheme. In minimal subtraction schemes
$u S_R(u)$ is an entire function (at least for one chain of fermion
loops) and -- when combined with the appropriate singular piece of
$B[R]_0(u)$ to be defined below, Eq.~(\ref{defsub}) -- $S_R(u)$ yields
an unambiguous
contribution to the bubble sum

\be
S_R(\alpha) \equiv \left(-\frac{1}{\beta_0}\right)
\intl_0^\infty d u \,e^{-u/(-\beta_0\alpha)}\,
\Big(S_R(u) + \mbox{singular term in } B[R]_0(u)
\Big) \,.
\ee

\n General expressions for $S_R(u)$ in minimal subtraction schemes
have been derived in Appendix A
of \cite{BB94}. In this Appendix we outline a simple method
to obtain $S_R(u)$ and calculate $S_R(\alpha)$ in minimal subtraction
schemes.
An essential simplification comes from the
fact that one needs only the dominant large-$\lambda$ behaviour
of dimensionally regularized Feynman amplitudes, where $\lambda$
is a mass for the gluon. As an illustration of the technique we
calculate the subtractions for the correlation function of two
heavy-light currents in heavy quark effective theory.

\subsection{Calculation of $S_R(\alpha)$}

As always we assume that the quantity $R$ has been
made dimensionless, is infrared finite and
such that the
order-$\alpha$ radiative correction comes from gluon exchange. We
shall also assume that the corresponding diagrams have one loop
and are at most logarithmically ultraviolet divergent. The last
assumption is actually unnecessary, see the explicit example in
the subsequent subsection. The
dimensionally regularized order-$\alpha$ correction can then be
represented as

\be \label{co}
r_0^{\rm bare}(\eps) \,\alpha = \alpha\mu^{2\eps}\int d^d k\,
F(k,Q)\frac{1}{k^2}\,.
\ee

\n We use $d=4-2\eps$ and denote by $Q$ a set of external momenta.
For gauge-dependent quantities, we take Landau gauge. The factor
$(k_\mu k_\nu/k^2-g_{\mu\nu})$ from the gluon propagator is
included in the integrand $F(k,Q)$ and we do not write Lorentz
indices explicitly.
We define the both, dimensionally and analytically regularized
coefficient by

\be\label{dimanalytic}
\hat{r}_0^{\rm bare}(s,\eps) \alpha = \alpha\mu^{2\eps}\int d^d k\,
F(k,Q)\frac{1}{k^2}\left(-\frac{\mu^2}{k^2}\right)^s\,,
\ee

\n where $\mu$ is a subtraction scale. Note that $B[R]_0(u)=
\hat{r}_0^{\rm bare}(u,\eps=0)\,e^{-u C}$, see Eq.~(\ref{gluonprop}).

Insert $n$ fermion loops into the gluon line of all diagrams
that generate $r_0$. The fermion loop integrations can be done.
Each loop gives a factor

\bdm
-\frac{\beta_0^f}{\eps} \frac{6\Gamma(1+\eps)\Gamma(2-\eps)^2}
{\Gamma(4-2\eps)} \left(-\frac{k^2}{4\pi\mu^2}\right)^{-\eps}\,,
\edm

\n where $\beta_0^f=T/(3\pi)$ is the fermionic contribution to the
beta-function ($T=1$ in QED and $T=1/2$ in QCD). Performing the
final integration over gluon momentum $k$, the result for the
coefficient of order $\alpha^{n+1}$ can be written as

\be
r_n^{\rm bare}(\eps) = \frac{\left(\beta_0^f\right)^n}{(n+1)
(-\eps)^{n+1}}\,G(-\eps,-(n+1)\eps)
\ee
\be
G(-\eps,-(n+1)\eps) = \left((4\pi)^\eps \frac{6\Gamma(1+\eps)
\Gamma(2-\eps)^2}{\Gamma(4-2\eps)}\right)^n (n+1) (-\eps)\,
\hat{r}_0^{\rm bare}(s=n\eps,\eps)\,,
\ee

\n where Eq.~(\ref{dimanalytic}) has been used. The function $G$
introduced above coincides with the one of Appendix A of
\cite{BB94} [Note that $d=4+2\eps$ has been used in
\cite{BB94}, which motivates the signs in the arguments
of $G$ above.]. To obtain the subtraction function $S_R(u)$ in
the $\MS$ scheme,
one has to add to $r_n^{\rm bare}(\eps)$ the diagrams with
fermion loops replaced by their $\MS$ counterterms
and then compute the finite part of the sum. For any quantity $R$,
the steps
of Appendix A in \cite{BB94} for the self-energy of a
heavy quark can be repeated with the general result

\be\label{olres}
S_R(u) = \frac{\tilde{G}_0(u)}{u}\qquad \tilde{G}_0(u) =
\sum_{n=0}^\infty \frac{g_n}{n!} u^n\,,
\ee

\n where $g_n$ are the expansion coefficients of

\be G_0(-\eps) \equiv G(-\eps,-(n+1)\eps=0) = \sum_{m=0}^\infty
g_m (-\eps)^m\,.
\ee

\vspace{0.2cm}
Since $G_0$ is related to subtractions, it should be unnecessary
to calculate the full diagram, encoded in the function $G$, in
order to obtain $G_0$. We prove
that $G_0$ can be deduced from the large-mass expansion
of the dimensionally regularized one-loop coefficient
$r_0^{\rm bare}(\lambda^2,\eps)$ computed
with a gluon mass $\lambda$. That is, $1/k^2$ in Eq.~(\ref{co})
is replaced by
$1/(k^2-\lambda^2)$ (in Landau or Feynman gauge). Since the
original four-dimensional integral was logarithmically
ultraviolet divergent, the asymptotic behaviour
at large $\lambda$ is given by

\be \label{as}
r_0^{\rm bare}(\lambda^2,\eps)\stackrel{\lambda^2\rightarrow\infty}{=}
-\frac{1}{\eps}\,r_\infty(\eps) \left(\frac{\mu^2}{\lambda^2}\right)^
\eps\,.
\ee

\n This equation defines the $\lambda^2$-independent function
$r_\infty(\eps)$. Its value at $\eps=0$ is denoted by $r_\infty$. By
definition

\be
G_0(-\eps) = -\frac{1}{(4\pi)^\eps} \frac{\Gamma(4-2\eps)}{6\Gamma(
1+\eps)\Gamma(2-\eps)^2}\,\lim_{s\rightarrow -\eps}\,(s+\eps)\,
\hat{r}_0^{\rm bare}(s,\eps)\,.
\ee

\n To evaluate the limit, we use that $\hat{r}_0^{\rm bare}(s,\eps)$
is related to the one-loop coefficient with finite gluon mass
$r_0^{\rm bare}(\lambda^2,\eps)$ by a Mellin transform:

\be
\hat{r}_0^{\rm bare}(s,\eps) = -\frac{\sin\pi s}{\pi} \intl_0^\infty
\frac{d \lambda^2}{\lambda^2} \left(\frac{\lambda^2}{\mu^2}
\right)^{-s} r_0^{\rm bare}(\lambda^2,\eps)
\ee

\n This represents the analytic continuation of the analogon of
Eq.~(\ref{final1}) into the $u$-interval $]-\eps,0[$. Integrating by
parts, we get

\be
\hat{r}_0^{\rm bare}(s,\eps) = -\frac{\sin\pi s}{\pi\,(s+\eps)}
\intl_0^\infty d\lambda^2 \left(\frac{\lambda^2}{\mu^2}
\right)^{-(s+\eps)} \! f^\prime(\lambda^2,\eps)\,,
\ee

\n where $f(\lambda^2,\eps)$ is defined by dividing
$\hat{r}_0^{\rm bare}(s,\eps)$ by $(\mu^2/\lambda^2)^\eps$.
The limit is now easily taken with the result

\be
\lim_{s\rightarrow -\eps}\,(s+\eps)\,
\hat{r}_0^{\rm bare}(s,\eps) = -\frac{\sin\pi\eps}{\pi\eps}\,r_\infty
(\eps)\,.
\ee

\n This yields the final result

\be\label{gzero}
G_0(u) = \frac{1}{(4\pi)^{-u}} \frac{\Gamma(4+2 u)}{6\Gamma(
1-u)\Gamma(2+u)^2}\,\frac{\sin\pi u}{\pi u}\,r_\infty
(-u)\,.
\ee

\n The computation of the subtraction function has been reduced to
calculation of the large mass limit of one-loop corrections with
non-zero gluon mass.

\vspace*{0.2cm}
Combining Eq.~(\ref{final1}) for the bare Borel transform with
Eq.~(\ref{olres}), the renormalized Borel transform is given by

\be
B[R](u) = -\frac{\sin\pi u}{\pi u} \intl_0^\infty
d \lambda^2 \left(\frac{\lambda^2}{\mu^2} e^C
\right)^{-u} r_0^\prime(\lambda^2) + \frac{\tilde{G}_0(u)}{u}\,.
\ee

\n The derivative $r_0^\prime(\lambda^2)$ is ultraviolet finite, which
allows to put $\eps=0$. However, the integral is not yet finite at
$u=0$ and the above expression is not suited to take derivatives at
$u=0$. Since, by Eq.~(\ref{as}), $r_0^\prime(\lambda^2)=r_\infty/\lambda^2$
for large $\lambda^2$, we obtain

\bea\label{bofin2}
B[R](u) &=& -\frac{\sin\pi u}{\pi u} \intl_0^\infty
d \lambda^2 \left(\frac{\lambda^2}{\mu^2} e^C
\right)^{-u} \left[r_0^\prime(\lambda^2)-\frac{r_\infty}{\lambda^2}
\,\Theta(\lambda^2-\mu^2 e^{-C})\right]\nonumber\\
&&\,+ \frac{1}{u}\left(\tilde{G}_0(u)-r_\infty\,\frac{\sin\pi u}
{\pi u}\right)\,.
\eea

\n In this form the integral exists for $u=0$ and the second line is
finite at $u=0$. We note that this expression is equivalent to, but
slightly different from the one given in \cite{BB94b}. In the
present form $\mu^2$ and $C$ appear only in their natural combination
$\mu^2 e^{-C}$.

To obtain the bubble sum, all steps that lead from Eq.~(\ref{final1})
to Eq.~(\ref{rBS}) can now be repeated on the first line of
Eq.~(\ref{bofin2}). The Borel integral of the second line is given by

\bea \label{defsub}
S_R(\alpha) &\equiv& \left(-\frac{1}{\beta_0}\right)
\intl_0^\infty d u \,e^{-u/(-\beta_0\alpha)}\,\frac{1}{u}
\left(\tilde{G}_0(u)-r_\infty\,\frac{\sin\pi u}
{\pi u}\,e^{-u C}\right)\\
&=& \left(-\frac{1}{\beta_0}\right)\intl_0^{-\beta_0\alpha}
\frac{d u}{u}\left(G_0(u)-r_\infty\right)-\frac{r_\infty}{\beta_0}
\left[\frac{\arctan(\pi\beta_0\alpha)}{\pi\beta_0\alpha}+\frac{1}{2}
\ln\left(1+\pi^2\beta_0^2\alpha^2\right) - 1\right]\nonumber
\,.
\eea

\n Thus, in case subtractions are required beyond coupling renormalization,
Eq.~(\ref{rBS}) is replaced by

\bea\label{rBSren2}
r_0 a_s M_\infty(a_s) &=&
\int_0^\infty d\lambda^2\, \Phi(\lambda^2)\,\left(r'_0(\lambda^2) -
\frac{r_\infty}{\lambda^2}\,\Theta(\lambda^2-\mu^2 e^{-C})\right)
+[r_0(\lambda_L^2)-r_0(0)]\nonumber\\
&&\hspace*{-2cm}
\,+ \intl_0^{\,a_s}\frac{d u}{u}\left(G_0(u)-r_\infty\right) + r_\infty
\left[\frac{\arctan(\pi a_s)}{\pi a_s}+\frac{1}{2} \ln\left(
1+\pi^2 a_s^2\right) - 1\right]\,,
\eea

\n with $G_0(u)$ given by Eq.~(\ref{gzero}).

\subsection{A sample calculation}

As a non-trivial example, we calculate the subtraction function for the
correlation function of heavy-light currents in heavy quark effective
theory,

\be\label{CF2}
\Pi_5(\omega)\,=\,i\int\dd^4 x\,e^{i\omega (v\cdot x)}\,
\langle 0|T\{j_5^\dagger(x) j_5(0)\}|0\rangle\qquad
j_5(x) = \bar{h}_v(x) i\gamma_5 q(x)\,.
\ee

\begin{figure}[t]
   \vspace{-1cm}
   \epsfysize=23.8cm
   \epsfxsize=17cm
   \centerline{\epsffile{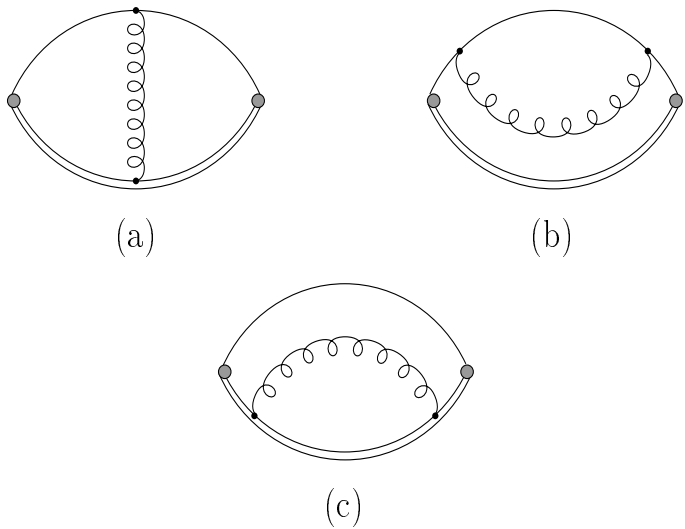}}
   \vspace*{-14cm}
\caption{\label{hqetI} Radiative corrections to the correlation
function of heavy-light currents. The shaded circle represents a
current insertion with momentum $q$ and the double line denotes
a heavy quark propagator.}
\end{figure}

\n The corresponding bare Borel transform has been given in \cite{BB94}.
At first sight the method exposed in the previous subsection appears
inapplicable, because the diagrams to be considered have two loops, see
Fig.~\ref{hqetI}, and, because the correlation function is quadratically
divergent, one also needs subtractions for the two-loop diagram itself.
To eliminate these, we shall consider the third derivative

\be
D(\omega)\equiv \omega\,\frac{\dd^3\Pi_5(\omega)}{\dd\omega^3}\,.
\ee

\n Then all subtractions originate from divergent one-loop subdiagrams,
which are subsequently inserted in the lowest order one-loop diagram
for $\Pi_5(\omega)$.
We remind the reader that the heavy-light current in the effective theory
is not conserved. We treat the three diagrams in turn and take Landau
gauge.\\

Let us start with diagram (a) and investigate the one-loop correction to
the heavy-light vertex with non-zero gluon mass. The numerator of the
corresponding Feynman integrand is proportional to

\be \gamma^\rho (\!\not\! q-\!\not\! k) \gamma_5 v^\tau (k^2 g_{\rho\tau}
- k_\rho k_\tau)\,,
\ee

\n where $v$ is the velocity of the heavy quark and the momentum
assignments to the lines of the diagram are evident. Since the contribution
proportional to $\lambda^{-2\eps}$ required by Eq.~(\ref{as}) comes
from $k\sim\lambda\gg q$, we can drop $\not\!\! q$ in the above
expression and the integral is simplified to

\be
\int\frac{d^d k}{(2\pi)^d} \frac{\!\not\! v\!\not\! k-v\cdot k}{
(q-k)^2 \,v\cdot(q^\prime-k) \,(k^2-\lambda^2)}\,.
\ee

\n The corresponding integral with numerator replaced by $k_\mu$ has
structure $F_1 v_\mu+F_2 q_\mu+F_3 q_\mu^\prime$. By dimensional counting,
only $F_1$ can behave as $\lambda^{-2\eps}$ for large $\lambda^2$,
the other structures being suppressed by powers of $\lambda$. But then
the potential term $\lambda^{-2\eps}$ is proportional to
$\!\not\! v\!\not\! v-v^2 = 0$ and therefore $r_\infty(\eps)$ vanishes
identically for this diagram (This property is specific to Landau gauge.).

For diagram (b) we need the self-energy insertion of a light (massless)
quark, given by

\be
-C_F g^2\mu^{2\eps} \int\frac{d^d k}{(2\pi)^d} \frac{\gamma^\rho
(\!\not\! p-\!\not\! k)\gamma^\tau}{(p-k)^2 (k^2-\lambda^2)}\,\left[
g_{\rho\tau}-\frac{k_\rho k_\tau}{k^2}\right]\,,
\ee

\n where $p$ is the external momentum for the subdiagram. Again the
term involving $\lambda^{-2\eps}$ in the large $\lambda^2$-expansion
comes from $k\sim\lambda\gg p$, which allows to expand the propagator
$1/(p-k)^2$. The integrand simplifies to

\be
\frac{\gamma^\rho
(\!\not\! p-\!\not\! k)\gamma^\tau}{(p-k)^2 (k^2-\lambda^2)}\,\left[
g_{\rho\tau}-\frac{k_\rho k_\tau}{k^2}\right]\longrightarrow
\frac{\eps \,(3-2 \eps)}{2-\eps} \frac{\!\not\! p}{k^2 (k^2-\lambda^2)}
+ O\left(\frac{1}{k^5}\right)\,,
\ee

\n where the neglected terms do not produce a $\lambda^{-2\eps}$
contribution. The result is

\be
(-i)\!\not\! p \left(\frac{4\pi\mu^2}{\lambda^2}\right)^\eps
 \frac{C_F\alpha}{
4\pi} \frac{(3-2 \eps)\Gamma(1+\eps)}{(1-\eps)(2-\eps)}\,.
\ee

\n Integration of the final quark loop is straightforward, and taking
three
subtractions we obtain

\be
G_0^{(b)}(u) = \frac{C_F N_c}{4\pi^3}\left[-\frac{u}{6}(3+2 u) \frac{
\Gamma(4+2 u)}{\Gamma(1-u)\Gamma(2+u)^2\Gamma(3+u)}\right]\,.
\ee

Turning to the self-energy of a heavy quark in heavy quark effective
theory in diagram (c), we encounter another
apparent obstacle to applying the
technique of the previous subsection: The subdiagram is linearly
ultraviolet divergent, and -- at large $\lambda^2$ -- behaves like
$\lambda^{1-2\eps}$. The remedy is simply to ignore this term and
pick up the term $\lambda^{-2\eps}$, since power divergences are
discarded in minimal subtraction schemes. The large mass expansion of
the heavy quark self-energy is

\be
(-i)\frac{1+\!\not\! v}{2}\left(\frac{4\pi\mu^2}{\lambda^2}\right)^\eps
\frac{C_F\alpha}{4\pi} \left[\Gamma\left(\frac{1}{2}\right)\Gamma\left(
\eps-\frac{1}{2}\right)\,\lambda+\frac{(3-2\eps)\Gamma(\eps)}{1-\eps}\,
v\cdot p + O\left(\frac{v\cdot p}{\lambda}\right)\right]\,.
\ee

\n Discarding the first term and inserting the remaining one into
the quark loop, we get

\be
G_0^{(c)}(u) = \frac{C_F N_c}{4\pi^3}\left[\frac{1}{6} (3+2 u) \frac{
\Gamma(4+2 u)}{\Gamma(1-u)\Gamma(2+u)^3}\right]\,.
\ee

\n Adding the contributions from all diagrams, the result is

\be
G_0(u) = \frac{C_F N_c}{4\pi^3}\left[\frac{1}{3} (3+2 u) \frac{
\Gamma(4+2 u)}{\Gamma(1-u)\Gamma(2+u)^2\Gamma(3+u)}\right]\,.
\ee


\mysection{Radiative corections to $R_{e^+e^-}$ and $R_\tau$ with
           finite gluon mass}
          {Radiative corections to $R_{e^+e^-}$ and $R_\tau$ with
           finite gluon mass}

We write the lowest order radiative corrections to the
total cross section of $e^+e^-$ annihilation and the hadronic $\tau$
decay width with finite gluon mass $\lambda$ as

\bea
R_{e^+e^-} &=& 3\left[1+\alpha_s(s)\left\{r_{e^+e^-}^{virt}(y)+\Theta(1-y)\,
r_{e^+e^-}^{real}(y)\right\}\right]\,,\\
R_\tau &=& 3\left[1+\alpha_s(m_\tau)\left\{r_\tau^{virt}(y)+\Theta(1-y)\,
r_\tau^{real}(y)\right\}\right]\,,
\eea

\n where $y=\lambda^2/s$ or $y=\lambda^2/m_\tau^2$, respectively.
Quarks are taken as massless.
$r^{virt}(y)$ are virtual and $r^{real}(y)$ real gluon emission
corrections:

\bea
r_{e^+e^-}^{virt}(y) &=&\frac{2}{3\pi}\Bigg[2(1+y)^2\Bigg\{-\frac{1}{2}
\ln^2 y +\ln y \ln(1+y) +\frac{\pi^2}{6}+ \mbox{Li}_2(-y)\Bigg\}
\nonumber\\
&&{} -\frac{7}{2}-2y-3\ln y -2y \ln y \Bigg]\\
r_{e^+e^-}^{real}(y) &=&\frac{2}{3\pi}\Bigg[2(1+y)^2\Bigg\{\frac{1}{2}
\ln^2 y -2\ln y \ln(1+y) -\frac{\pi^2}{6}-2 \mbox{Li}_2(-y)\Bigg\}
\nonumber\\
&&{} +5-5y^2+3\ln y +4y \ln y +3 y^2\ln y\Bigg]\,,
\eea

\n and

\bea
r_\tau^{virt}(y)&=&\frac{1}{324\pi}\Bigg[-2577+72 \pi^2+(2392-240 \pi^2)\,
y+(828-432 \pi^2)\,y^2+144 (1-\pi^2)\,y^3\nonumber\\
&&\hspace*{-1.5cm} -24 \pi^2 y^4-1332\ln y+
(2208+288\pi^2)\,y\ln y+792 y^2\ln y+144 y^3\ln y-216 \ln^2 y
\nonumber\\[0.3cm]
&&\hspace*{-1.5cm}
+720 y \ln^2 y+1296 y^2\ln^2 y+432 y^3\ln^2 y+72 y^4\ln^2 y+864 y\ln^3 y
\nonumber\\[0.3cm]
&&\hspace*{-1.5cm} +
(432-1440 y-2592 y^2-864 y^3-144 y^4)\,(\ln y\ln(1+y)+\mbox{Li}_2(-y))
\nonumber\\
&&\hspace*{-1.5cm}
+
1728 y \ln y\,\mbox{Li}_2(-y)-3456 y\,\mbox{Li}_3\left(-\frac{1}{y}\right)
\Bigg]\\
r_\tau^{real}(y)&=&\frac{1}{324\pi}\Bigg[2901-72 \pi^2+(-9736+240 \pi^2+5184
\zeta(3))\,
y+(6120+432 \pi^2)\,y^2\nonumber\\
&&\hspace*{-1.5cm} +(456+144\pi^2)\,y^3+(259+24 \pi^2)\, y^4+1332\ln y-
(1776+864\pi^2)\,y\ln y-4176 y^2\ln y\nonumber\\[0.3cm]
&&\hspace*{-1.5cm} -288 y^3\ln y+216 \ln^2 y
-720 y \ln^2 y-1296 y^2\ln^2 y-864 y^3\ln^2 y-144 y^4\ln^2 y
\nonumber\\[0.3cm]
&&\hspace*{-1.5cm} -1440 y\ln^3 y+
(-864+2880 y+5184 y^2+1728 y^3+288 y^4)\,(\ln y\ln(1+y)+\mbox{Li}_2(-y))
\nonumber\\
&&\hspace*{-1.5cm}
-
3456 y \ln y \,\mbox{Li}_2(-y)+6912 y\,\mbox{Li}_3\left(-\frac{1}{y}\right)
\Bigg]
\eea


\mysection{Abelian five-loop diagrams to the hadronic vacuum polarization}
{Abelian five-loop diagrams to the hadronic vacuum polarization}

\begin{figure}[t]
   \vspace{-1cm}
   \epsfysize=24cm
   \epsfxsize=20cm
   \centerline{\epsffile{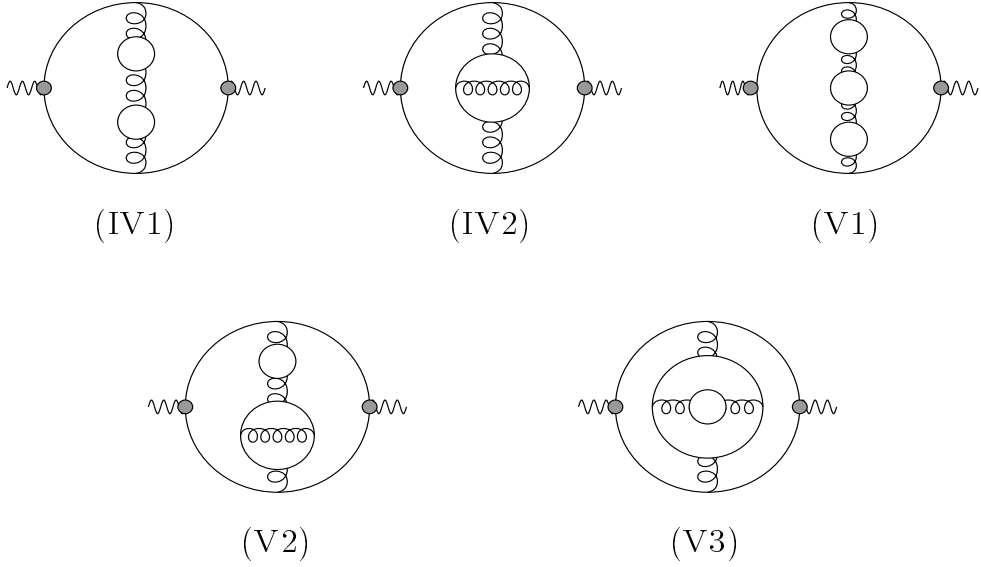}}
   \vspace*{-13.8cm}
\caption{Representative abelian four- and five-loop
diagrams for the hadronic vacuum polarization. It is understood that
all self-energy type diagrams are added to each class of diagrams.}
\label{fiveloop}
\end{figure}

\phantom{\ref{fiveloop}}

In this Appendix we collect the abelian five-loop diagrams to
the hadronic vacuum polarization that are
included in the scale-setting in next-to-leading order as described in
Sect.~5. For completeness, we also list the four-loop diagrams. Numerical
values are given for

\be Q^2\frac{d\Pi(Q^2)}{dQ^2} = \frac{1}{4\pi^2}\sum_{n=0}^\infty
D_{n+1}\left(\frac{\alpha(Q)}{\pi}\right)^n\,.
\ee

\n The diagrams are shown in Fig.~\ref{fiveloop}. Each diagram shown
stands for the class of diagrams obtained, when one adds the diagrams
where gluons form self-energy type insertions rather than exchange type
topologies. Diagram (V2) also includes the symmetric diagram, where
the two inner loops are interchanged. We obtain:

\bea
D_4^{\mbox{\tiny (IV1)}} &=& C_F (T N_f)^2\left[\frac{151}{54}-
\frac{19}{9} \zeta(3)
\right]\\
D_4^{\mbox{\tiny (IV2)}} &=& C_F (C_F T N_f)\left[-\frac{101}{64}+
\frac{3}{2} \zeta(3)
\right]\\[0.2cm]
D_5^{\mbox{\tiny (V1)}} &=& C_F (T N_f)^3
\left[-\frac{6131}{972}+\frac{203}{54}\zeta(3)
+\frac{5}{3} \zeta(5)\right]\\
D_5^{\mbox{\tiny (V2)}} &=& C_F (T N_f) (C_f T N_f)
\left[\frac{3571}{576}-\frac{59}{8}
\zeta(3)+ 2 \zeta(3)^2\right]\\
D_5^{\mbox{\tiny (V3)}} &=& C_F (T N_f) (C_f T N_f)
\left[\frac{10199}{3456}-\frac{7}
{2}\zeta(3)+\zeta(3)^2\right]
\eea

\n Here $T=1/2$,
$C_F=4/3$ for $SU(3)$ with fermions in the fundamental
representation and $T=1$, $C_F=1$ for $U(1)$.
The sum of these terms yields $d_2^{\rm NLO}$ and $d_3^{\rm NLO}$
in QED (after proper normalization).

\end{appendix}



\end{document}